\documentclass[11pt]{article}
\pdfoutput=1 
\usepackage{savesym}
\usepackage[makeroom]{cancel}
\usepackage{slashed, tikz}
\usepackage{graphicx,array}
\usepackage{booktabs}
\usepackage{amssymb}
\usepackage{dsfont}
\usepackage{stackrel}
\usepackage{setspace}
\usepackage[normalem]{ulem}
\usepackage[T1]{fontenc} 
\usepackage{xcolor}
\usepackage{changepage}
\usepackage[most]{tcolorbox}
\usepackage{float}
\usepackage{caption}
\usepackage{subfigure}
\usepackage{jheppub} 
\usepackage{stackrel}
\usepackage{hyperref}
\usepackage{physics}
\usepackage{tensor}
\usepackage[compat=1.1.0]{tikz-feynman}
\usepackage{graphicx}
\usepackage{slashed}
\usepackage{multirow}
\usepackage{feynmf}
\tikzfeynmanset{warn luatex=false}

\usepackage[T1]{fontenc}
\usepackage[utf8]{inputenc}

\makeatletter\g@addto@macro\bfseries{\boldmath}\makeatother
\DeclareMathSymbol{\shortminus}{\mathbin}{AMSa}{"39}
\def\figureautorefname~#1\null{fig.\,#1\null}
\def\equationautorefname~#1\null{eq.\,(#1)\null}

\makeatletter
\DeclareFontFamily{OMX}{MnSymbolE}{}
\DeclareSymbolFont{MnLargeSymbols}{OMX}{MnSymbolE}{m}{n}
\SetSymbolFont{MnLargeSymbols}{bold}{OMX}{MnSymbolE}{b}{n}
\DeclareFontShape{OMX}{MnSymbolE}{m}{n}{
    <-6>  MnSymbolE5
   <6-7>  MnSymbolE6
   <7-8>  MnSymbolE7
   <8-9>  MnSymbolE8
   <9-10> MnSymbolE9
  <10-12> MnSymbolE10
  <12->   MnSymbolE12
}{}
\DeclareFontShape{OMX}{MnSymbolE}{b}{n}{
    <-6>  MnSymbolE-Bold5
   <6-7>  MnSymbolE-Bold6
   <7-8>  MnSymbolE-Bold7
   <8-9>  MnSymbolE-Bold8
   <9-10> MnSymbolE-Bold9
  <10-12> MnSymbolE-Bold10
  <12->   MnSymbolE-Bold12
}{}

\let\llangle\@undefined
\let\rrangle\@undefined
\DeclareMathDelimiter{\llangle}{\mathopen}%
                     {MnLargeSymbols}{'164}{MnLargeSymbols}{'164}
\DeclareMathDelimiter{\rrangle}{\mathclose}%
                     {MnLargeSymbols}{'171}{MnLargeSymbols}{'171}
\makeatother

\definecolor{orange}{rgb}{1,0.5,0}
\definecolor{brown}{rgb}{0.59, 0.29, 0.0}

\definecolor{note^fontcolor}{rgb}{0.80078125, 0.80078125, 0.80078125}

\newtcbox{\mymath}[1][]{%
    nobeforeafter, math upper, tcbox raise base,
    enhanced, colframe=blue!30!black,
    colback=white, boxrule=1pt,
    #1}

\def\beq{\begin{equation}}
\def\eeq{\end{equation}}
\def\bea{\begin{eqnarray}}
\def\eea{\end{eqnarray}}

\makeatletter
\newcommand*{\Relbarfill@}{\arrowfill@\Relbar\Relbar\Relbar}
\newcommand*{\xeq}[2][]{\ext@arrow 0055\Relbarfill@{#1}{#2}}
\makeatother

\newcommand{\colvec}[1]{\left(\begin{array}{c}#1\end{array}\right)}

\newcommand{\colmatt}[1]{\left(\begin{array}{cc}#1\end{array}\right)}

\newcommand{\colmatf}[1]{\left(\begin{array}{cccc}#1\end{array}\right)}

\definecolor{MHBlue}{RGB}{156, 156, 252}

\def\e{\epsilon}
\def\f{\Phi}

\def\lpar#1#2#3#4{\rlap{\raise#3\hbox{$\hskip#4#1\left\{\mbox{\phantom{\rule[0mm]{0mm}{#2}}}\right.$}}}
\def\rpar#1#2#3#4{\rlap{\raise#3\hbox{$\hskip#4\left\}#1\mbox{\phantom{\rule[0mm]{0mm}{#2}}}\right.$}}}
\renewcommand{\subsubsection}[1]{\addtocounter{subsubsection}{1}
\par\nobreak
\medskip
\nobreak
\noindent{\it \thesubsubsection.  #1 }
\par\nobreak\medskip\nobreak}

\definecolor{MHPink}{RGB}{255, 227, 253}
\definecolor{MHOrange}{RGB}{255, 232, 194}
\definecolor{MHCyan}{RGB}{209, 241, 255}
\definecolor{MHGreen}{RGB}{222, 255, 231}

\newcommand{\nn}{\nonumber}

\title{\boldmath Abelian Instantons and Monopole Scattering}

\author[a]{Csaba Cs\'aki,}
\affiliation[a]{Department of Physics, LEPP, Cornell University, Ithaca, NY 14853, USA}
\emailAdd{csaki@cornell.edu}

\author[b]{Rotem Ovadia,}
\affiliation[b]{Racah Institute of Physics, Hebrew University of Jerusalem, Jerusalem 91904, Israel}
\emailAdd{rotemov@gmail.com
}
\author[b]{Ofri Telem,}
\emailAdd{t10ofrit@gmail.com}

\author[c]{John Terning}
\emailAdd{jterning@gmail.com}
\affiliation[c]{QMAP, Department of Physics, University of California, Davis, CA 95616, USA}

\author[d]{and Shimon Yankielowicz}
\affiliation[d]{The Raymond and Beverly Sackler School of Physics and Astronomy,
Tel Aviv University, Ramat Aviv 69978, Israel}
\emailAdd{shimonya@tauex.tau.ac.il}

\begin{abstract}
{It is usually assumed that $4D$ instantons can only arise in non-Abelian theories. In this paper we re-examine this conventional wisdom by explicitly constructing instantons in an Abelian gauge theory: $QED_4$ with $N_f$ flavors of Dirac fermions, in the background of a Dirac monopole. This is the low-energy effective field theory for fermions interacting with a 't~Hooft-Polyakov monopole, in the limit where the monopole is infinitely heavy (hence pointlike) and static. 
This theory, whose non-topological sectors were studied by Rubakov and Callan, has a far richer structure than previously explored. We show how to calculate the topological instanton number, demonstrate the existence of 't~Hooft zero modes localized around such instantons, and show how instantons in the path integral provide the underlying mechanism for the Callan-Rubakov process: monopole-catalyzed baryon decay with a cross section that saturates the unitarity bound. Our computation relies on correctly identifying the relevant $2D$ EFT for monopole catalysis as \textit{Axial} $QED_2$ in an effective $AdS_2$ metric.} 
\end{abstract}

\begin{document}

\maketitle

\section{Introduction}
Instantons are conventionally associated with winding configurations in non-Abelian pure Yang-Mills (YM) theories, where they appear as finite action solutions of the Euclidean equations of motion (EOM) in the vacuum. Such configurations are absent in $U(1)$ theories, and so it is usually assumed that no $4D$ $U(1)$ configuration can have a non-vanishing winding number.
In this paper we provide a counterexample to this conventional wisdom by studying $QED_4$ in the background of a magnetic monopole. As we will show, this theory does, in fact, support winding configurations which we call \textit{Abelian instantons}. In these configurations, the winding in the $(\theta,\varphi)$ directions comes from the field of the monopole, while the winding in the $(t,r)$ directions is the result of an \textit{Abelian vortex}. Interestingly, the $4D$ instanton number is the product of an integer and a half-integer: (a) the integer $2q$, which is twice the monopole flux integrated over the $(\theta, \phi)$ directions (i.e. twice the monopole charge); and (b)  
the half-integer $n/2$, which is the winding number of the Abelian vortex on the half-plane $(t,r\geq 0)$ with an ``electric conductor'' boundary condition at $r=0$. The latter is quantized in half-integer units as it can always be ``doubled'' into an integer vortex on the full $\mathbb{R}^2$ spanned by $(t,r)$ (not just the $r\geq 0$ half plane). 

Some of the properties of Abelian instantons are analogous to their usual non-Abelian cousins, however some aspects are quite different. 

\begin{itemize}

\item Similarly to the non-Abelian case, Abelian instantons possess fermionic zero modes, leading to the generation of 't~Hooft vertices in the path integral.

\item Most notably, our instanton number is half-integer compared to the integer instanton number of standard non-Abelian instantons. This is consistent with the topology effectively being $\mathbb{R}^4-\{{\rm 't\,Hooft\,line}\}$ rather than $\mathbb{R}^4$ (more precisely $S^4$ for standard non-Abelian instantons). Furthermore, the number of zero modes is twice the instanton number, i.e. an integer.

\item Unlike non-Abelian instantons, the vortex part of Abelian instantons is not a vacuum solution of the Euclidean EOMs. Rather, it is the saddle-point of the gauge path integral \textit{in the presence of external sources}. The latter appears in correlators over electrically charged fermions in a monopole background. Moreover, Abelian instantons have no moduli space and no collective coordinates, as they are ``pinned'' to the sources that generate them.

\item While the total action (including the static monopole background) is infinite, the contribution to the action from the additional vortex (on top of the action of the monopole background) is finite, and in fact approaches zero as the sources are moved to infinity. Consequently, there is no exponential suppression of monopole catalyzed scattering processes \cite{Kazama:1976fm, Rubakov:1982fp, Callan:1982ah, Callan:1982au, Nair:1983ps, Craigie:1984pc, Affleck:1993np}. 
\end{itemize}

The appearance and properties of the 2D vortex are quite interesting. To understand its origin, it is helpful to keep only the lowest partial waves of the photon and the fermions in the monopole background, as these are the only states relevant for vortex generation. The resulting EFT turns out to be axial $QED_2$ (a.k.a the Schwinger model), in an effective AdS$_2$ metric. It is well-known that the photon in the Schwinger model becomes massive through a fermion loop effect \cite{Schwinger:1962tp}, which is also the case in ${AdS}_2$. From the 4D point of view, the monopole background explicitly breaks the classical $4D$ conformal symmetry $SO(3,2)$ to $SL(2,R)\times SO(3)$, the isometry of $AdS_2\times S_2$ (see, e.g. \cite{Aharony:2023amq}). This breaking allows for a non-local and gauge-invariant photon mass term, which is generated at one loop. Due to this photon mass generation mechanism, vortex configurations exist even in the absence of any scalar fields that could Higgs the $U(1)$ \cite{Nielsen:1977aw}. Instead of a scalar it is the fermion field insertions in the correlator that source the vortex. 

There are several important physical consequences of our results. Most importantly, we clarify the on-going confusion regarding the fermion boundary condition at the location of the monopole from first principles. As already pointed out in \cite{Rubakov:1982fp}, the only boundary condition consistent with the 't~Hooft Polyakov UV completion of the monopole is a \textit{charge depositing} one, which conserves a $SU(N_F)$ global symmetry and makes the boundary variation of the fermions vanish \cite{Besson:1980vu, Callan:1982ah,Callan:1982au, Lam:1984sm, Isler:1987xn}. This boundary condition is essential to the counting of the fermionic zero modes, and the form of the 't~Hooft vertex, but it does not lead to charge deposition on the monopole at low energies \cite{Rubakov:1982fp,Callan:1982ah,Callan:1982au}. Instead, it allows for the appearance of Abelian instantons in monopole catalysis correlators. The latter violate chirality and baryon number, but conserve charge.

In this paper we perform the full Euclidean path integral calculation of multifermion correlators in monopole-background $QED_4$. These correlators have been calculated in the past either (a) indirectly using cluster decomposition in a non-winding sector of the theory \cite{Rubakov:1982fp,Callan:1982ah,Callan:1982au,Craigie:1984pc,Nair:1983ps}; or (b)
using models with a prescribed boundary condition and no dynamical photon\footnote{In some of these there is a dynamical photon, but it does not play a key role in the catalysis.} \cite{Polchinski:1984uw, Affleck:1993np, Maldacena:1995pqn, Smith:2019jnh, Smith:2020nuf, vanBeest:2023dbu,vanBeest:2023mbs}. In all of our monopole catalysis correlators, our full EFT correlators are identical to the ones of the ``Dyon Boundary Condition'' (DBC) model of \cite{Affleck:1993np, Maldacena:1995pqn, Smith:2019jnh, Smith:2020nuf, Hamada:2022eiv,vanBeest:2023dbu,vanBeest:2023mbs} \footnote{By monopole catalysis correlators we mean $N\rightarrow N$ processes like Callan-Rubakov, \textit{not} the ``Monopole Unitarity Puzzle''.}. This is not surprising, since all of these correlators are fixed to their lowest-partial-wave unitarity bound, by the combined requirement of angular momentum and charge conservation \cite{Csaki:2020inw, Csaki:2020yei}. The extra insight we provide in this paper is \textit{how} the full QFT is able to saturate these angular momentum and charge conservation requirements; the key is the appearance of Abelian instantons. 

One immediate application of our explicit path integral formalism is the derivation of the correlators responsible for Kazama-Yang-Goldhaber \cite{Kazama:1976fm} (KYG) helicity flip process. This process is the result of a 't~Hooft vertex with $2$ fermionic legs. As we will see, the number of legs on the 't~Hooft vertex is $2|q n| N_f$, where $q$ is the pairwise helicity \cite{Csaki:2020inw, Csaki:2020yei, Csaki:2022tvb, Mouland:2024zgk} of the charge-monopole system, $n/2$ is the vortex number and $N_f$ is the number of (Dirac) fermion flavors. For a single flavor and $|q|=1/2$, the KYG helicity flip is induced by an $n/2=1$ vortex in the correlator. We also derive a helicity-flip correlator also in the $N_f=2,\,|q|=1/2$ theory, where it is driven by an $n/2=1/2$ vortex.

Finally, we would like to note, that while the presence of such Abelian instantons has been indirectly inferred in \cite{Rubakov:1982fp, Callan:1982ah, Callan:1982au, Nair:1983ps, Craigie:1984pc} using cluster decomposition arguments, we present them here in their bare form with full, non-trivial winding configurations. From another angle, the interplay between magnetic monopoles and abelian instanton densities has been thoroughly investigated in \cite{Heidenreich:2020pkc}, showing that the latter imply the existence of a dyon collective coordinate on the monopole. To the best of our knowledge, the only reference to explicitly consider Abelian vortices as the origin of monopole catalysis can be found in the exhaustive and underrated lectures by Craigie \cite{Craigie:1985ti}. It did so in an effective $2D$ theory without the uplift to $4D$ Abelian instantons, and without computing the general path integral over winding sectors. Furthermore, we think that \cite{Craigie:1985ti}, as well as the original \cite{Rubakov:1982fp}, might have included a factor of $2$ mistake in the expression for their vortex number which made it look like an integer, rather than a half-integer. We expect that our investigation of the winding sectors of monopole-back ground $QED_4$ may eventually allow us to fully resolve the remaining paradoxes in the field of magnetic scattering/monopole catalysis \cite{Callan:1983tm}.

Our paper is structured as follows. In Section~\ref{sec:Abelian Instantons} we investigate the path integral for fermions \textit{as a functional} of a gauge field configuration involving the field of a Dirac monopole and a vortex in $(t,r)$. In this section we do not yet assume that the vortex is a saddle point of the gauge path integral. Neglecting irrelevant higher partial waves for the fermions, we evaluate the fermionic path integral and show that it involves fermionic zero modes and 't~Hooft vertices. In Section~\ref{sec:2D reduction} we go one step further and consider the full path integral over the gauge field and the fermions. Keeping only the relevant s-wave for the gauge field, we reduce the path integral to a Gaussian integral, obtaining the true EFT for monopole catalysis including its topologically non-trivial sectors. Our results for monopole catalysis correlators are presented in Section~\ref{sec:corr} -- and they all saturate their respective unitarity bounds in the lowest partial wave. In Section~\ref{sec:DBC} we explicitly show the agreement between our result for the Callan-Rubakov process and the amplitude derived within the DBC model. The agreement between the two is not surprising, since monopole catalysis correlators are uniquely fixed by angular momentum and charge conservation \cite{Csaki:2020inw, Csaki:2020yei}. Finally, we conclude in Section~\ref{sec:conc}, where we emphasize some of the potential implications of our investigations to the ``Monopole Unitarity Puzzle'' \cite{Callan:1983tm}, but we leave a full explication for future work.  We also include some useful appendices: Appendix~\ref{app:ads2} provides a basic guide for QFT in AdS$_2$, while Appendix~\ref{app:chvbc} is a derivation of the fermionic boundary condition at $r=0$ from the 't~Hooft-Polyakov UV completion of our Dirac monopole. Finally, Appendix~\ref{app:partial wave decomp} contains a detailed Kaluza-Klein (KK) decomposition of our theory over the $S^2$ in $(\theta,\varphi)$, or in other words, the partial wave decomposition in the monopole background.

\section{Abelian Instantons around a Dirac Monopole}\label{sec:Abelian Instantons}
%
It is well known that 4 dimensional non-Abelian gauge theory allows for field configurations with non-zero \textit{2nd Chern number}---or in physics parlance \textit{instanton}, or \textit{winding} number--- 
\begin{eqnarray}\label{eq:ch2non}
    Ch_2=\frac{e^2}{16\pi^2}\int d^4x\,{\rm tr}\left[F_{\mu\nu} \tilde{F}^{\mu\nu}\right]\,,
\end{eqnarray}
where $Ch_2$ is an integer. These field configurations, known as instantons, extremize the Yang-Mills path integral. When the non-Abelian gauge theory is coupled to fermions, the latter acquire (hypersphere)-normalizable\footnote{By \textit{hypersphere-normalizable}, we mean $\int d^4x\,|\psi|^2$ is at most log divergent, as is the case for zero modes of $2D$ vortices \cite{Nielsen:1976hs,Nielsen:1977aw,Hortacsu:1979fg}. Formally, we would need to IR-regulate the problem by compactifying it to $S^4$ (along the lines of \cite{tHooft_Instantons}) to regulate this log divergence, and then show that all observables are independent of the IR cutoff. We skip this formality here, as it is inconsequential to our results. We thank David Tong for turning our attention to the normalization of our zero-modes.} zero mode solutions to the Dirac equations in the instanton background:
\begin{eqnarray}
    \slashed{D}\psi=0\,.
\end{eqnarray}
These zero modes are chiral, and their number is constrained by the Atiyah-Singer index theorem \cite{Atiyah:1963zz}:
\begin{eqnarray}\label{eq:AS theorem}
    Ch_2=n_+-n_-\,,
\end{eqnarray}
where $n_+$ ($n_-$) is the number of solutions for $\slashed{D}\psi=0$ ($\slashed{D}^\dagger\psi=0$). The presence of fermionic zero modes in the path integral leads to far-reaching consequences---in particular to the non-perturbative generation of a \textit{'t~Hooft} vertex, an interaction vertex linking all fermions in the theory.

Using Stokes' theorem, one usually relates \eqref{eq:ch2non} to the integral of the Chern-Simons current over the $S^3$ at Euclidean spacetime infinity:
\begin{eqnarray}\label{eq:ch2nonCS}
    Ch_2=-\frac{e^2}{8\pi^2}\int_{S_3} {\rm tr}\,\left(A\wedge {\rm d} A + \frac{2e}{3}A\wedge A\wedge A\right)\,,
\end{eqnarray}
where\footnote{The $i$ factor is a convention, see \cite{Nakahara:2003nw}.} $A\equiv iA_{\mu}dx^\mu$. For non-Abelian BPST instantons, the first term vanishes, as $dA$ vanishes at spacetime infinity faster than $1/r^{2}$. The second term then gets integer values corresponding to elements of the homotopy group $\pi_3(G)$, which maps the $S_3$ at spacetime infinity to the gauge group $G$. Since the homotopy group $\pi_3(G)$ is nontrivial for $G\supseteq SU(2)$, one conventionally infers that winding field configurations with nonzero $Ch_2$ are necessarily non-Abelian. The caveat in this logic is the exclusion of gauge configurations which approach pure gauge at spacetime infinity with field strengths that scale like $1/r^2$. This indicates, by Gauss' theorem, that they are generated by \textit{charged sources}. For these configurations, the first term in \eqref{eq:ch2nonCS} does not vanish. We will explicitly construct \textit{Abelian} field configurations with this property whose $Ch_2$ is a nonzero integer. By similar arguments as the conventional ones for instantons, our Abelian winding configurations result in zero modes and generate 't~Hooft vertices, and in fact are the underlying mechanism for \textit{monopole catalysis of baryon decay}.

Before presenting our explicit construction, let us emphasize two key differences between Abelian instanton configurations and conventional instantons. The first one is that the former are not extrema of the \textit{vacuum} Maxwell equations. Instead, they are extrema of a path integral involving both (a) a background  \textit{Dirac monopole}; and (b) charged fermionic fields. As such, they appear as saddle points in \textit{monopole-background correlators of charged fermions}. The second key difference is that our Abelian instantons have a boundary condition $A_t=0$ on the worldline of the monopole ('t~Hooft line), which is simply $r=0$ for our static monopole. This effectively changes our topology, allowing for Abelian instantons.

In this section, we will not concern ourselves yet with the path integral over Abelian gauge fields and its saddle points. Instead, we shall first consider the fermionic path integral as a function of some external Abelian field configuration $A^{\rm tot}_{\mu}$. We will explicitly see that when $A^{\rm tot}_{\mu}$ has nonvanishing $Ch_2$, it leads to fermionic zero modes. We also use the terms ``Abelian winding configurations'' and ``Abelian instantons'' interchangeably, though strictly speaking we not show that they arise in the path integral until the next section. Throughout this paper, we work in Euclidean signature, where topology is more manifest.

Without further ado, consider the following field configuration
\begin{eqnarray}\label{eq:fcon}
    A^{\rm tot}_{\mu}=A^{\rm mon}_{\mu}+A^{{\rm vor}, n/2}_{\mu}\,,
\end{eqnarray}
where 
\begin{eqnarray}\label{eq:dirac potential}
    A^{\rm mon}_{t}=0~~~,~~~\vb{A}^{\rm mon}=\frac{g}{e}\frac{1-\cos\theta}{r\sin\theta}\,\hat{\varphi}\,,
\end{eqnarray}
is the gauge field of a Dirac magnetic monopole at $r=0$, with magnetic field
\begin{eqnarray}
    \vb{B}^{\rm mon} = \frac{g}{e}\frac{\vu{r}}{r^{2}}\,.
\end{eqnarray}
Note that we have defined the coupling constant $e$ so that fundamental electric charges have charge $Q=1$ in units of $e$. By Dirac quantization, $q=gQ=g$ is half-integer. 
Furthermore, we consider $A^{{\rm vor}, n/2}$ of the form
\begin{eqnarray}\label{eq:aE}
    A^{{\rm vor}, n/2}_{t}=\frac{1}{\sqrt{4\pi R^2}}\,a_t(t,r)~~~,~~~\vb{A}^{{\rm vor}, n/2}=\frac{1}{\sqrt{4\pi R^2}}\,a_r(t,r)\,\vu{r}\,.
\end{eqnarray}
Here the somewhat arbitrary $(4\pi R^2)^{-1/2}$ normalization is chosen for consistency with Section~\ref{sec:2D reduction}, while $R$ is an arbitrary length scale that drops out of all correlators. With this normalization, $a_\alpha$ is dimensionless, which is natural for $2D$ gauge fields.
Using the expression \eqref{eq:ch2nonCS} for the winding number, all we need to know to compute the second Chern number is $a_\alpha$ at Euclidean spacetime infinity. To do this, we switch to polar coordinates in $(t,r)$ as $\lambda=\sqrt{t^2+r^2}, \kappa=\arctan(r/t)$, and consider the $\lambda\rightarrow\infty$ behavior of $a_\alpha$. Let us now focus on \eqref{eq:aE} with the prescribed asymptotic behavior
\begin{eqnarray}\label{eq:aEas}  &&\lim_{\lambda\rightarrow\infty}a_{\alpha}= - \frac{n}{e_{\rm 2D}}\frac{\epsilon_{\alpha\beta}x^\beta}{x^2}~~,~~\alpha,\beta\in\{t,r\}\,,
\end{eqnarray}
where $e_{\rm 2D}=e/\sqrt{4\pi R^2}$ is an effective $2D$ coupling constant of mass dimension $1$. In terms of $\kappa$, we have
\begin{eqnarray}\label{eq:aEas2}  \lim_{\lambda\rightarrow \infty}\,a_\alpha dx^\alpha &=&\frac{n}{e_{\rm 2D}} d\kappa\,.
\end{eqnarray}
At $\lambda\rightarrow\infty$, $a_{\alpha}$ looks like a Euclidean vortex in $(t,r\geq 0)$, while at $r=0$ it has the boundary condition of an ``electric conductor'' $a_t=0$. Euclidean vortices with this $\lambda\rightarrow\infty$ behavior are ubiquitous in massless $QED_2$ on $\mathbb{R}^2$ (the Schwinger model). There, the $1/e$ dependence arises from a coupling, $e$, and a propagator factor for the massive photon $1/m_{\rm Sch}^2$, where $m_{\rm Sch} \propto e$ is the Schwinger mass \cite{Schwinger:1962tp}. As we shall see, in the presence of a monopole, the effective theory in $(t,r)$ is $QED_2$ (albeit in an effective ${AdS}_2$ metric), and so vortices like \eqref{eq:aEas} play a dominant role. We note, however, an important difference from $QED_2$ on $\mathbb{R}^2$. In the latter, the asymptotic behavior \eqref{eq:aEas} leads to \textit{integer} vortex number $n$, calculated as
\begin{eqnarray}\label{eq:aEas3}  Ch_1({QED_2\,{\rm on}\,R^2})=\frac{e_{\rm 2D}}{2\pi}\,\int_{S_1\,{at}\,\lambda\rightarrow\infty}\,a_\alpha dx^\alpha=\frac{e_{\rm 2D}}{2\pi}\,\int_0^{2\pi}\,d\kappa\,\frac{n}{e_{\rm 2D}} d\kappa\,=\,n\,.
\end{eqnarray}
On the other hand, gauge field is defined on the half plane $(t,r\geq 0)$, we must also prescribe a boundary condition at $r=0$. We focus on configurations with the asymptotic behavior \eqref{eq:aEas} as well as the boundary condition
\begin{eqnarray}\label{eq:bca}
    a_t(t,r=0)=0\,.
\end{eqnarray}
As we shall see in the next section, this is the boundary condition relevant to the full path integral of monopole-background $QED_4$. Differently from \eqref{eq:aEas3}, our half-plane gauge configurations with the boundary condition \eqref{eq:bca} have winding number
\begin{eqnarray}\label{eq:aEas4}  Ch_1({QED_2\,{\rm on}\,R^2_{r\geq 0}})=\frac{e_{\rm 2D}}{2\pi}\,\int_0^{\pi}\,d\kappa\,\frac{n}{e_{\rm 2D}} d\kappa\,+\,\cancel{\frac{e_{\rm 2D}}{2\pi}\,\int_{-\infty}^\infty\,a_t(t,r=0)\,dt}\,=\,\frac{n}{2}\,,
\end{eqnarray}
so our winding number is generically \textit{half-integer}, and we label $A^{{\rm vor},\, n/2}_\mu$ with the superscript $n/2$.
Our half-integer winding number is consistent with the general topological argument for the quantization of vortices on the half-plane with the boundary condition \eqref{eq:bca}. This can be seen as follows: every vortex in $(t,r\geq 0)$ with an $a_t=0$ boundary condition can be ``doubled'' into a vortex on the entire $(t,r)$ plane using an opposite image charge. Since the latter is quantized in integer units\footnote{The winding number on $\mathbb{R}^2$ is $n\in \pi_1[S^1]=\mathbb{Z}$.}, and the t-axis does not contribute to the winding, the original $(t,r\geq 0)$ vortex must be quantized in half-integer units \footnote{We would like to thank David Tong for his illuminating questions regarding the quantization of our vortex number.}.
To compute the second Chern number\footnote{The formal definition of the second Chern number over the 4-manifold $M$ is \cite{Nakahara:2003nw} $Ch_2=\int_M\,ch_2$, where $ch_2$ is the second Chern character,
\begin{eqnarray}
ch_2\equiv -\frac{e^2}{8\pi^2}{\rm tr}\left(F\wedge F\right)=\frac{1}{2}\left[ch_1\wedge ch_1-2c_2\right]\,.
\end{eqnarray}
The first Chern \textit{character} $ch_1$ and second Chern class $c_2$ are defined by 
\begin{eqnarray}
ch_1\equiv \frac{ie}{2\pi}{\rm tr}F~~,~~c_2\equiv -\frac{e^2}{8\pi^2}\left[{\rm tr}F\wedge {\rm tr}F-{\rm tr}(F\wedge F)\right]\,.
\end{eqnarray}
For non-Abelian instantons in 4D, $ch_1=0$ and $ch_2=-c_2$. For our Abelian instantons, $c_2=0$ and $ch_2=\frac{1}{2}ch_1\wedge ch_1$.
}, we use the Abelian version of \eqref{eq:ch2nonCS},
\begin{eqnarray}\label{eq:ch2nonCSab}
    Ch_2=-\frac{e^2}{8\pi^2}\int_{S_3}\,A\wedge F\,,
\end{eqnarray}
where $F=dA=\frac{i}{2}F_{\mu\nu}dx^\mu\wedge dx^\nu$.
We can compute this integral in 4D Euclidean hyperspherical coordinates, $\lambda,\,\kappa,\,\theta$ and $\varphi$, with the latter two defined as usual in spherical coordinates. In these coordinates the $S_3$ at spacetime infinity is parametrized by the angles $(\kappa,\theta,\varphi)$ at $\lambda\rightarrow\infty$. One can explicitly check that
\begin{eqnarray}\label{eq:gf}
    \lim_{\lambda\rightarrow \infty}\,A^{{\rm vor}, n/2}&=&i\frac{n}{\sqrt{4\pi R^2}\,e_{\rm 2D}} d\kappa=i\frac{n}{e} d\kappa\nonumber\\[5pt]
    \lim_{\lambda\rightarrow \infty}\,F^{\rm mon}&=&i\frac{2q}{e}\sin\theta d\theta d\varphi\,.
\end{eqnarray}
Plugging this in \eqref{eq:ch2nonCSab} and integrating, we have
\begin{eqnarray}\label{eq:ch2res}
    Ch_2=qn\,,
\end{eqnarray}
which is quantized in \textit{half-integer} units.

The second Chern number is also given as a volume integral by
\begin{eqnarray}
    Ch_2 &=& \frac{e^2}{16\pi^2} \int d^4x\, F_{\mu\nu}
    \widetilde{F}^{\mu\nu}=\frac{1}{2}\,\int_{\mathbb{R}^4}\,\underset{ch_1}{\underbrace{\pqty{\frac{ie}{2\pi} \frac{iF_{\mu\nu}}{2}dx^\mu\wedge dx^\nu}}}\wedge \underset{ch_1}{\underbrace{\pqty{\frac{ie}{2\pi} \frac{iF_{\rho\sigma}}{2} dx^\rho \wedge dx^\sigma}}}
    =\nonumber\\[5pt]
    &=&\underset{Ch_1(\theta,\varphi)}{\underbrace{\pqty{\frac{e}{2\pi}\int\,d\Omega\,r^2\,\vb{B}^{\rm mon}}}}\cdot\underset{Ch_1(t,r)}{\underbrace{\pqty{\frac{e}{2\pi}\int\,dtdr\,\vb{E}^{{\rm vor},\,n/2}}}}\,=(2q)(n/2)
\end{eqnarray}
Here the $ch_1$ stands for the first Chern \textit{character}, which gives the first Chern \textit{number} $Ch_1$ when integrated on a 2-manifold. In other words, the overall winding number is the product of the winding number $2q$ of $\vb{B}^{\rm mon}$ around the $\vu*{\theta}, \vu*{\varphi}$ directions, and the winding number $n/2$ of the Abelian vortex $\vb{E}^{{\rm vor}, n/2}$ field around the $\vu*{\kappa}$ direction. Thus we can identify the winding number $n/2$ with the 1st Chern number in the $t,r$ subspace and $2q$ with the 1st Chern number in the $\theta,\varphi$ subspace. 
In summary, an Abelian field configuration which is the sum of the field of a Dirac monopole and a vortex in $(t,r)$ has half-integer 2nd Chern number.

\subsection{Zero modes}\label{sec:ZM}
%
For standard non-Abelian instantons on $\mathbb{R}^4$ (more precisely its one-point-compactification $S^4$), the Atiyah-Singer index theorem \eqref{eq:AS theorem} guarantees that winding field configurations have normalizable chiral zero modes. Furthermore, standard non-Abelian instantons with $Ch_2>0$ ($Ch_2<0$) have exactly $|Ch_2|$ LH (RH) normalizable zero modes, and no RH (LH) zero modes at all.
Our situation is slightly different from the standard non-Abelian story. For one, the topology of our space is not really $\mathbb{R}^4$ but effectively $\mathbb{R}^4$ minus the 't~Hooft line of the monopole, where we enforce the boundary condition 
\eqref{eq:bca}. Correspondingly, our overall 2nd Chern number is the \textit{half-integer} $n/2$. In this section, we present an explicit construction of fermionic zero modes in the background of our field configuration \eqref{eq:fcon}. Similarly to the gauge field, our fermions would also require imposing a boundary condition at $r=0$. As we shall see below, the correct boundary condition descending from the UV completion of our monopole-background $QED_4$ is the  \textit{bag boundary condition}, stated below in \eqref{eq:bcr0YL}. Our results for the zero modes are as follows. In the background \eqref{eq:fcon}, each fermionic flavor has exactly $2|Ch_2|=2|qn|$ zero modes. Each one of those zero modes resides both in $\psi$ and $\overline{\psi}$ as the two are connected by the boundary condition at $r=0$. The normalization of the latter zero modes $\int d^4x |\psi|^2$ is at most \textit{logarithmically divergent}, and can be regularized by putting the system in a $4D$ ``hyperspherical box'' of size $\lambda_f$, as was done in \cite{tHooft_Instantons,Nielsen:1976hs,Nielsen:1977aw,Hortacsu:1979fg}. One can then explicitly check the all observables are independent of $\lambda_f$ as $\lambda_f\rightarrow\infty$. Though this is indeed the formally correct way to deal with logarithmically divergent zero modes, we ignore this formality in our calculations, as it does not influence any of our results.

It would be very interesting to prove an index theorem relevant for our situation, i.e. to formalize the relation between the half-integer $Ch_2$ for an Abelian configuration on $\mathbb{R}^4-\{{\rm 't\,Hooft\, line}\}$, and the $2|Ch_2|=2|qn|$ zero modes that we find with the ``bag'' boundary condition \eqref{eq:bcr0YL}. We leave the investigation of this conjectured index theorem for future work.

\subsection{Explicit Construction of Zero Modes}
First, we define the 4D massless \textit{Dirac} spinors $\psi$ and $\overline{\psi}$ with charges $1$ and $-1$ respectively, in units of $e$. Note that $\psi$ and $\overline{\psi}$ are independent fields in the fermionic path integral, and are only related on-shell. 
Since we consider massless fermions the $\theta$ term is unphysical and thus omitted (c.f. \cite{Fan:2021ntg} for the way the Witten effect generates a potential for the vacuum angle in the absence of massless fermions). 
Our fermionic 4D action in spherical coordinates is defined on the half-infinite interval $r=0$
\begin{eqnarray}\label{eq:Diracr}
    S_{\psi}=\int dt dr d\Omega\,r^2\,i \overline{\psi}\overleftrightarrow{\slashed{D}}\psi\,,
\end{eqnarray}
where $\overleftrightarrow{\slashed{D}}\equiv\frac{1}{2}\left(\overrightarrow{\slashed{D}}-\overleftarrow{\slashed{D}}\right)$ and
\begin{eqnarray}\label{eq:dsl}
    \slashed{D}\equiv\gamma^\mu\left(\nabla_\alpha- ieA^{\rm mon}_{\mu}- ieA^{{\rm vor}, n}_{\mu}\right)\,.
\end{eqnarray}
Here $\nabla_\alpha$ is the derivative in spherical coordinates and $\mu\in\{t,\theta,\varphi,r\}$. Taking the bulk variations of \eqref{eq:Diracr} with respect to $\psi$ and $\overline{\psi}$, we get the the Dirac equations
\begin{eqnarray}\label{eq:direq}   \slashed{D}\psi=0~~~,~~~\overline{\psi}\overleftarrow{\slashed{D}}=0\,.
\end{eqnarray}
Each of these has exactly $|Ch_2|$ \textit{normalizable} solutions, i.e. solutions that  vanishe at spacetime infinity $\lambda\equiv\sqrt{t^2+r^2}\rightarrow\infty$. We note that since the action in spherical coordinates is defined on the half-infinite interval $r\geq0$, one should be careful about the \textit{boundary variation} as well by imposing a boundary condition that makes it vanish. The implications of such boundary condition are discussed below.
\subsubsection{Monopole-Background-Only Solutions}
We first focus on solutions for $\psi$. As a first step in finding normalizable solutions for \eqref{eq:direq}, we consider solutions for a related equation
\begin{eqnarray}\label{eq:direqb}
    \gamma^\mu\left(\nabla_\alpha- ieA^{\rm mon}_{\mu}\right)\widetilde{\psi}=0\,,
\end{eqnarray}
which is simply \eqref{eq:direq} with $A^{{\rm vor}, n}_\mu=0$. The solutions $\widetilde{\psi}$ are known, though they are not normalizable. For the purpose of finding the zero modes $\psi$ of  \eqref{eq:dsl}, its enough to focus on the \textit{lowest partial wave} of $\widetilde{\psi}$, namely in addition to \eqref{eq:direq} they also satisfy
\begin{eqnarray}\label{eq:lowest}
    J^2_q\left[\widetilde{\psi}\right] = j_{\rm min}\,(j_{\rm min}+1)\,\widetilde{\psi}\,,
\end{eqnarray}
where $J^2_q$ is the angular momentum operator in the monopole background
\begin{eqnarray}\label{eq:angmom}
    \vec{J}_q=\vb{r} \times \pqty{\vb{p} - e \vb{A}^{\rm mon}} - q \, \vu{r} + \vb{S} \, . 
\end{eqnarray}
whose explicit form is given in \eqref{eq:L ang-mom}. Here $j_{\rm min}=|q|-\frac{1}{2}$ is its minimal\footnote{This is reminiscent of the non-trivial $j_{\rm min}\neq 1/2$ which appears from the restriction of the Hilbert space due to Wess-Zumino terms \cite{Rabinovici:1984mj}.} eigenvalue \cite{Kazama:1976fm}. For the lowest partial wave \eqref{eq:lowest}, $\widetilde{\psi}$ has the form
\begin{eqnarray}\label{eq:psitillowest}
    \widetilde{\psi}_{j_{\rm min}m}&=&\colvec{\chi(t,r)\,\Omega^{(3)}_{qj_{\rm min}m}(\theta,\varphi)\\\eta^{\dagger}(t,r)\,\Omega^{(3)}_{qj_{\rm min}m}(\theta,\varphi)}\,.
\end{eqnarray}
Substituting this ansatz in \eqref{eq:direqb}, we can explicitly solve for $\chi(t,r),\,\eta^\dagger(t,r)$ and get
\begin{eqnarray}\label{eq:psitilswave}
    \widetilde{\psi}_{j_{\rm min}m}&=&\sqrt{\frac{\mu}{2\pi r^2}}\colvec{f_L(t+ir)\,\Omega^{(3)}_{qj_{\rm min}m}(\theta,\varphi)\\f_R(t-ir)\,\Omega^{(3)}_{qj_{\rm min}m}(\theta,\varphi)}\, ,
\end{eqnarray}
where $-j_{\rm min}\leq m\leq j_{\rm min}\equiv|q|-\frac{1}{2}$ and $\Omega^{(3)}_{qj_{\rm min}m}$ is a monopole spinor harmonic defined in Appendix~\ref{app:Omegas}. In \eqref{eq:psitilswave}, $\mu$ is an arbitrary scale which gives the fermion the correct mass dimension:  $3/2$. The form \eqref{eq:psitilswave} solves \eqref{eq:direqb} for \textit{arbitrary} dimensionless functions $f_L(t+ir),\,f_R(t-ir)$. The dependence on $t\pm ir$ (equivalently $r\mp i t$) is the Euclidean version of left/right movers in 2D Lorentzian signature.

For future reference, we also present here the Green's function for $\widetilde{\psi}_{j_{\rm min}}$, which we denote by $\widetilde{G}_{j_{\rm min}}$. This Green's function satisfies
\begin{eqnarray}\label{eq:gff}
    &&i\left(\slashed{\partial}-ie\slashed{A}^{\rm mon}\right) \, \widetilde{G}_{j_{\rm min}}(x,x') = \frac{1}{\sqrt{g}}\delta^{4}(x-x') I_{2} \,\nonumber\\[5pt]
    &&J^2_q\left[\widetilde{G}_{j_{\rm min}}\right] = j_{\rm min}(j_{\rm min}+1)\,\widetilde{G}_{j_{\rm min}} \,,
\end{eqnarray}
where $I_2$ is the identity matrix on the spinor indices, and $g$ is the metric for $\mathbb{R}^4$ in spherical coordinates. For reasons outlined below in subsection~\ref{sec:halving}, we enforce on this Green's function the boundary condition
\begin{eqnarray}\label{eq:gffbc}
\eval{\widetilde{G}_{j_{\rm min}}(x,x')}_{r=0} = \eval{\overline{\widetilde{G}}_{j_{\rm min}}(x,x')}_{r=0} \,.
\end{eqnarray}
Similarly to other Greens' functions in spherical coordinates \cite{Jackson:1998nia}, the angular part of $\widetilde{G}_{j_{\rm min}}$ is an outer product of the angular eigenfunctions $\Omega^{(3)}$ for $x$ and $x'$, or explicitly
\begin{eqnarray}
    \mathcal{W}_m&\equiv&\Omega^{(3)}_{qj_{\rm min}m}(\theta,\varphi)\otimes \Omega^{(3)\,\dagger}_{-qj_{\rm min}m}(\theta',\varphi's)\,.
\end{eqnarray}
The $(t,r)$ part of the Green's function is typical for 2D massless fermions, since the $j_{min}$ partial wave is essentially a massless fermion on the 2D space $(t,r)$. The Green's function for a 2D left (right-) moving massless fermion is
\begin{eqnarray}
    \widetilde{G}^{2D}_{\chi\chi^\dagger}=\frac{1}{2\pi}\frac{1}{\Delta t + i\Delta r}~~,~~\widetilde{G}^{2D}_{\eta^\dagger\eta}=\frac{1}{2\pi}\frac{1}{\Delta t - i\Delta r}\,.
\end{eqnarray}
where $\Delta x = x - x'$ and $\chi$ and $\eta^\dagger$ refer to the Wey components in \eqref{eq:psitillowest}. The boundary condition \eqref{eq:gffbc} results in off-diagonal terms, corresponding to reflection of the fermions
\begin{eqnarray}
    \widetilde{G}^{2D}_{\chi\eta^\dagger}=-\frac{i}{2\pi}\frac{1}{\Delta t + i\Sigma r}~~,~~\widetilde{G}^{2D}_{\eta^\dagger\chi}= - \frac{i}{2\pi}\frac{1}{\Delta t - i\Sigma r}\,.
\end{eqnarray}
where $\Sigma x = x + x'$.
Note that the Green's functions for $2D$ fermions have dimension $1$ while $4D$ fermionic Green's functions have dimension $3$; this is compensated for by an additional factor of $1/rr'$ in 
$\widetilde{G}_{j_{\rm min}}$.
Overall, the explicit expression for $\widetilde{G}_{j_{\rm min}}$ is
\begin{eqnarray}\label{eq:DOmega3G}
    \widetilde{G}_{j_{\rm min}} &=&\sum_m\, \frac{1}{2\pi} \frac{1}{rr'}\,
    \pmqty{\frac{-i\mathcal{W}_m}{\Delta t + i \Sigma r} & \frac{\mathcal{W}_m}{\Delta t + i \Delta r}\\ 
    -\frac{\mathcal{W}_m}{\Delta t - i \Delta r} & \frac{-i\mathcal{W}_m}{\Delta t - i \Sigma r} }
    = \sum_m \frac{1}{2\pi} \frac{1}{r r'} \pmqty{\widetilde{G}_{\chi\chi^{\dagger}} & \widetilde{G}_{\chi\eta^{\dagger}} \\ \widetilde{G}_{\eta\chi^{\dagger}} & \tilde{G}_{\eta\eta^{\dagger}} }\,.
\end{eqnarray}

\subsubsection{Log-Normalizable Zero Modes}
%
To solve \eqref{eq:direq}, we need to relate $\psi$ to $\widetilde{\psi}$, and find $f_L,\,f_R$ so that $\psi$ is normalizable. For this, we use a well known strategy borrowed from $QED_2$ \cite{Adam:1994by, Hortacsu:1979fg, Abdalla:1991vua, Sachs:1991en}.
Define the solution $\psi$ as a position dependent axial rotation of $\widetilde{\psi}_{j_{\rm min}m}$ from 
\eqref{eq:psitilswave}:
\begin{eqnarray}\label{eq:4Dpsipsitil}
    \psi&=&\exp\pqty{-e\,\gamma_5\,\partial^{-2}E_r(t,r)}~\widetilde{\psi}_{j_{\rm min}m}\, ,
\end{eqnarray}
where $E_r=\partial_tA^{tot}_r-\partial_rA^{tot}_t$ is the radial electric field. The non-local operator $\partial^{-2}$ is defined as
\begin{eqnarray}
    \partial^{-2}f(t,r)\equiv \int dt'dr' \,{\cal D}(t,r;t',r')f(t',r')\,,
\end{eqnarray}
where ${\cal D}$ is the Green's function of $\partial^2\equiv\partial^2_t+\partial^2_r$ satisfying
\begin{eqnarray}
    \partial^{2}{\cal D}(t,r;t',r')=\delta(t-t')\,\delta(r-r')\,.
\end{eqnarray}
Explicitly, 
\begin{eqnarray}\label{eq:sch2G2}
    {\cal D}(t,r;t',r')\,=\,-\frac{1}{4\pi}\log\left(\frac{1}{\mu^2(\Delta t^2+\Delta r^2)}\right)\,,
\end{eqnarray}
where, again, $\Delta x = x-x'$. Our job is to show that the $\psi$ defined in \eqref{eq:4Dpsipsitil} is a solution for \eqref{eq:direq}. To see this, we substitute it into \eqref{eq:direq} and get
\begin{eqnarray}\label{eq:direqsub}
e^{e\,\gamma_5\,\partial^{-2}E_r}\gamma^\mu\left(\partial_\mu- ieA^{\rm mon}_{\mu}- ieA^{{\rm vor}, n}_{\mu}\right)\psi & = & \gamma^\mu\left(\partial_\mu- ieA^{\rm mon}_{\mu}-ieA^{{\rm vor}, n}_{\mu} - e\,\gamma_5\,\partial^{-2}\partial_\mu E_r\right)\widetilde{\psi}_{j_{\rm min}m}\,\nonumber\\[5pt]
    &=&\gamma^\mu\left(\partial_\mu- ieA^{\rm mon}_{\mu}\right)\widetilde{\psi}_{j_{\rm min}m}=0\,.
\end{eqnarray}
The cancellation between the $A^{{\rm vor}, n}_\mu$ and $E_r$ terms is by virtue of the relations
\begin{eqnarray}\label{eq:canc}
&&A^{{\rm vor}, n}_t=~ -\partial_r\,\partial^{-2}\,E_r~~~,~~~A^{{\rm vor}, n}_r=\partial_t\,\partial^{-2}\,E_r\,\nonumber\\[5pt]
&&\left(\gamma_\alpha\gamma_5-i\epsilon_{\beta \alpha}\gamma^\beta\right)\,\widetilde{\psi}_{j_{\rm min}m}=0~~,~~\alpha,\beta\in\{t,r\}\,.
\end{eqnarray}
We then see explicitly that $\psi$ defined by \eqref{eq:4Dpsipsitil} indeed solves the full Dirac equation \eqref{eq:direq}. We are not fully done, though, as we still need to find $f_L,\,f_R$ in \eqref{eq:psitilswave} so that $\psi$ is normalizable, and in particular decays at least as $1/r$ at spacetime infinity. To do that, we first note the spacetime asymptotic behavior of the exponential prefactor in \eqref{eq:4Dpsipsitil}. Given the background vortex \eqref{eq:aEas}, we have
\begin{eqnarray}\label{eq:norlim}
    \lim_{\lambda\rightarrow \infty}\psi&=&\lim_{\lambda\rightarrow \infty}\,\sqrt{\frac{\mu}{2\pi r^2}}\,\exp\pqty{-e\,\gamma_5\,\partial^{-2}E_r(t,r)}\colvec{f_L(t+ir)\,\Omega^{(3)}_{qj_{\rm min}m}(\theta,\varphi)\\f_R(t-ir)\,\Omega^{(3)}_{qj_{\rm min}m}(\theta,\varphi)}\nonumber\\[5pt]
    &=&\lim_{\lambda\rightarrow \infty}\,\sqrt{\frac{\mu}{2\pi r^2}}\colvec{f_L(t+ir)\left[\mu^2(t^2+r^2)\right]^{-n/2}\,\Omega^{(3)}_{qj_{\rm min}m}(\theta,\varphi)\\f_R(t-ir)\left[\mu^2(t^2+r^2)\right]^{n/2}\,\Omega^{(3)}_{qj_{\rm min}m}(\theta,\varphi)}\,.
\end{eqnarray}
As explained in the beginning of this section, our normalizability condition is that $\int d^4x |\psi|$ is at most \textit{logarithmically} divergent, as even logarithmically divergent zero modes contribute to the path integral in a 4D box of ``radius'' $\lambda_f\rightarrow\infty$ \cite{tHooft_Instantons,Nielsen:1976hs,Nielsen:1977aw,Hortacsu:1979fg}. By examining \eqref{eq:norlim}, we can directly infer the constraints that it puts on $f_L,\,f_R$. When $n>0$, normalizabilty requires $f_R=0$, while the allowed $f_L(t+ir)$ is spanned by $0\leq l < |n|$ basis elements $[\mu(t+ir)]^l$. Conversely, for $n<0$, $f_L=0$, while the allowed $f_R(t-ir)$ is spanned by $0\leq l < |n|$ basis elements $[\mu(t-ir)]^l$. Summing up, we have $2|qn|$ normalizable zero modes of the form
\begin{eqnarray}\label{eq:4Dzm}
    \psi^{(0)}_{ml}&=&\sqrt{\frac{\mu}{2\pi}}\,\exp\pqty{- e\,\gamma_5\,\partial^{-2}E_r(t,r)}~\frac{[\mu(t+ir)]^l}{ r}\colvec{\Omega^{(3)}_{qj_{\rm min}m}(\theta,\varphi)\\0}~~,~~n>0\nonumber\\[5pt]
    \psi^{(0)}_{ml} & = & \sqrt{\frac{\mu}{2\pi}}\,\exp\pqty{- e\,\gamma_5\,\partial^{-2}E_r(t,r)}~\frac{[\mu(t-ir)]^l}{ r}\colvec{0\\\Omega^{(3)}_{qj_{\rm min}m}(\theta,\varphi)}~~,~~n<0\,.
\end{eqnarray}
where $0\leq l < |n|$ and $-j_{\rm min}\leq m\leq j_{\rm min}$. Similarliy to $\psi$, $\overline{\psi}$ in \eqref{eq:direq} has $|Ch_2|$ normalizable zero modes
\begin{eqnarray}\label{eq:4Dzmb}
    \overline{\psi}^{(0)}_{ml}&=&\sqrt{\frac{\mu}{2\pi}}\,\exp\pqty{-e\,\gamma_5\,\partial^{-2}E_r(t,r)}~\frac{[\mu(t-ir)]^l}{r}\colvec{\Omega^{(3)}_{-qj_{\rm min}m}\\0}~~,~~n>0 \, , \nn \\[5pt]
    \overline{\psi}^{(0)}_{ml}&=&\sqrt{\frac{\mu}{2\pi}}\,\exp\pqty{-e\,\gamma_5\,\partial^{-2}E_r(t,r)}~\frac{[\mu(t+ir)]^l}{r}\colvec{0\\\Omega^{(3)}_{-qj_{\rm min}m}}~~,~~n<0 \,.
\end{eqnarray}
Note that $\psi$ and $\overline{\psi}$ are independent fields in the path integral. Only the \textit{non-zero mode} solutions to the Dirac equation are identified via $\overline{\psi}=\psi^\dagger\gamma^0$.
\subsubsection{The Boundary Condition and Zero Mode ``Halving''}\label{sec:halving}
Previously, we got the Dirac equation \eqref{eq:direq} from the bulk variation of the action \eqref{eq:Diracr}, leaving the boundary variation for later. We now consider this boundary variation, as well as the boundary condition that it implies. The boundary variation is the boundary term from the integration by parts of \eqref{eq:Diracr} to get the Dirac equation. It is given by
\begin{eqnarray}\label{eq:Diracrr0}
    S^{r=0}_{\psi}= i \int dt \, d\Omega \, \left(\delta\overline{\psi}^{\,r}\gamma_r\psi^r-\overline{\psi}^{\,r}\gamma_r \delta\psi^r\right)\,,
\end{eqnarray}
where $\psi^r=\lim_{r\rightarrow 0}r\psi$ and $\overline{\psi}^{\,r}=\lim_{r\rightarrow 0}r\overline{\psi}$.
In the absence of a monopole, $\psi^r=\overline{\psi}^{\,r}=0$ for all of the solutions of the bulk Dirac equation, and so this boundary term automatically vanishes. For this reason, we never need to bother about boundary conditions in $QED_4$ without monopoles. In the presence of the monopole, however, $\psi^r$ and $\overline{\psi}^{\,r}$ are finite for zero modes, and so we need to impose a boundary condition on them. There are two possible \textit{charge conserving} boundary conditions we could consider, namely
\begin{itemize}
\item $\psi^r=\delta\psi^r=0$\,.
\item $\overline{\psi}^r=\delta\overline{\psi}^r=0$\,.
\end{itemize}
The first one ``kills'' the zero modes in $\psi$, while the second one kills the ones in $\overline{\psi}$. In either case, no charge-conserving monopole-catalysis processes can be generated, as we will show explicitly below. Furthermore, these boundary conditions do not correctly reflect the non-Abelian physics at the core of the monopole, under which $\psi,\,\overline{\psi}$ (more precisely their LH components) are members of the same $SU(2)$ doublet and should be treated symmetrically. More specifically the charge conserving boundary conditions above respect an enhanced $SU(N_f)^2$ symmetry while any UV completion groups the fermion into $SU(2)$ doublets transforming only under the diagonal $SU(N_f)$. By starting from a full 't~Hooft-Polyakov monopole, Rubakov in his foundational work \cite{Rubakov:1982fp} on monopole catalysis derived the correct boundary condition
\begin{eqnarray}\label{eq:bcr0}
    \overline{\psi}^{\,r}=\psi^r~~\rightarrow~~\delta\overline{\psi}^{\,r}=\delta\psi^r\,,
\end{eqnarray}
which also makes the boundary variation vanish. Let us make a few comments about this boundary condition.
\begin{itemize}
\item It doesn't conserve the electric charge of the fermions. This reflects the possibility of depositing charge on the monopole, a process suppressed by the huge mass gap between the monopole and its first dyonic excitation \cite{Besson:1980vu,Rubakov:1982fp, Callan:1982ah,Callan:1982au, Isler:1987xn, Hook:2024als}.
\item A more accurate (and formally gauge invariant) statement of \eqref{eq:bcr0}, as emphasized in \cite{Lam:1984sm, Isler:1987xn}, would be 
\begin{eqnarray}\label{eq:bcr0YL}
    \overline{\psi}^r=e^{-i\phi}\psi^r\,,
\end{eqnarray}
where $\phi$ is the \textit{dyonic collective coordinate} of the monopole. In this way fermions with high enough energies can deposit charge on the monopole by exciting the collective coordinate. In this work we focus on instanton processes which do not involve any interaction with the collective coordinate, and so we let go of the more correct \eqref{eq:bcr0YL} in favor of the simpler \eqref{eq:bcr0}, which is simply a gauge-fixed version of \eqref{eq:bcr0YL}. This has no bearing on our calculations.
\item It is the only boundary condition consistent with the non-Abelian UV completion of the monopole \cite{Besson:1980vu, Callan:1982ah,Callan:1982au, Lam:1984sm, Isler:1987xn}. See Appendix~\ref{app:chvbc} for a derivation of this fact.
\item The ``dyon boundary condition'' (DBC) model of refs.~\cite{Affleck:1993np, Maldacena:1995pqn, Smith:2019jnh, Smith:2020nuf, Hamada:2022eiv,vanBeest:2023dbu,vanBeest:2023mbs} is a model simulating the instanton physics derived in this paper by prescribing a specific boundary condition for the fermions. In fact, the monopole catalysis correlators derived from our instantons are exactly equal to the ones derived from the DBC theory, as had to be the case based on standard model charge and angular momentum conservation. Naively, one would worry that our \textit{charge-depositing} boundary condition \eqref{eq:bcr0} leads to different correlators from those of the DBC model. This is not the case, for the following reason: in all of the monopole catalysis correlators derived in our full EFT, every flavor appears only once as a LH (RH) fermion for $n>0$ ($n<0$). As the boundary condition \eqref{eq:bcr0} only affects Green's functions between LH and RH fermions of \textit{the same flavor}, it never enters into any monopole-catalysis correlators in our theory. We elaborate more on the relation between our full EFT and the DBC model in Section~\ref{sec:DBC}. 
\end{itemize}
Finally, imposing \eqref{eq:bcr0} has one significant implication, which is the fact that the modes in \eqref{eq:4Dzm} and \eqref{eq:4Dzmb} are no longer independent. Instead, they are collectively given by
\begin{eqnarray}\label{eq:4Dzmnew}
    \colvec{\psi^{(0)}_{ml}\\\overline{\psi}^{(0)}_{ml}}&=&\sqrt{\frac{\mu}{2\pi}}\,e^{-e\,\partial^{-2}E_r(t,r)}~\frac{1}{ r}\colvec{{[\mu(t+ir)]^l}\colvec{\Omega^{(3)}_{qj_{\rm min}m}\\0}\\{[\mu(t-ir)]^l}\colvec{\Omega^{(3)}_{-qj_{\rm min}m}\\0}}~~,~~n>0\nonumber\\[5pt]
    \colvec{\psi^{(0)}_{ml}\\\overline{\psi}^{(0)}_{ml}}&=&\sqrt{\frac{\mu}{2\pi}}\,e^{e\,\partial^{-2}E_r(t,r)}~\frac{1}{r}\colvec{{[\mu(t-ir)]^l}\colvec{0\\\Omega^{(3)}_{qj_{\rm min}m}}\\{[\mu(t+ir)]^l}\colvec{0\\\Omega^{(3)}_{-qj_{\rm min}m}}}~~,~~n<0
\end{eqnarray}
with $0\leq l < |n|$ and $-j_{\rm min}\leq m\leq j_{\rm min}$. 
The boundary condition \eqref{eq:bcr0}, and the resulting combined $(\psi^{(0)},\overline{\psi}^{(0)})$ zero modes have far reaching consequences. For one, it explains why our 2nd Chern number is allowed to be half-integer. Clearly, a half-integer 2nd Chern number would be nonsensical on a compact manifold without a boundary, where the Atiyah-Singer index theorem \eqref{eq:AS theorem} links it to the difference between left- and right-moving zero modes. However, in our case the boundary conditions \eqref{eq:bca} and \eqref{eq:bcr0YL} make our manifold effectively $\mathbb{R}^4-\{\rm 't\,Hooft\,line\}$, a manifold with a boundary where a ``bag'' boundary condition is imposed\footnote{Since the boundary condition is \textit{local}, the Atiyah-Patodi-Singer (APS) \cite{Atiyah:1975jf} theorem is of no use here either. The setup in Section 4 of \cite{Beneventano:2002im} is close to a $2D$ reduction of our setup. Note, However, that they consider \textit{vector} $QED_2$, while the correct $2D$ EFT for us is \textit{Axial} $QED_2$. Consequently, they have no zero modes.}. Consequently, we observe that in every flavor $f\in\{1,\ldots,N_f\}$, there are now exactly $2|qn|$ zero modes, each one residing both in $\psi_f$ and in $\overline{\psi}_f$.

As an aside, we note that our counting of zero modes is actually \textit{consistent} with the standard one for non-Abelian instantons. For an $SU(2)$ BPST instanton, the d.o.f of $\psi_f$ and $\overline{\psi}_f$ are conveniently packaged in a single $SU(2)$-doublet 4D Weyl fermion. Since each Weyl fermion has $n$ zero modes (in this case $q=1/2$), the zero mode counting matches between $SU(2)$ instatons and our Abelian ones. It is tempting to push this further in the future, and try to realize our Abelian instantons as constrained versions of non-Abelian instantons.

As in the case of non-Abelian instantons, the existence of zero modes in the background \eqref{eq:fcon} has non-perturbative dynamical consequences. As we shall see below, these zero modes non-perturbatively generate a \textit{'t~Hooft Vertex}: a vertex connecting all fermions in the theory. This 't~Hooft vertex is responsible for charge-conserving, baryon-number and chirality violating processes called \textit{Monopole Catalysis} processes.

\subsection{'t~Hooft Vertex and Monopole Catalysis}\label{sec:intuition MC}
%
To see the emergence of a 't~Hooft vertex, we consider the Euclidean fermionic path-integral for $N_f$ Dirac fermions in the background \eqref{eq:fcon}:
\begin{eqnarray}\label{eq:4DfPI}
    Z^{(n/2)}_{\psi}[A^{\rm tot}_{\mu},\overline{\eta},\eta]=\int\,\prod_{f=1}^{N_f}\,{\cal D}\overline{\psi}_f {\cal D}\psi_f\,\exp\left\{\sum_f\,\int d^4x\,\left[i\overline{\psi}_f\slashed{D}\psi_f+\overline{\eta}_f\psi_f+\overline{\psi}_f\eta_f\right]\right\}\,.
\end{eqnarray}
Note that we added fermionic sources $\overline{\eta}_f,\eta_f$ since we are ultimately interested in fermionic correlators. 

To derive the 't~Hooft vertex, we expand $\overline{\psi}_f,\,\psi_f$ as
\begin{eqnarray}
    \colvec{\psi_f\\\overline{\psi}_f}&=&\sum_{ml}c_{mlf}\,\colvec{\psi^{(0)}_{ml}\\\overline{\psi}^{(0)}_{ml}}~+~\sum\,{{\rm nonzero}~{\rm  eigenmodes}}\,,
\end{eqnarray}
where the $c_{mlf}$ are Grassman variables, and the ``nonzero eigenmodes'' term stands for a sum over Grassman coefficents times nonzero eigenmodes of $\slashed{D}$. 
Changing integration variables in the path integral \eqref{eq:4DfPI} to $c_{mlf}$ and the nonzero eigenmodes, we get (c.f. \cite{Adam:1994by, Hortacsu:1979fg, Abdalla:1991vua, Sachs:1991en})
\begin{eqnarray}\label{eq:4DfPI2rep2}
    Z^{(n/2)}_{\psi}\,[A^{\rm tot}_{\mu},\overline{\eta},\eta]&=&\mathcal{F}_{n/2}[a_\alpha,\overline{\eta},\eta]\,\exp{-\Gamma'[a_\alpha,\overline{\eta},\eta]}\,\bqty{{\rm det}^{\prime}(i \slashed{D})}^{N_f}\,,
    \end{eqnarray}
    where $\mathcal{F}_{n/2},\,\Gamma'$ and $\bqty{{\rm det}^{\prime}(i \slashed{D})}^{N_f}$ are the 't~Hooft vertex, generating functional for fermionic non-zero modes, and the fermionic determinant (with zero modes omitted), respectively. Explicitly,
    \begin{eqnarray}\label{eq:comps2}
\mathcal{F}_{n/2}[a_\alpha,\overline{\eta},\eta]&=&\int\,\prod_{mlf}\, d c_{mlf}\,e^{\int d^4x\,\left[c_{mlf}\left(\overline{\eta}_f\psi^{(0)}_{ml}+\overline{\psi}^{(0)}_{ml}\eta_f\right)\right]}\nonumber\\[5pt]
     \Gamma'[a_\alpha,\overline{\eta},\eta]&=&\int d^4xd^4y\,\overline{\eta}_f(x)G'(x,y)\eta_f(y)\,,
\end{eqnarray}
and ${\rm det}^{\prime}\left[\slashed{D}\right]^{N_f}$ will be calculated below.
First we will focus on the 't~Hooft vertex, which involves a Grassmannian integral over the $c_{mlf}$. We note that the fermion bilinear term vanishes for zero modes, by definition. This allows us to perform the Grassmannian integration directly and get
\begin{eqnarray}\label{eq:4DfPI2}
    \mathcal{F}_{n/2}[A^{\rm tot}_{\mu},\overline{\eta},\eta]&=&\epsilon_{N_f}\left(X_1,\ldots,X_{N_f}\right)\,,
\end{eqnarray}
where
\begin{eqnarray}
X_f&\equiv&\prod_{ml}\left[\left(\overline{\eta}_f\,\psi^{(0)}_{ml}\right)+\left(\overline{\psi}^{(0)}_{ml}\,\eta_f\right)\right]\nonumber\\[5pt]
(fg)&\equiv&\int d^4x f(x)g(x)\,.
\end{eqnarray}
In our shorthand notation,  
\begin{eqnarray}\label{epsilonnotation}   \epsilon_{N_f}\left(X_1,\ldots,X_{N_f}\right)&\equiv& \epsilon^{f_1,\ldots,f_{N_f}}\,X_{f_1}\ldots X_{f_{N_f}}\,,
\end{eqnarray}
generating an $SU(N_f)$ \textit{invariant}. The 't~Hooft vertex is the $\epsilon_{N_f}$ term in \eqref{eq:4DfPI2}. It is inherently non-perturbative, and generates flavor-invariant fermionic correlators involving $Ch_2=|2q|n$ insertions for each of the $N_f$ flavors. As a consequence of the ``halving'' of zero modes discussed in Section~\ref{sec:halving}, each of these insertions can be either $\psi_f$ or $\overline{\psi}_f$. This is consistent with the 't~Hooft vertex from standard $SU(2)$ instantons, where each leg on the 't~Hooft vertex can be either the top or the bottom parts of each $SU(2)$ doublet. Note that it is only sensible to extract gauge invariant correlators from \eqref{eq:4DfPI2}. Thus correlators with an even number of fermionic insertions, so that half of them are $\psi$ and half are $\overline{\psi}$, are non-vanishing. For the rest of this paper, we focus exclusively on these correlators, since other correlators involve the UV process of charge deposition on the monopole core through the charge-depositing boundary condition \eqref{eq:bcr0}.

To illustrate this result let us consider an example where $N_f=4,\,n/2=1/2$ and $q=1/2$. In this case there is only one insertion for each of the 4 flavors. For a charge conserving correlator, two of the insertions have to be $\overline{\psi}$, while the other two are $\psi$.
We then have
\begin{eqnarray}\label{eq:cor4}
    &&\left\langle\,\tfrac{1}{N_f!}\,\epsilon_{N_f}\left(\overline{\psi}_{L1}(x_1),\overline{\psi}_{L2}(x_2),{\psi}_{L3}(x_3),{\psi}_{L4}(x_4)\right)\right\rangle_{A^c}=\nonumber\\[5pt]
    &&\tfrac{1}{N_f!}\,\epsilon_{N_f}\left(P_L\frac{\partial}{\partial\eta_1(x_1)},P_L\frac{\partial}{\partial\eta_2(x_2)},P_L\frac{\partial}{\partial\overline{\eta}_3(x_3)},P_L\frac{\partial}{\partial\overline{\eta}_4(x_4)}\right)\,\log Z^{(n/2)}_{\psi}\,[A^{\rm tot}_{\mu},\overline{\eta},\eta]=\nonumber\\[5pt]
    &&\left({\rm det}^\prime[\slashed{D}]\right)^{4}\,\overline{\psi}^{(0)}(x_1)\overline{\psi}^{(0)}(x_2)\psi^{(0)}(x_3)\psi^{(0)}(x_4)\,.
\end{eqnarray}
Since we took $n/2=1/2>0$, the zero modes \eqref{eq:4Dzmnew} are always in the top (left-handed) part of $\psi_f\,,\overline{\psi}_f$. For this reason, the left-handed projectors $P_L=(1-\gamma_5)/2$ simply drop out. Had we used a right-handed projector $P_R=(1+\gamma_5)/2$, the correlator would have vanished.
We also note in passing that a different boundary condition for the fermions, e.g. $\psi=0$ or $\overline{\psi}=0$, would have led to 't~Hooft vertices that involve only $\overline{\psi}$ or $\psi$ fields, respectively. But the latter are not gauge invariant, and would be set to zero when considering the gauge path integral. In other words, the boundary condition \eqref{eq:bcr0} is imperative for non-vanishing monopole catalysis correlators. This is good news, since it is also the correct boundary condition that descends from the UV completion.

Of course the correlator \eqref{eq:cor4} is not quite gauge-invariant yet since we did not include Wilson lines between oppositely-charged fermions. It is also a correlator in the fixed background field $A^c$ rather than a path integral over the dynamical photon. Nevertheless, this correlator illustrates the central physical process responsible for monopole catalysis: in the presence of the Abelian instanton \eqref{eq:fcon}, there are non-perturbative, flavor and chirality changing processes like \eqref{eq:cor4}. In particular, for the minimal $SU(5)$ GUT monopole, the standard model fields exactly embed\footnote{The conventional normalization of the coupling constant assigns electric charges $\pm1/2e$ for the fermions and a magnetic charge $e^{-1}$ to the monopole. Here we normalized $e\rightarrow 2e$ so that the fermions have charge $\pm e$ and the monopole has charge $1/2e^{-1}$.} into the $N_f=4,\,q=\frac{1}{2}$ case as
\begin{eqnarray}\label{eq:emb}
    \psi_1=\colvec{-\overline{u}^2\\(u^1)^\dagger}~,~\psi_2=\colvec{\overline{u}^1\\(u^2)^\dagger}~,~\psi_3=\colvec{e\\-(\overline{d}_3)^\dagger}~,~\psi_4=\colvec{d^3\\(\overline{e})^\dagger}\, ,
\end{eqnarray}
and the correlator \eqref{eq:cor4} contains
\begin{eqnarray}\label{eq:cor5}
    &&\left\langle\,u^1(x_1)u^2(x_2)e(x_3)d^3(x_4)\right\rangle_{A^c}\,.
\end{eqnarray}
This means that the Abelian instanton background induces monopole catalysis:
\begin{eqnarray}\label{eq:cor6}
    &&u^1+u^2\rightarrow e^\dagger+(d^3)^\dagger~~~,~~~{\rm in}~A^c~{\rm background}\,,
\end{eqnarray}
also known as the \textit{Callan-Rubakov} effect \cite{Rubakov:1982fp, Callan:1982au, Callan:1982ah}. We emphasize that the embedding \eqref{eq:emb} and the boundary condition \eqref{eq:bcr0}, which are ingredients inherited from the UV theory at the monopole core, only determine the particle content of this process. The non-perturbative dynamics are realized completely in the Abelian effective theory in the IR.

\subsection{Fermion Determinant and the  Gauge-Invariant Mass Term for \texorpdfstring{$a_{\alpha}$}{}}\label{sec:intuition schwinger mass}
%
Looking at the expression \eqref{eq:4DfPI} for the path integral or the particular correlator \eqref{eq:cor4}, we note that there is a missing piece of the story that we haven't yet computed: the fermionic determinant ${\rm det}^\prime(i\slashed{D})$. This determinant provides an Euler-Heisenberg-like dependence on the gauge field $A^{{\rm vor}, n}_\mu$. We are interested in computing only the leading, non-perturbative contribution to this determinant from resumming all insertions of the monopole field $A^{\rm mon}_\mu$. Below we outline a diagrammatic computation of this leading piece, in a manifestly 4D language. In Section~\ref{sec:2D reduction} we will revisit this computation in an effective 2D theory, where we will be able to compute ${\rm det}^\prime(i\slashed{D})$ using heat-kernel methods.

In this section, we calculate the leading contribution to ${\rm det}^\prime(i\slashed{D})$. This contribution is generated at 1-loop from the fermions, and is 1-loop exact as we show explicitly below. This contribution is of the form \cite{Rubakov:1982fp, Callan:1982au, Callan:1982ah}
\begin{eqnarray}\label{eq:sch}
    \frac{{\rm det}^\prime(i\slashed{D})}{{\rm det}^\prime(i\slashed{\partial})}= \exp{- \,\int dt\,  dr\,\left[\frac{m^2}{2}\,a_{\alpha}\left(\eta^{\alpha\beta}-\frac{\partial^{\alpha}\partial^{\beta}}{\partial^2}\right)a_{\beta}+\ldots\right]}\,,
\end{eqnarray}
where $\alpha=\{t,r\}$, and $m^2=\frac{2|q|e^2_{\rm 2D}}{\pi}$ is a 1-loop generated 2D photon mass. Here the ``$\ldots$'' stand for higher dimensional operators in $a_\alpha$ as well as the field strength renoramlization of $a_\alpha$. All of these are not relevant for us and we will omit from now on. Let us make a few comments about the result \eqref{eq:sch}.
\begin{itemize}
\item It is gauge invariant, as can be checked by taking $a_\alpha\rightarrow a_\alpha+\partial_\alpha\upsilon$.
\item It is not $4D$ Lorentz invariant, and for this reason it does not appear in $QED_4$ without the monopole background. The latter breaks $4D$ Lorentz to $U(1)\times SO(3)$, i.e boosts in $(t,r)$ $\times$ spatial rotations. In fact, the theory is classically conformal, so the breaking is really $SO(3,2)\rightarrow SL(2,R)\times SO(3)$, i.e. $4D$ conformal symmetry is explicitly broken to the isometry of $AdS_2\times S_2$.
\item For future reference we find it useful to recast \eqref{eq:sch} into an effective $AdS_2$ metric. Do to so we define the effective ${AdS}_2$ metric $g^{\rm 2D}_{\alpha\beta}=(R/r)^2\,{\rm diag}(1,1)$, and write
\begin{eqnarray}\label{eq:schAds}
    \frac{{\rm det}^\prime(i\slashed{D})}{{\rm det}^\prime(i\slashed{\nabla})}=\exp{-\int d^2x\,\sqrt{g_{\rm 2D}}\,\left[\frac{m^2}{2} a_{\alpha}\left(g^{\alpha\beta}_{\rm 2D}-\frac{\nabla^{\alpha}\nabla^\beta}{\square}\right)a_{\beta}\right]}\,,
\end{eqnarray}
where $\nabla^{\alpha}$ is the ${AdS}_2$ covariant derivative, and $\square=\nabla^2$. In this effective ${AdS}_2$ metric, \eqref{eq:schAds} has a distinct meaning as the \textit{Schwinger} mass, which is generated from the 2D chiral anomaly in $QED_2$, albeit in our case, this $QED_2$ set up in an effective ${AdS}_2$ metric. 
Indeed, in Appendices \ref{app:partial wave decomp}-\ref{app:fermdet}, we explicitly derive \eqref{eq:schAds} by KK decomposing $QED_4$ into (axial) $QED_2$ in ${AdS}_2$.
\item Naively, a pole mass for the radial photon would result in screening, as is the case in massless $QED_2$ \cite{Schwinger:1962tp}. Here, however, the pole mass is of the same scale as the ${AdS}_2$ curvature. For this reason, the radial photon $a_\alpha$ is not screened, and can form vortices of the form \eqref{eq:aEas}. This is consistent with the behavior of the 4D photon, which is not screened.
\end{itemize}

Now that we have gotten more acquainted with the 1-loop exact contribution \eqref{eq:sch}, it's time to derive it from an explicit 1-loop calculation of ${\rm det}^\prime(i\slashed{D})$. The latter is simply given at 1-loop by the collection of all Feynman diagrams involving 1 fermion loop and an arbitrary number of legs with the gauge field $A^{tot}_{\mu}$.
The two-photon diagram gives the field-strength renormalization of QED. Note that in our case we can take each one of the photon legs to be either the monopole background $A^{\rm mon}_\mu$ or the ``radial photon''\footnote{Note here that we are limiting our scope to an effective 1-loop action for the ``radial'' photon \eqref{eq:aE}. We comment on this choice in the next section.} $A^{{\rm vor}, n}_\mu$. However, note that since $A^{\rm mon}_\mu$ includes a $e^{-1}$ factor, we have to sum over all insertions of $A^{\rm mon}_\mu$ in our results---there is no sense in which an additional insertion of $A^{\rm mon}_\mu$ makes the diagram smaller. In other words, we need to take into account the monopole background \textit{non-perturbatively}. To do so efficiently, we simply use the ``monopole-dressed'' propagator in Figure ~\ref{fig:DP}. Note that this propagator is nothing but the propagator for the Dirac equation in the pure monopole background, i.e. the propagator for the fermion $\widetilde{\psi}$ from \eqref{eq:direqb}.
%
\begin{figure}[ht]
\begin{center}
\begin{tikzpicture}
    \begin{feynman}
      \vertex (a00);
      \vertex [right=of a00] (a01);
      \vertex [right=1ex of a01] (c1) {$=$};
      
      \vertex [right=2.5ex of c1] (a10);
      \vertex [right=of a10] (a11);
      \vertex [right=1ex of a11] (c2) {$+$};
      
      \vertex [right=2.5ex of c2] (a20);
      \vertex [right=5ex of a20] (a21);
      \vertex [above=7.5ex of a21] (b21);
      \vertex [right=5ex of a21] (a22);
      \vertex [right=1ex of a22] (c2) {$+$};
      
      \vertex [right=2ex of c2] (a30);
      \vertex [right=5ex of a30] (a31);
      \vertex [above=7.5ex of a31] (b31);
      \vertex [right=5ex of a31] (a32);
      \vertex [above=7.5ex of a32] (b32);
      \vertex [right=5ex of a32] (a33);

      \vertex [right=1ex of a33] (c3) {$\cdots$};
      \diagram* {
        (a00) -- [double distance=0.5ex, line width=1pt] (a01);
        
        (a10) -- [plain, line width=1pt] (a11);

        (a20) -- [plain, line width=1pt] (a21);
        (a21) -- [plain, line width=1pt] (a22);
        (a21) -- [boson, insertion=0.98] (b21);

        (a30) -- [plain, line width=1pt] (a31);
        (a31) -- [plain, line width=1pt] (a32);
        (a31) -- [boson, insertion=0.98] (b31);
        (a32) -- [plain, line width=1pt] (a33);
        (a32) -- [boson, insertion=0.98] (b32);
      };
    \end{feynman}
\end{tikzpicture}
\caption{The ``monopole dressed'' fermion propagator resums all insertion of the monopole background $A^{\rm mon}_\mu$.}\label{fig:DP}
\end{center}
\end{figure}
%
Using the dressed propagator, we can organize our 1-loop effective action into ``dressed diagrams''. Specifically, we will be interested in the first of these diagrams, which generates a non-trivial, gauge invariant bilinear for $a_\alpha$,
\begin{eqnarray}\label{eq:scha}
&& {\rm det}^\prime(i \slashed{D})=\exp{e^2_{\rm 2D}\,\int dt dr\,r^2\,\int dt' dr'\,r'^2\,a_{\alpha}(t,r)\,\Pi^{\alpha\beta}(t,r;t',r')\,a_{\beta}(t',r')+\ldots}\,\nonumber\\[5pt]
&&\Pi^{\alpha\beta}(t,r;t',r')=
\begin{tikzpicture}[baseline=(current bounding box.center)]
    \begin{feynman}
        \vertex (z);
        \vertex [right=7.5ex of z] (a);
        \vertex [above right=7.5ex of a] (b);
        \vertex [large, dot, below right=7.5ex of b] (c);
        \vertex [below right=7.5ex of a] (d);
        \vertex [right=7.5ex of c] (y);
    
        \diagram* {
            (a) -- [double distance=0.5ex, line width=1pt, quarter left] (b) -- [double distance=0.5ex, line width=1pt, quarter left] (c) -- [double distance=0.5ex, line width=1pt, quarter left] (d) -- [double distance=0.5ex, line width=1pt, quarter left] (a);
            (z) -- [boson] (a);
            (c) -- [boson] (y);
        };
    \end{feynman}
    \draw[fill=black] (a) circle(0.66mm);
    \draw[fill=black] (c) circle(0.66mm);
\end{tikzpicture}
+ O(e^2_{\rm 2D})
\end{eqnarray}
The other diagrams are suppressed by powers of $e^2_{\rm 2D}$ with respect to this diagram, and only lead to a small perturbative modification of our picture.

In fact, we can go one step further. Instead of considering the full loop of the fermion $\widetilde{\psi}$, all we need to consider in the diagram \eqref{eq:scha} is a loop of the lowest partial wave $\widetilde{\psi}_{j_{\rm min}}$ given in \eqref{eq:psitillowest}.
This is because higher partial waves never contribute to \eqref{eq:sch}, as can be shown by explicit calculation \cite{Rubakov:1982fp}. For this reason, we exclusively focus on the lowest partial wave Green's function $\widetilde{G}_{j_{\rm min}}$ from \eqref{eq:DOmega3G}. The photon self energy diagram shown in \eqref{eq:scha} is then given by
\begin{eqnarray}
    \Pi^{\alpha\beta}(t,r;t',r') 
    &=& \int \sin\theta d\theta d\varphi\, \sin\theta' d\theta' d\varphi'\,{\rm tr}\Bqty{\sigma^{\alpha} \widetilde{G}_{j_{\rm min}}(x,x') \, \sigma^{\beta} \widetilde{G}_{j_{\rm min}}(x',x)} \, . 
\end{eqnarray}
Substituting \eqref{eq:DOmega3G} and noting the the angular eigenfunctions $\Omega^{(3)}$ are eigenstates of $\sigma^{t}, \sigma^{r}$, we can directly perform the angular integrals and obtain \eqref{eq:sch}. Divergences are regulated using differential regularization \cite{Freedman:1991tk, Adam:1993fy, Chen:1999jg} where the only difference in our case is the underlying ${AdS}_2$ metric.

Finally, we argue that the term \eqref{eq:sch} is 1-loop-exact. To see this, we present an alternative derivation of \eqref{eq:sch} related to the 4D \textit{chiral anomaly}, inspired by \cite{Nair:1983ps}. Since the latter is famously 1-loop-exact \cite{Adler:1969er}, this implies that \eqref{eq:sch} is, too. The trick is to change variables in the fermionic path integral so that the new fermions decouple from $A^{{\rm vor}, n}_
\mu$. We argued above that only the \textit{lowest partial wave} of the fermions actually contributes to \eqref{eq:sch}, a fact that will be further justified in Section \ref{sec:2D reduction}. 
Thus, we need a redefinition that decouples the lowest partial wave fermions from $A^{{\rm vor}, n}_\mu$. Such a field redefinition is the \textit{axial rotation}\footnote{The attentive reader may be worried about the missing factor of $i$ here. This is because we are working in Euclidean space. In Minkowski space the transformation has an $i$ and is a proper axial transformation. Here the chiral anomaly is implicitly defined by analytical continuation from Minkowski space.}
\begin{eqnarray}\label{eq:chirot}
    \psi\rightarrow \psi^{new}=e^{-s_qe\gamma_5\partial^{-2}E_r}\psi \, ,
\end{eqnarray}
where $s_q={\rm sign}(q)$. In terms of the new fermions, the Dirac operator acting on the lowest partial wave reads
\begin{eqnarray}
    \overline{\psi}_{j_{\rm min}}\slashed{D}\psi_{j_{\rm min}}=\overline{\psi}^{new}_{j_{\rm min}}e^{s_qe\gamma_5\partial^{-2}E_r}\slashed{D}e^{s_qe\gamma_5\partial^{-2}E_r}\psi^{new}_{j_{\rm min}}=\overline{\psi}^{new}_{j_{\rm min}}\gamma^{\mu}(\partial_\mu-ieA^{\rm mon}_\mu)\psi^{new}_{j_{\rm min}}\,.
\end{eqnarray}
In other words, the chiral rotation \eqref{eq:chirot} \textit{decouples} $\psi_{j_{\rm min}}$ from $A^{{\rm vor}, n}_\mu$ and so these gauge fluctuations can no longer contribute to \eqref{eq:sch}, while the other partial waves only contribute to the photon field strength renormalization \cite{Rubakov:1982fp}. So where does \eqref{eq:sch} come from in this picture? The mystery is lifted once we remember that the axial rotation \eqref{eq:chirot} suffers from the ABJ anomaly \cite{Adler:1969er,Bell:1969ts}. In other words, the non-invariance of the path integral measure under \eqref{eq:chirot} means that the axial symmetry is anomalous. Thus there is a (partially non-local) three gauge boson vertex generated:
\begin{eqnarray}\label{eq:gtr}
    {\rm log\, det}^\prime(i\slashed{D})=\int d^4x\,\frac{e^2}{16\pi^2}\,s_qe\,\partial^{-2}E_r\,F_{\mu\nu}\widetilde{F}^{\mu\nu}=\int d^4x\,\frac{e^2}{4\pi^2}\,E_r\partial^{-2}E_r\,s_qeB_r+\ldots 
\end{eqnarray}
Note that while this vertex comes from a triangle diagram, it does not indicate a gauge anomaly, it is perfectly gauge invariant. This seems to contradict our standard intuition that gauge triangle diagrams necessarily violate gauge symmetry, i.e. constitute a gauge anomaly. In fact, this statement is only true when we impose $4D$ \textit{Lorentz symmetry}. Since the monopole background explicitly broke $4D$ Lorentz (in fact conformal) symmetry to $SL(2,R)\times SO(3)$, the isometry of $AdS_2\times S_2$, there are now more terms that we are allowed to write, that would normally be forbidden by $4D$ Lorentz invariance. One of these terms is the gauge invariant \eqref{eq:gtr}. Furthermore, the local and non-local pieces in \eqref{eq:gtr} conspire to produce a gauge invariant term, in analogy with the Schwinger mass term for the photon in $QED_2$. The latter is usually (and correctly) attributed to the $2D$ axial anomaly. However one could just as well attribute the Schwinger mass to a non-anomalous gauge boson ``self-energy'' diagram generating a term similar to \eqref{eq:sch}, in which the local and non-local term combine into a gauge invariant result.

Substituting the monopole magnetic field $B_r=g/er^2$ and integrating over $(\theta,\varphi)$, we get 
\begin{eqnarray}
    {\rm det}^\prime(i\slashed{D})=\exp{ \frac{m^2}{2}\int dt dr\,\,E_r\partial^{-2}E_r+\ldots}\,,
    \label{eq:triangle}
\end{eqnarray}
which is exactly equal to \eqref{eq:sch}. Since the chiral anomaly is 1-loop exact, this proves the 1-loop exactness of \eqref{eq:sch}.

\subsection{Summary}
%
To summarize the current state of our investigation, we note the following points
\begin{itemize}
\item In the background of a Dirac monopole, there are Abelian field configurations with nonzero winding number $Ch_2$. We call them \textit{Abelian instantons}.
\item In the background of Abelian instantons, charged fermions have $2\,|Ch_2|$ zero modes per flavor.
\item the zero modes non-perturbatively induce an 't~Hooft vertex, which leads to \textit{monopole-catalysis} correlators like \eqref{eq:cor4}.
\item The leading contribution to ${\rm det}^\prime(i\slashed{D})$ is the Schwinger-like ``mass term'' for the radial photon \eqref{eq:sch}-\eqref{eq:schAds} (equivalently the three gauge boson term \eqref{eq:triangle}), generated by a loop of lowest partial-wave fermions in the monopole background. This term will be crucial in obtaining Abelian instantons as the saddle point of the full $QED_4$ path integral in the background of a monopole.
\end{itemize}
We also note the remaining parts of the story required for a full understanding and calculation of the monopole catalysis processes:
\begin{enumerate}
\item We need to perform the \textit{gauge part} of the path integral for $QED_4$ and show that it is dominated by Abelian instantons. Unlike the non-Abelian variety, Abelian instantons appear in the leading vacuum of the path integral, rather than in exponentially suppressed higher vacua. As such, they are not \textit{tunnelling} processes \cite{Rubakov:1982fp}.
\item We need to add Wilson lines to the fermionic correlators to make them gauge invariant.
\item We need to reproduce all known monopole catalysis correlators \cite{Kazama:1976fm, Rubakov:1982fp, Callan:1982au, Callan:1982ah, Polchinski:1984uw, Craigie:1984pc, Affleck:1993np} to show that they are induced by Abelian instantons.
\end{enumerate}

These points are addressed in the next two sections.

\section{\texorpdfstring{$QED_4$}{} in a Monopole Background}\label{sec:2D reduction}
%
In the previous section, we have encountered some of the key elements in the derivation of monopole catalysis correlators. In particular, we considered the fermionic path integral \eqref{eq:4DfPI} in the presence of the Abelian instanton \eqref{eq:fcon}, and derived the resulting 't~Hooft vertex which leads to monopole catalysis. 
What we have yet to show is how Abelian instantons of the form \eqref{eq:fcon} are actually generated in the full path integral of $QED_4$ in the background of a Dirac monopole. In this section we explicitly show how Abelian instantons of the form \eqref{eq:fcon} are not only an option, but actually the dominant (and only) field configurations in monopole-catalysis correlators. 
To understand the relation between our derivation and those of Rubakov \cite{Rubakov:1982fp} and Callan \cite{Callan:1982ah,Callan:1982au}, note that the latter were only strictly valid in the $n/2=0$ sector of the theory. For this reason, Rubakov and Callan could only infer the existence of monopole-catalysis correlators \textit{indirectly}, by considering non-winding correlators that are secretly a product of an instanton and an anti-instanton, and invoking cluster decomposition. Our derivation, on the other hand, is valid in sectors with non-zero winding $n\neq 0$, and so it allows us to directly compute monopole catalysis correlators without appealing to cluster decomposition, highlighting their Abelian instanton nature.

The picture that emerges is then of a synergistic effect: in the background of Dirac monopole, the fermions insertions in monopole-catalysis correlators \textit{source} Abelian instantons. The Abelian instantons generate 't~Hooft vertices among the inserted fermions. We now derive this picture in detail, and in the next section use it to reproduce all known monopole-catalysis correlators in the literature.

\subsection{Setup of the Theory and Truncation of Higher Partial Waves}\label{sec:setup and trunc}
%
Consider $QED_4$ with $N_f$ Dirac fermions of charge 1 (in units of $e$).  We wish to investigate this theory in the presence of an infinitely massive monopole residing at $r=0$, i.e. in the background of $A^{\rm mon}_{\mu}$ defined in Eq.~\eqref{eq:dirac potential}. The partition function of this theory in Euclidean space is
\begin{eqnarray}\label{eq: vanilla 4D action}
    Z[J^\mu, \overline{\eta},\eta] &=& \int {\cal D} A_\mu \,\prod_{f=1}^{N_f}\,{\cal D}\overline{\psi}_f {\cal D}\psi_f\, e^{-S_{mQED_4}}\nonumber\\[5pt]
S_{mQED_4}&=&-\int d^4 x\,\left\{-\frac{1}{4} F^{\mu\nu}F_{\mu\nu} + J^\mu A_\mu\,+\,\sum_f\left(i\overline{\psi}_f\slashed{D}\psi_f+\overline{\eta}_f\psi_f+\overline{\psi}_f\eta_f\right)\right\}\,,
\end{eqnarray}
where the covariant derivative on the fermions depends on $A^{\rm tot}_{\mu}=A^D_\mu+A_\mu$. The subscript $mQED_4$ stands for $QED_4$ in the background of the monopole potential $A^D_\mu$. We would like to use the path integral \eqref{eq: vanilla 4D action} to compute non-perturbative correlators over fermions in the background of the monopole. Following \cite{Rubakov:1982fp, Callan:1982ah,Callan:1982au}, we consider the partial-wave expansions for the photon and the fermions (see Appendix~\ref{app:partial wave decomp} for details)
\begin{eqnarray}\label{eq:photon PWDmain}
    A^t(x) &=&  \frac{1}{\sqrt{4\pi R^2}}\,a^t(t,r)+ \sum_{j>0,m} {\rm h.p.w}~~,~~
    \vb{A}(x) = \frac{1}{\sqrt{4\pi R^2}}\,a^r(t,r)\,\vu{r} + \sum_{j > 0,m,\ell}\,{\rm h.p.w}\nonumber\\[5pt]
    \psi_f(x) &=& \frac{1}{R}\,{\left(\frac{R}{r}\right)}^{\frac{3}{2}}\,\sum_{m=-j_{\rm min}}^{j_{\rm min}} \colvec{\chi_{fm,L}(t,r)\,\Omega^{(3)}_{qj_{\rm min}m}(\theta,\varphi)\\\chi_{fm,R}(t,r)\,\Omega^{(3)}_{qj_{\rm min}m}(\theta,\varphi)} + \sum_{j > j_{\rm min},\,m}\,{\rm h.p.w} \, ,
\end{eqnarray}
where $j_{\rm min}=|q|-\frac{1}{2}$, h.p.w stands for higher partial waves and $R$ is a length scale which drops out of all physical observables. We also define the $2D$ Dirac spinors $\chi_{fm}=(\chi_{fm,L},\chi_{fm,R})$. Since the higher partial wave are obtained via dimensional reduction on the $S^2$ spanned by $(\theta,\varphi)$, we will also refer to them as Kaluza-Klein (KK) modes in the following. We can substitute \eqref{eq:photon PWDmain} back into \eqref{eq: vanilla 4D action}, and explicitly perform the $\theta,\varphi$ integral. The result, as expected, is a $2D$ action for $a_\alpha$ and $\chi_f$ plus $n_f$ towers of massive $2D$ fermions $\xi^{(i)}_{fjm}\,,\,i\in\{1,2\}$ and a tower of massive 2D gauge bosons $W^{(i)}_{jm\,\alpha},\,i\in\{1,2\}$, which are linear combinations of the higher partial waves. We can go one step further; the normalization in \eqref{eq:photon PWDmain} was chose so that the effective $2D$ theory can be cast into the form of \textit{axial $QED_2$ in} $AdS_2$. Defining the ${AdS}_2$ metric $g_{\rm 2D}=(R/r)^2\,{\rm diag}(1,1)$, the $AdS_2$ formulation rids us of the awkward $r$ dependence from the spherical Jacobian and encapsulates it in the $AdS_2$ metric (see \cite{Cuomo:2021rkm, Aharony:2023amq} for a related discussion). All-in-all, we get
\begin{eqnarray}\label{eq:2deftfull}
    &&S_{mQED_4}=S_a+S_\chi+S_{KK}\nonumber\\[5pt]
    &&S_a=\int dt dr\,\sqrt{g_{\rm 2D}}\,\left(\frac{1}{4} f^{\alpha\beta}f_{\alpha\beta} -J^\alpha_a a_\alpha\right)\nonumber\\[5pt]
    &&S_\chi=\int dt dr\,\sqrt{g_{\rm 2D}}\,\left[\sum_{f=1}^{N_f}\,\sum_{m=-j_{\rm min}}^{j_{\rm min}}\left(i\bar{\chi}_{fm}\slashed{D}\chi_{fm}+\overline{\eta}_{\chi_{fm}}\chi_{fm}+\overline{\chi}_{fm}\eta_{\chi_{fm}}\right)\right]\,,
\end{eqnarray}
where $f_{\alpha\beta} = \partial_\alpha a_{\beta} - \partial_\beta a_\alpha$ and $J^\alpha_a,\,\eta_{\chi_{fm}}\,\overline{\eta}_{\chi_{fm}}$ are the lowest partial waves of the sources $J_\mu,\,\eta_f,\,\overline{\eta}_f$. Here the covariant derivative on the fermions is 
\begin{eqnarray}
    \slashed{D} \chi=\gamma^\alpha\left(\nabla_\alpha-e_{2D}a_\alpha\gamma_5\right) \chi\,.
\end{eqnarray}
Here $e_{\rm 2D}=e/\sqrt{4\pi R^2}$ is the effective 2D gauge coupling, which has mass dimension $1$. Our conventions for $AdS_2$, including the $AdS_2$ covariant derivative $\nabla_\alpha$ and $2D$ gamma matrices, are given in Appendix~\ref{app:ads2}. Finally, $S_{KK}$ is the action involving all gauge and fermion KK modes, as well as their interactions with $a_\alpha$ and $\chi_f$,
\begin{eqnarray}
    S_{KK}&=&\int dt dr\,\sqrt{g_{\rm 2D}}\,\left\{\sum_{j>0}\sum_{m=-j}^j\sum_{i=1,2}\,\left(-\frac{1}{4} F^{(i)\,\alpha\beta}_{jm}F^{(i)}_{jm\,\alpha\beta}-\frac{1}{2}\,\mu^2_{W_j}\,W^{(i)\,\alpha}_{jm}W^{(i)}_{jm\,\alpha}-J^\alpha_{W^{(i)}_{jm}} W^{(i)}_{jm\,\alpha}\right)+\right.\nonumber\\[5pt]
    &&\left.~~~~~~~~~~~~~~~~~~~\sum_{f}\sum_{j>j_{\rm min}}\sum_{m=-j}^j\sum_{i=1,2}\,i\bar{\xi}^{(i)}_{fjm}\,(\slashed{D}-\mu_{\xi^{(i)}_j})\,\xi^{(i)}_{fjm}+{\rm interactions}\right\}\,
\end{eqnarray}
where $F^{(i)}_{j m \, \alpha \beta} = \partial_\alpha W^{(i)}_{jm\, \beta}  - \partial_\beta W^{(i)}_{jm\, \alpha}$, $\mu^2_{W_j}=j(j+1)$ and $\mu_{\xi^{(i)}_j}$ is the fermion KK mass. The full KK Lagrangian including the interaction terms (i.e. the covariant coupling of the fermions to the photon KK modes and the coupling of the fermion KK modes to $a_\mu$) is given in Appendix~\ref{app:partial wave decomp}.

The standard practice in the literature \cite{Rubakov:1982fp, Callan:1982ah, Callan:1982au} is to truncate the higher partial waves in \eqref{eq:2deftfull} by simply omitting $S_{KK}$ from the action, leaving just the s-wave photon $a_\alpha$ and the lowest-partial-wave fermions in the theory. The underlying logic is that monopole catalysis is generated non-perturbatively in the truncated theory and that the KK modes can at most lead to a perturbative modification of that statement. Recently, however, the validity of this truncation was scrutinized more rigorously in \cite{vanBeest:2023dbu}. Evidently, the above argument for truncating \eqref{eq:2deftfull} is only valid when all of the effective operators generated by integrating out the KK modes are \textit{irrelevant}, and so cannot modify the IR phase of the theory. Unfortunately, for non-minimal monopoles $(q>1/2)$, integrating out the photon KK modes results in the generation of marginal quartic (as well as higher-point) interaction terms among the $\chi_f$. Depending on the sign of these marginally relevant operators\footnote{Since we have the complete 4D theory, it should be possible, at least in principle, to calculate the relevant signs. We leave this for future work.}, the theory may generate a gap in the IR \cite{Coleman:1976uz, Cherman:2022ecu, Dempsey:2023gib}, effectively destroying any would-be zero-modes\footnote{We would like to thank Zohar Komargodski for explaining this important subtlety to us.}. For example, consider the case with $N_f=1$ and $q=1$. In this case, $j_{\rm min}=\frac{1}{2}$, and there are two fermionic lowest partial waves $\chi_{m=\pm1/2}$. Integrating out the KK modes then generates the dimension 2 quartic
\begin{eqnarray}\label{eq:marginal}
\mathcal{O}_{\rm quartic}&=&\,\sum_{i=1}^3\,j^i_{\,\alpha} j^{i\,\alpha}~~~,~~~j^i_{\,\alpha}\equiv\,\frac{1}{2}\,\sum_{mm'=\pm\frac{1}{2}}\,\bar{\chi}_{m}\sigma^i_{mm'}\gamma_\alpha\chi_{m'}\,,
\end{eqnarray}
where the $\sigma^i_{mm'}$ are Pauli matrices. This is the interaction term of the $SU(2)$ Thirring model -- famously dual\footnote{In our case the theory is also coupled to axial $QED_2$, so further analysis is required.} to the theory of a $2D$ free massless scalar and a \textit{Sine-Gordon} field \cite{BANKS1976}. Depending on the uncalculable sign of $\mathcal{O}_{\rm quartic}$, the theory may or may not exhibit a Berezinskii–Kosterlitz–Thouless (BKT) phase transition \cite{Kosterlitz1973, DJAmit_1980, Dempsey:2023gib}, possibly invalidating our zero mode-calculation. For the minimal monopole, angular momentum selection rules forbid the coupling between the lowest partial wave fermions $\chi$ and the photon higher partial waves $W^{(i)}_{jm}$. In the sector involving $\chi$ and $W^{(i)}_{jm}$, this leads to an accidental enhanced global symmetry $SU(N_f) \times SU(N_f)$ acting independently on the $\chi_f$ and the $\bar{\chi}_f$. When integrating our the photon KK modes $W^{(i)}_{jm}$, the latter symmetry remains unbroken and forbids the generation of quartics of the form \eqref{eq:marginal}, as well as all other marginally relevant terms. For non-minimal monopoles the $\chi$ couple to $W_{jm}$, and global symmetry is just the diagonal $SU(N_f)_V$, which allows for marginally relevant terms of the form \eqref{eq:marginal}. A similar conclusion was also reached in \cite{vanBeest:2023dbu}. For this reason, for the rest of the paper we focus on the minimal monopole case with $q=1/2$ and $j_{\rm min}=0$, and truncate the theory by omitting $S_{KK}$ and the KK modes from our path integral. To show the explicit dependence on $q$ we retain it as an explicit parameter in our derivation, but we caution the reader that our results show only be fully trusted when $q=1/2$. Nevertheless, the elegance of our results hints they may be valid even for generic values of $q$.

\subsection{Topological Sectors and Explicit Results for the Path Integral}
%
After truncating all KK modes, we have
\begin{eqnarray}\label{eq: vanilla 4D action s}  &&Z[J^\alpha_a,\overline{\eta}_\chi,\eta_\chi] = \int {\cal D} a_\alpha\, e^{-S_a}\,Z_\chi[a_\alpha,\overline{\eta}_\chi,\eta_\chi]~~,~~Z_\chi[a_\alpha,\overline{\eta}_\chi,\eta_\chi]=\int \prod_{f}\prod_m\,{\cal D}\overline{\chi}_{fm} {\cal D}\chi_{fm}\,e^{-S_\chi}\,.\nonumber\\
\end{eqnarray}
Note that we keep this expression in its full generality for arbitrary $q$, but it should only be fully trusted for $q=1/2$ due to the aforementioned marginal operators from integrating out the KK modes. In the $q=1/2$ there is no sum over $m$ since $j_{\rm min}=0$.
We note that $Z_\chi[a_\alpha,\overline{\eta}_\chi,\eta_\chi]$ is nothing but the truncation of 
$Z_{\psi}[A^{\rm tot}_{\mu},\overline{\eta},\eta]$ from \eqref{eq:4DfPI} to the lowest partial wave of the fermions. As a matter of fact, in the rest of the paper we will only be concerned with correlators involving these lowest partial waves, so we may as well replace $Z_\chi[a_\alpha,\overline{\eta}_\chi,\eta_\chi]$ in \eqref{eq: vanilla 4D action s} with its already-computed $4D$ uplift $Z_{\psi}[A^{\rm tot}_{\mu},\overline{\eta},\eta]$. This is a mere convenience, as we could have easily just reproduced the derivation of Section~\eqref{sec:intuition MC} for $Z_\chi[a_\alpha,\overline{\eta}_\chi,\eta_\chi]$.

In accordance with the discussion in Section~\ref{sec:Abelian Instantons}, the integration over $a_\alpha$ is limited to field configurations whose behaviour at infinity gives rise to half-integer winding number $n/2$. The asymptotic behavior of $a_\alpha$ in each sector is given by \eqref{eq:aEas}. We can, therefore, write the path integral as
\begin{eqnarray}\label{eq: vanilla 4D action sect}
    Z[J^\alpha, \overline{\eta},\eta] &= & \sum_{n=-\infty}^{\infty}\,Z^{(n/2)}[J^\alpha, \overline{\eta},\eta] \nonumber\\[5pt]
    Z^{(n/2)}[J^\alpha, \overline{\eta},\eta]&=&\int\limits_{Ch_1=n/2} {\cal D} a_\alpha \, e^{-S_a}\,Z^{(n/2)}_{\psi}[A^{tot}_{\mu},\overline{\eta},\eta]\,,
\end{eqnarray}
where $A^{tot}_{\mu}=A^{D}_{\mu}+A^{{\rm vor}, n/2}_{\mu}$ depends on $a_\alpha$, via \eqref{eq:aE}, which parametrizes the most general $2D$ gauge field with winding number $n/2$ (determined by its behaviour at infinity), which leads to $2|qn|$ fermionic zero modes. To dispel any potential misunderstanding, when we say zero modes we exclusively mean log-normalizable, zero eigenvalue solutions to the Dirac eqaution \eqref{eq:direq} in the background of the \textit{monopole}+\textit{gauge configuration with winding number} $n/2$. The zero modes are, in particular, lowest partial waves of the KK decomposition over $(\theta,\varphi)$. 
The explicit evaluation of $Z^{(n/2)}_{\psi}[A^{\rm tot}_{\mu},\overline{\eta}_f,\eta_f]$ was carried out in Section~\ref{sec:intuition MC}, and yielded \eqref{eq:4DfPI2} which we repeat here for completeness, 
\begin{eqnarray}\label{eq:4DfPI2rep}
    Z^{(n/2)}_{\psi}\,[A^{\rm tot}_{\mu},\overline{\eta},\eta]&=&\mathcal{F}_{n/2}[a_\alpha,\overline{\eta},\eta]\,\exp{-\Gamma'[a_\alpha,\overline{\eta},\eta]}\,{\rm det}^{\prime}\left(i\slashed{D}\right)^{N_f}\,,
\end{eqnarray}
where $\mathcal{F}_{n/2},\,\Gamma'$ and ${\rm det}^{\prime}\left(i\slashed{D}\right)^{N_f}$ are the 't~Hooft vertex, the generating functional for the fermionic non-zero modes, and the fermion determinant, which includes the 1-loop generated Schwinger mass for the (radial part of the) photon, respectively. Explicitly,
\begin{eqnarray}\label{eq:comps}
\mathcal{F}_{n/2}[a_\alpha,\overline{\eta},\eta]&=&\epsilon_{N_f}\left(X_1,\ldots,X_{N_f}\right)~~~,~~~X_f=\prod_{ml}\left[\left(\overline{\eta}_f\,\psi^{(0)}_{ml}\right)+\left(\overline{\psi}^{(0)}_{ml}\,\eta_f\right)\right]\nonumber\\[5pt]
     \Gamma'[a_\alpha,\overline{\eta},\eta]&=&\int d^4xd^4y\,\overline{\eta}_f(x)G'(x,y)\eta_f(y)\nonumber\\[5pt]
\bqty{{\rm det}^{\prime}(i \slashed{D})}^{N_f}&=&\bqty{{\rm det}^{\prime}(i \slashed{\partial})}^{N_f}\,\exp{-\int d^2x\,\sqrt{g_{\rm 2D}}\,\left[\frac{m^2_a}{2} a_{\alpha}\left(g^{\alpha\beta}_{\rm 2D}-\frac{\nabla^{\alpha}\nabla^\beta}{\square}\right)a_{\beta}+\ldots\right]}\,,\nonumber\\
\end{eqnarray}
where $m^2_a=2|q|N_fe^2_{2D}/\pi$ is the ``Schwinger mass'' in ${AdS}_2$. Furthermore, $\nabla_\alpha$ is the ${AdS}_2$ covariant derivative, and $\square=\nabla_\alpha\nabla^\alpha$. The last expression for $\bqty{{\rm det}^{\prime}(i \slashed{D})}^{N_f}$ is given in its $AdS_2$ form from \eqref{eq:schAds}, and the $\ldots$ stand for field-strength renormalization and higher terms in the effective Euler-Heisenberg Lagrangian. All of these operators are irrelevant, and we omit them. The $\epsilon_{N_f}$ is defined in \eqref{epsilonnotation}, and amounts to antisymmetrization in the flavor indices. Recall that the 't~Hooft vertex, $\mathcal{F}_{n/2}$ in \eqref{eq:4DfPI2rep}, has $2Ch_2=2|qn|$ legs for each one of the $N_f$ flavors. As the zero modes making up $\mathcal{F}_{n/2}$ are exclusively in the \textit{lowest partial wave}, this gives further justification to our above replacement of $Z_{\chi}[a_\alpha,\overline{\eta},\eta]$ by its $4D$ uplift $Z_{\psi}[A^{\rm tot}_{\mu},\overline{\eta},\eta]$; the two give the exact same results for correlators induced by the 't~Hooft vertex $\mathcal{F}_{n/2}$.
Gathering terms, we get
\begin{eqnarray}\label{eq: vanilla 4D action sect 2D}
    Z^{(n/2)}[J^\alpha, \overline{\eta},\eta]=\int\limits_{Ch_1=n/2} {\cal D} a_\alpha \, \mathcal{F}_{n/2}[a_\alpha,\overline{\eta},\eta]\,\exp{-\Gamma'[a_\alpha,\overline{\eta},\eta]}\exp{-S[a_\alpha,J_\alpha]}\,,
\end{eqnarray}
where
\begin{eqnarray}\label{eq:gaugeactads}
    S[a_\alpha,J_\alpha]=\int dt dr\,\sqrt{g_{\rm 2D}}\,\Bqty{\frac{1}{4} f^{\alpha\beta}f_{\alpha\beta} + \frac{1}{2}m^2_a\,a_{\alpha}\left(g_{\rm 2D}^{\alpha\beta} -\frac{\nabla^{\alpha}\nabla^\beta}{\square}\right)a_{\beta} -J^\alpha a_\alpha}\,.
\end{eqnarray}
Lastly, we use the generic property that any 2D gauge field can be defined by the derivatives of two scalar functions $\upsilon ,\,\sigma$ \cite{Schwinger:1962tp,Rubakov:1982fp}
\begin{eqnarray}\label{eq:2D vec identity}
    a_\alpha(x) = \frac{1}{e_{\rm 2D}} \bqty{\epsilon_{\alpha\beta} \nabla^\beta \sigma(x) + \nabla^{\alpha} \upsilon(x)} \, ,
\end{eqnarray}
and $\epsilon_{\alpha\beta}$ is the ${AdS}_2$ Levi-Civita tensor, which has an implicit factor of $\sqrt{g_{\rm 2D}}$. The scalar $\upsilon$ is gauge dependent, and we will set it to zero for the rest of the discussion. The scalar $\sigma$ is gauge invariant, and is related to the electric field $E\equiv f_{tr}={\sqrt{4\pi R^2}}\frac{r^2}{R^2}E_r$ via
\begin{eqnarray}\label{eq:boxeq}
    E = -\frac{\sqrt{g_{\rm 2D}}}{e_{\rm 2D}} \square\sigma\,.
\end{eqnarray}
This, in turn, feeds back into the fermion zero mode profiles in $\mathcal{F}_n$ via \eqref{eq:4Dzmnew}, and will be crucial in deriving vortices as the dominant and only saddle point contributions to the path integral.
In terms of $\sigma$, the path integral now reads
\begin{eqnarray}\label{eq: vanilla 4D action sect 2Dbeta}
    Z^{(n/2)}[\lambda, \overline{\eta},\eta]&=&\int\limits_{Ch_1=n/2} {\cal D} \sigma \, \mathcal{F}_{n/2}[\sigma,\overline{\eta},\eta]\,\exp{-\Gamma'[\sigma,\overline{\eta},\eta]}\exp{-S[\sigma,\lambda]}\nonumber\\[5pt]
    S[\sigma,\lambda]&=&\int d^2 x \sqrt{g_{\rm 2D}} \, \pqty{ \frac{1}{2 e_{\rm 2D}^2} \sigma(\square - m_a^2) \square \sigma - \lambda \sigma}\,,.   
\end{eqnarray}
Note that for simplicity we traded the source $J^\alpha$ for $a_\alpha$ with the source $\lambda$ for $\sigma$. Equation \eqref{eq: vanilla 4D action sect 2Dbeta}  will be our master equation for computing monopole-catalysis correlators. One can readily see that the path integral is Gaussian in $\sigma$, and can be solved explicitly.

The path integral \eqref{eq: vanilla 4D action sect 2Dbeta} very closely resembles the one for $QED_2$, i.e. the Schwinger model. The full path integral of the latter was explored in\footnote{For related work, see \cite{Rothe:1978hx,Rothe:1978hx,Adam:1994by} and the finite temperature version \cite{Sachs:1991en}, as well as the book \cite{Abdalla:1991vua}.} \cite{Hortacsu:1979fg} following the discovery of topological vortices in $QED_2$ in \cite{Nielsen:1976hs,Nielsen:1977qk}. The difference from the known $QED_2$ path integral is in (a) our $2D$ metric is ${AdS}_2$ (b) our zero modes and $G'$ differ from $QED_2$ due to the boundary condition. The non-chiral structure of the zero modes also reflects the fact that our $2D$ EFT for the lowest fermionic partial wave is in fact \textit{Axial} $QED_2$ in ${AdS}_2$.

Finally, we comment on the boundary condition for $\sigma$. A simple variation of the bulk action leads to the requirement that
\begin{eqnarray}\label{eq:gbc}
    \partial_r\sigma=0~~~,~~~r=0\,,
\end{eqnarray}
which in our Lorentz gauge implies $a_t=0$. This is also the gauge boundary condition considered in \cite{Rubakov:1982fp}. This boundary condition leads to the quantization of vortex number in half-integer units, as explained in Section~\ref{sec:Abelian Instantons}.

\subsection{Green's Functions}
%
\subsubsection{Green's Function for $\sigma$}
The Green's function $G_{\sigma}$ is the inverse of the quadratic term in the gauge action \eqref{eq: vanilla 4D action sect 2Dbeta}, satisfying
\begin{eqnarray}
    (\square - m^2_a)\, \square \,G_\sigma= \frac{e^2_{2D}}{\sqrt{g_{\rm 2D}}}\delta(x-x')\, .
\end{eqnarray}
To find its explicit form, we note the relation
\begin{eqnarray}\label{eq:betarel}
    \bqty{(\square - m^2_a)\, \square }^{-1} = \frac{1}{m_a^2}\bqty{(\square - m_a^2)^{-1}-\square^{-1}} \, ,
\end{eqnarray}
which is the ${AdS}_2$ analog of the flat space relation (in momentum space),
\begin{eqnarray}
    \frac{1}{ (p^2 - m^2_a) \,p^2} = \frac{1}{m_a^2}\bqty{ \frac{1}{p^2 - m_a^2}-\frac{1}{p^2} } \, .
\end{eqnarray}
Using \eqref{eq:betarel}, the Green's function for $\sigma$ is now given by
\begin{eqnarray}\label{eq:gbeta}
    G_\sigma(x,x')= \frac{e_{\rm 2D}^2}{m_a^2} \left[\mathcal{P}(x,x';m_a)-\mathcal{D}^N(x,x')\right]\,.
\end{eqnarray}
Here ${\cal P}(x,x';m_a)$ is the bulk-to-bulk propagator for a massive scalar in ${AdS}_2$, given explicitly in \eqref{eq:Dprop}, and $\mathcal{D}^N(x,x') = {\cal D}(t, r; t', r') + {\cal D}(t,r;t', -r')$ is the massless scalar propagator in ${AdS}_2$ with Neumann boundary conditions at $r=0$, in accordance with \eqref{eq:gbc}. Here ${\cal D}(t,r;t',r')$ is the propagator for a massless 2D scalar defined in Eq.~\eqref{eq:sch2G2}.  From here it is straightforward to get the Green's functions of $a^\alpha$ and the electric field $ E\equiv f_{tr}$
\begin{eqnarray}\label{eq:2D GF fields}
    G_{a_{\alpha}}(x,x') = \frac{1}{e_{\rm 2D}} \epsilon_{\alpha\beta} \nabla^{\beta} G_{\sigma}\, , ~~~~ G_E(x,x') = - \frac{\sqrt{g_{\rm 2D}}}{e_{\rm 2D}} \square G_{\sigma}(x,x') = - e_{\rm 2D}\sqrt{g_{\rm 2D}} {\cal P}(x,x';m_a) \, .\nonumber\\
\end{eqnarray}

In practice, we are only interested in the leading, non-perturbative contributions to the correlators responsible for monopole catalysis. To this end, we can take the $e\rightarrow 0$ limit of the massive propagator, which is the massless scalar propagator ${\cal D}^D(x,x')$ in ${AdS}_2$ with Dirichlet boundary conditions at $r=0$. 
Consequently,
\begin{eqnarray}\label{eq:Dpropm0}
    \lim_{e\rightarrow 0} G_\sigma &=& \frac{\pi}{2 \abs{q} N_f} \left[\mathcal{D}^D-\mathcal{D}^N\right] = -\frac{1}{4 \abs{q} N_f} \log\left(\frac{(t-t')^2 + (r+r')^2}{R^2}\right)\, .
\end{eqnarray}
This is the Green's function we will use to calculate correlators.
\subsubsection{Lowest Partial wave of the Fermion Green's Function for $n/2=0$}
In the non-winding sector $n/2=0$ there are no normalizable fermionic zero modes, and so $G'(x,x')=G(x,x')$, the full Green's function for fermions in the background of the monopole and $\sigma$. This Green's function satisfies the defining relation
\begin{eqnarray}\label{eq:gffbeta}
    &&i \left(\slashed{\partial}-ie\slashed{A}^{\rm mon}-ie\slashed{A}^{{\rm vor},\,n/2=0}\right) \, G(x,x') = \frac{1}{\sqrt{g_{4D,spherial}}}\delta^{4}(x-x') I_{2}\,.
\end{eqnarray}
Specifically, we will be interested in the lowest partial wave of this Green's function, with $j=j_{\rm min}$. The explicit expression for $G_{j_{\rm min}}$ is then given by 
\begin{eqnarray}\label{eq:grlowbet}
    G_{j_{\rm min}}(x,x')=e^{-\gamma_5\sigma(x)}\,\widetilde{G}_{j_{\rm min}}(x,x')\,e^{ \gamma_5\sigma(x')}\,,
\end{eqnarray}
where $\widetilde{G}_{j_{\rm min}}(x,x')$ was given in \eqref{eq:DOmega3G} (remember that $E=(r/R)^2E_r/{\sqrt{4\pi R^2}}$). In other words, the fermionic Green's function in the background of the $2D$ photon is related to the one without it, multiplied on the left (right) by $e^{-\gamma_5\sigma(x)}$ ($e^{ \gamma_5\sigma(x')}$). The intuition is that the lowest partial waves of the fermion can be decoupled from the s-wave of the photon by the rotation \eqref{eq:4Dpsipsitil}, which is related to $\sigma$ by \eqref{eq:boxeq}. Equation~\eqref{eq:grlowbet} can be checked by substituting \eqref{eq:grlowbet} into \eqref{eq:gffbeta}, noting the relation \eqref{eq:direqsub}.

\subsection{Topology}
%
We now consider the implications of a zero-mode insertion at $x'$ which provides a charge density $\rho(x,x') = 2 \abs{q} n N_f\delta(x-x')$. Using Eq.~\eqref{eq:2D GF fields} we find
\begin{eqnarray}
    a^\alpha & = & \frac{\pi n}{e_{\rm 2D}} \epsilon^{\alpha\beta} \nabla_{\beta} \bqty{{\cal P}(x,x';m_a) - {\cal D}^N(x,x')} \, , ~~~~~
    E  =   - 2 \abs{q} n N_f  \, e_{\rm 2D} \, \sqrt{g_{\rm 2D}}  {\cal P}(x,x';m_a) \,. \nn\\
\end{eqnarray}
First note that
\begin{eqnarray}\label{eq:2D a inf}
    \lim_{\abs{x} \rightarrow \infty} a^\alpha(x) = -\frac{n}{e_{\rm 2D}} \frac{\epsilon^{\alpha\beta} x_\beta}{x^2} + O\pqty{\frac{1}{\abs{x}^3}}
\end{eqnarray}
consistently with \eqref{eq:aEas}. Furthermore, we calculate the contribution of each insertion to the first Chern number
\begin{eqnarray}\label{eq:2D exact ch1}
    Ch_1 & = & \frac{e_{\rm 2D}}{4 \pi} \int d^2 x \sqrt{g_{\rm 2D}} \epsilon^{\alpha\beta} f_{\alpha\beta} = \frac{e_{\rm 2D}}{2\pi} \int d^2 x \, E = -
    \frac{n \, 2 \abs{q} N_f e_{\rm 2D}^2}{2\pi} \int d^2 x \, \sqrt{g_{\rm 2D}} {\cal P}(x,x';m_a) = n/2 \, ,~~~~~~\nn\\[5pt]
\end{eqnarray}
where the last equality is found by direct calculation. As such, we see that for $\abs{n} \times 2\abs{q} \times N_f$ such zero-mode insertions, as we have in the 't~Hooft vertices \eqref{eq:comps}, we get a $Ch_1 = n/2$ winding configuration. For this reason, we say the instanton is induced by the fermionic zero-mode insertions.

\subsection{Wilson Lines}
%
Wilson lines (WL), defined as 
\begin{eqnarray}
    W(x, x') = \exp(ie_{\rm 2D} \int_x^{x'} dy^\alpha a_\alpha) \, .
\end{eqnarray}
are used to enforce gauge invariance of the correlators in the theory. As can be seen from \eqref{eq:2D vec identity} this corresponds to sources for $\sigma$ of the form $\epsilon_{\alpha\beta} \nabla_{x'}^\beta \delta(x-x')$. Such sources do not contribute to the winding, as can be seen via direct computation of their contribution
\begin{eqnarray}\label{eq:2D WL winding}
    && \frac{m_a^2}{i\,  e^2_{\rm 2D}} \int d^2 x \, E^{\rm WL} = \int d^2 x \, \int dx^{\prime \alpha} \epsilon_{\alpha\beta} \nabla^{\beta}_{x'} {\cal P}(x,x';m_a) 
    \nn\\[5pt] & &~~~~
    = \int dx^{\prime \alpha} \epsilon_{\alpha\beta} \nabla^{\beta}_{x'} \, \int d^2 x \, {\cal P}(x,x';m_a) = 0 \, ,
\end{eqnarray}
where $E^{\rm WL}$ is the electric field sourced by the WL and we used Eq.~\eqref{eq:2D exact ch1} in the last step.

\section{Instanton-Mediated Correlators}\label{sec:corr}
We are now in the position to reap the fruit of our conceptual and quantitative understanding of the monopole catalysis EFT. Throughout this section we will calculate various correlators in the $e\rightarrow 0$ limit, reproducing all of the known correlators in the literature. Some of these correlators have only been computed in a model with an effective boundary condition \cite{Affleck:1993np, Maldacena:1995pqn, Smith:2019jnh, Smith:2020nuf, Hamada:2022eiv,vanBeest:2023dbu,vanBeest:2023mbs} (see Section~\ref{sec:DBC}), while some of them have been calculated in the full EFT but only indirectly using cluster decomposition of non-winding correlators \cite{Rubakov:1982fp, Craigie:1984pc}. It is encouraging to see that all of these correlators can be calculated directly from the full path integral \eqref{eq: vanilla 4D action sect 2Dbeta}. All of these correlators get their dominant (in fact only) contribution from a \textit{winding gauge field configuration}---an Abelian instanton, exactly of the form discussed in Section~\ref{sec:Abelian Instantons}. Importantly and in radical contrast with non-Abelian instantons, the vortices appearing in our correlators \textit{do not have collective coordinates}, and they cannot be moved or rotated. They are simply the (winding) electric field generated by the charged fermionic insertions in the correlator. Furthermore, unlike non-Abelian instantons, their contribution to the path integral is \textit{not exponentially suppressed} \cite{Rubakov:1982fp}. To understand this heuristically, note that the vortex contribution always contributes a factor
\begin{eqnarray}\label{eq:faction}
    \sim e^{-\sum_{ij}\,G_{\sigma}(x^{source}_{i},x^{source}_{j})}\,.
\end{eqnarray}
However the Green's function $G_\sigma$ given in \eqref{eq:gbeta} is essentially a logarithm of the Euclidean distance between the insertion points. For this reason, the factor \eqref{eq:faction} never leads to an exponential suppression of monopole catalysis correlators, unlike the parallel situation for non-Abelian instantons.

We remind the reader that $n$ is the winding number of $a^\alpha(\abs{x} \rightarrow \infty)$, $N_f$ is the number of flavors of fermions in the 4D theory and $q = Q g$ is a half-integer determined by Dirac quantization. Without loss of generality, we rescale all of our electric charges to be $|Q|= 1$ in units of $e$, so that $g=q$ is half-integer. Furthermore, in all but one case we will only consider a minimal monopole with magnetic charge $q=1/2$, so that the overall instanton number is $Ch_2=qn=n/2$. Every correlator will then have $2Ch_2N_f=nN_f$ fermions, $n$ from each flavor. For charge conservation, we only consider correlators where half of the fermions are $\psi$ and half are $\overline{\psi}$.

\subsection{Chirality Flip}
%
The first indication for surprising effects in monopole fermion scattering was in the classic work of ref.~\cite{Kazama:1976fm}. In this work, the authors considered a single fermion flavor in the background of a monopole with \textit{no dynamical photon}. Here we will analyze this situation, as well as a related 2-flavor process, reformulating the result of \cite{Kazama:1976fm} is a more modern language. Then we will show how the same setup unfolds in the true EFT defined by \eqref{eq: vanilla 4D action sect 2Dbeta}. As we already know, the lowest partial wave of the solution to the monopole-background Dirac equation is of the form \eqref{eq:psitilswave}. Focusing on the $q=1/2$ case and working in the Fourier representation with respect to time, we have
\begin{eqnarray}
    \widetilde{\psi}(x) & = & \int\,dk\,e^{ikt}\,\left[\chi^L_k(t,r)\,e^{ikr}\colvec{\Omega^{(3)}_{1/2,0,0}\\0}+\chi^R_k(t,r)\,e^{-ikr}\colvec{0\\\Omega^{(3)}_{1/2,0,0}}\right]\,\nonumber\\[5pt]
    \overline{\widetilde{\psi}}(x)&=&\int\,dk\,e^{ikt}\,\left[\overline{\chi}^L_k(t,r)\,e^{-ikr}\colvec{\Omega^{(3)}_{-1/2,0,0}\\0}+\overline{\chi}^R_k(t,r)\,e^{ikr}\colvec{0\\\Omega^{(3)}_{-1/2,0,0}}\right]\,.
\end{eqnarray}
This is already surprising. Since probability and angular momentum are conserved, it seems that there are only two options for the scattering of a single flavor off a monopole: either charge violation (more correctly charge-deposition on the monopole) or chirality violation. Which one of these is realized depends on the boundary condition enforced at $r=0$. We can summarize the two possibilities as
\begin{itemize}
    \item $\chi^L_k=\chi^R_k$ and $\overline{\chi}^L_k=\overline{\chi}^R_k$. This is (in our language) the possibility enforced in \cite{Kazama:1976fm}, by including a fictitious small anomalous magnetic moment for the fermions. The implication for scattering processes in that incoming LH fermions in the lowest partial wave change their chirality and come back as RH fermions, conserving charge in the process. As we shall see, the outcome of this option is also the correct one as dictated by the true EFT, but the underlying physics is very different. Instead of a fictitious anomalous magnetic moment / boundary condition enforcing $\chi^L_k=\chi^R_k$ and $\overline{\chi}^L_k=\overline{\chi}^R_k$, the true EFT involves a 't~Hooft vertex conserving charge and violating chirality.
    \item The fermions have the boundary condition
    \begin{eqnarray}\label{eq:chv}
        \widetilde{\psi}=\overline{\widetilde{\psi}}\,,
    \end{eqnarray}
    at $r=0$, leading to $\chi^L_k=\overline{\chi}^L_k$ and $\chi^R_k=\overline{\chi}^R_k$. This option superficially conserves chirality but does not conserve the electric charge of the fermions, which reflects the possibility of depositing charge on the monopole \cite{Besson:1980vu,Rubakov:1982fp, Callan:1982ah,Callan:1982au, Isler:1987xn, Hook:2024als}. In the purely QM setting of \cite{Kazama:1976fm}, this option is out of the question. The picture is more subtle in the true EFT. While the correct boundary condition reflecting the UV physics is indeed \eqref{eq:chv}, in practice it never actually plays a role in monopole cataylsis correlators, since they never involve left- and a right-moving fermions of the same flavor \cite{Rubakov:1982fp, Craigie:1984pc}. 
\end{itemize}
All-in-all, Ref.~\cite{Kazama:1976fm} shows that the chirality-flip process
\begin{eqnarray}\label{eq:kaz}
    \psi_L + {\rm mon} \rightarrow \psi_R + {\rm mon} \, ,
\end{eqnarray}
where $\psi_{L,R}=\tfrac{1}{2}(1\pm\gamma_5)\,\psi$, saturates the s-wave unitarity bound\footnote{The unitarity bound for 2-point functions of spin 1/2 fermions in CFTs is $\ev{\psi_1 \psi_2} \sim x^{-(d-1)/2}$ \cite{Poland:2018epd}.} for the fermions. This is consistent with the full EFT, though for different reasons than the fictitious anomalous magnetic moment of \cite{Kazama:1976fm}. We note in passing that for $|q|>1$, the process \eqref{eq:kaz} claimed in \cite{Kazama:1976fm} is inconsistent with the true EFT, this is easily seen from the fact that the 't~Hooft vertex in the true EFT has $2|qn|N_f>2$ legs for $|q|>1$. We now pass to the EFT~\eqref{eq: vanilla 4D action sect 2Dbeta} in order to compute the correlator responsible for \eqref{eq:kaz}.

\subsubsection{Two Flavor Case, $q=1/2$}
%
First, we compute a correlator in a slightly different theory from the one considered in \eqref{eq:kaz}. Nevertheless it also gives rise to a chirality-flip process, so we include it here for completeness. The theory we consider here has 2 flavors, $N_f=2$, and also take $q = 1/2$. The relevant chirality-flip process is
\begin{eqnarray}\label{eq:kaz2f}
    \psi_{1L} + {\rm mon} \rightarrow \psi_{2R} + {\rm mon} \, ,
\end{eqnarray}
where $1$ and $2$ are flavor indices. There is also an equivalent process in which $\psi_{2L}$ is incoming and $\psi_{1R}$ is outgoing. Note that in this case, flavor conserving processes of the form \eqref{eq:kaz} cannot happen, since the 't~Hooft vertex has legs of \textit{every} flavor in the theory.

We derive the correlator $C^{flip}_{(q=1/2,N_f=2)}$ leading to \eqref{eq:kaz2f} by taking the appropriate derivatives of the partition function\\
\begin{eqnarray}
    C^{flip}_{1/2,2} &= &\frac{1}{2} \ev{\epsilon^{fg}\overline{\psi}_{f\, L}(x_1) W(x_1, x_2) \psi_{g \, L}(x_2)}\nn \\[5pt]
    &=& \frac{1}{2}\,\epsilon^{fg}\,P_L \pdv{}{\eta_f} \exp{i\int_{x_1}^{x_2}  dy^\alpha \epsilon_{\alpha\beta} \nabla^\beta \pdv{}{\lambda}} P_L\pdv{}{\overline{\eta}_g} \log Z~~~{\rm at }~~~{\eta_{f,g}=\lambda =0} \, .
\end{eqnarray}
Note that the relation between the correlator and the process \eqref{eq:kaz2f} is that to get from one to the other you have to take the hermitian conjugate of the out state. For this reason, the 2nd flavor is represented in the correlator by $\psi_{2L}$, which becomes right-handed in the process \eqref{eq:kaz2f}. By direct calculation
\begin{eqnarray}
    && C^{flip}_{1/2,2} = \frac{F^{\theta\varphi}_{12}}{2\pi r_1 r_2 R} \frac{1}{Z[0,0,0]}\int {\cal D}\sigma \, \exp\pqty{- S_\sigma + \int d^2 x \, \sqrt{g_{\rm 2D}}\,  \sigma(x)\, \rho(x) } \, , \nn \\[5pt] 
    && \int d^2 x  \sqrt{g_{\rm 2D}} \, f(x) \rho(x) = - f(x_1)  - f(x_2) + i \int_{x_1}^{x_2} d y^\alpha \epsilon_{\alpha\beta} \nabla^\beta f(y)) \, ,
\end{eqnarray}
where $F^{\theta\varphi}_{12}=\Omega^{(3)}_{\frac{1}{2},\, 0}(\theta_1, \varphi_1)\otimes\Omega^{(3)\, \dagger}_{-\frac{1}{2},\, 0}(\theta_2, \varphi_2)$.
This is a Gaussian integral which can be solved \emph{exactly}
\begin{eqnarray}
    C^{flip}_{1/2,2}  = \frac{F^{\theta\varphi}_{12}}{2\pi r_1 r_2 R}\exp\pqty{\frac{1}{2} \int d^2 x \sqrt{g_{\rm 2D}(x)} \, d^2 y \sqrt{g_{\rm 2D}(y)} \rho(x) G_{\sigma}(x,y)\rho(y) } \, .
\end{eqnarray}
Using the leading order Green's function for $\sigma$ \eqref{eq:Dpropm0} yields
\begin{eqnarray}\label{eq:CF 2D flat corr}
    C^{flip}_{1/2,2}  = \frac{F^{\theta\varphi}_{12}}{2\pi r_1 r_2 (\Delta t_{12}-i\Sigma r_{12})}\, .
\end{eqnarray}
where we denote $ \Delta x_{ij}\equiv x_i - x_j$ and $\Sigma x_{ij}\equiv x_i + x_j $. Note that this correlator is defined up to an arbitrary phase.

\begin{figure}[t]
    \begin{center}
    \includegraphics[width=0.48\linewidth]{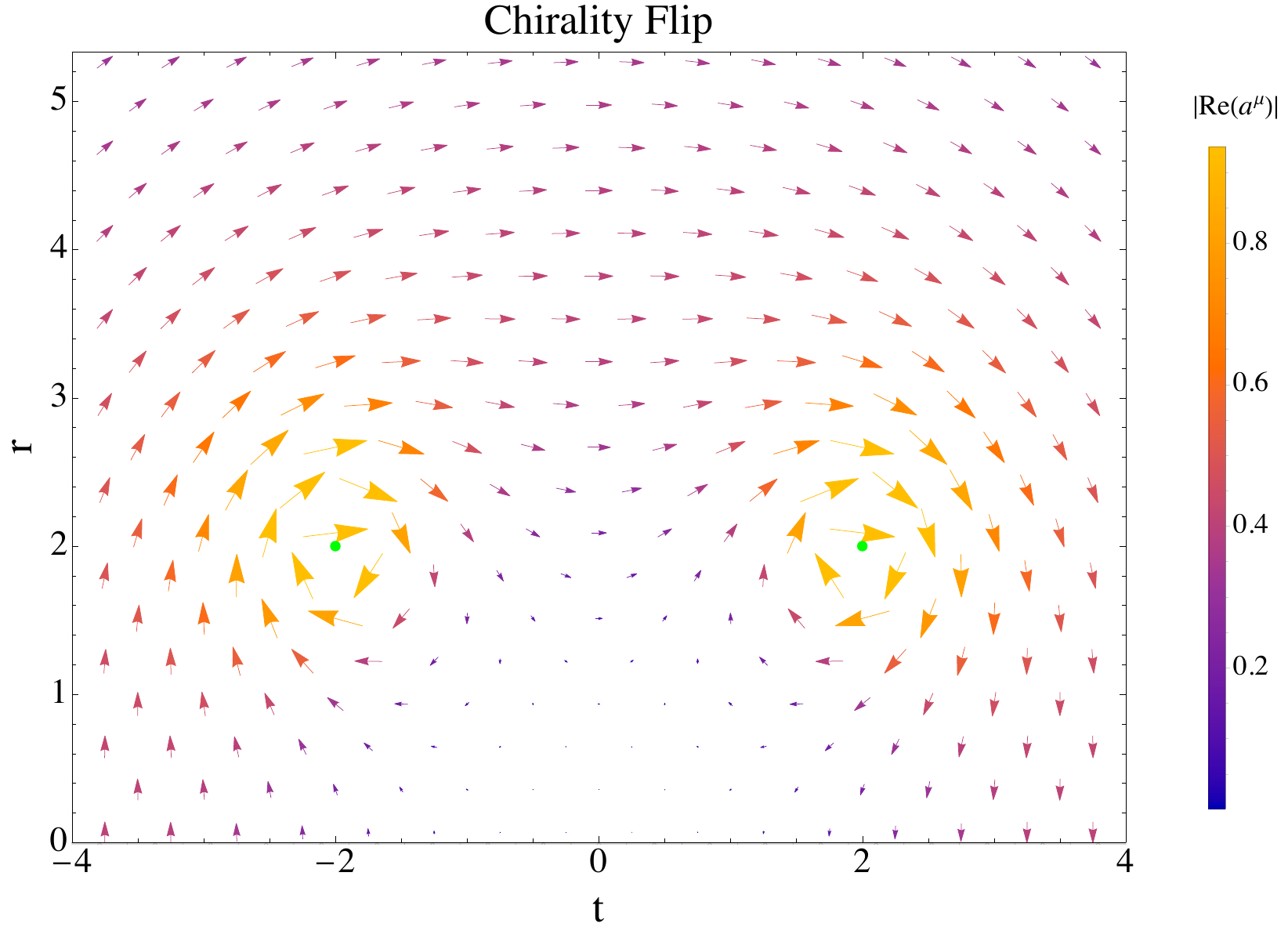} ~
    \includegraphics[width=0.48\linewidth]{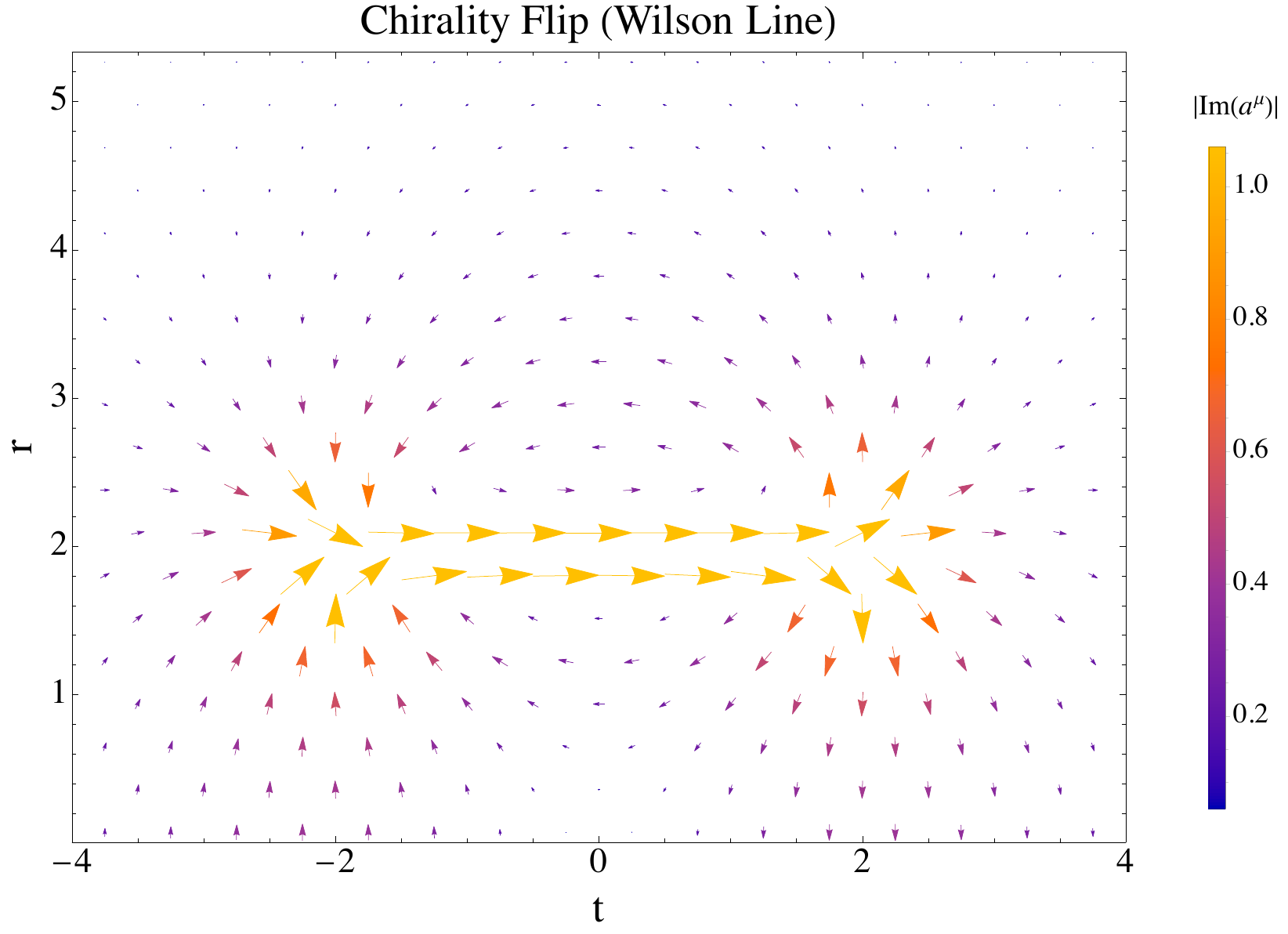}
    \caption{Vector plots of $a^\alpha$ for a chirality flip process. The sources are placed at $x_1 = (-2,2)$ and $x_2 = (2,2)$ indicated by green dots. 
    {\rm Left.} The (real) contribution to $a^\alpha$ from the $\delta$ function sources. We see that the charges have equal contributions to the vorticity of $a^\alpha$ leading to an overall $n/2=1/2$ vortex. {\rm Right.} The (imaginary) contribution to $a^\alpha$ from the WL sources. We see that the WL does not produce any winding, which is consistent with Eq.~\eqref{eq:2D WL winding}.}\label{fig:CF vecplot}
    \end{center}
\end{figure}

\subsubsection{One Flavor Case, $q=1/2$}
Here we compute the correlator responsible for \eqref{eq:kaz} in the true EFT~\eqref{eq: vanilla 4D action sect 2Dbeta} for $q=1/2,\,N_f=1$:
\begin{eqnarray}
    C^{flip}_{1/2,1} & = & \ev{\overline{\psi}(x_1) W(x_1,x_2) \psi(x_2)} \, .
\end{eqnarray}
As we shall see, the gauge saddle point of this correlator has vortex number $n/2 = 1$, which is consistent with $2|qn|N_f=2$ legs on the 't~Hooft vertex. By a completely analogous computation to the $N_f=2$ case, we get
\begin{eqnarray}\label{eq:nf1}
    C^{flip}_{1/2,1} = C^{flip}_{1/2,2}\, ,
\end{eqnarray}
which is expected, given that the two correlators saturate their s-wave unitarity bound in their respective theories. Amusingly, the saddle-point gauge field configuration which gives the dominant and only contribution to the gauge path integral for \eqref{eq:nf1} has vortex number $n/2=1$. This is consistent with having $2|qn|N_f=2$ legs on the 't~Hooft vertex.

\subsection{Cluster Decomposition}
%
Cluster decomposition allows us to study non-perturbative topological effects in the system by considering only the topologically trivial, non-winding sector $n/2=0$. This method was originally applied in the study of the Schwinger model \cite{Rothe:1978hx} and later for QED in a monopole background \cite{Rubakov:1982fp, Affleck:1993np}. The idea is to consider correlators, comprised of two 'clusters' of fermionic insertions. Counter to perturbative expectations, when the $4D$ distance between the two clusters goes to infinity, the correlator does not vanish. Instead, it is dominated by the product of an instanton for the first cluster and an anti-instanton for the second cluster.

To see this explicitly, we consider the case $N_f = 2, q=1/2$ and compute the correlator
\begin{eqnarray}
    C^{\rm CD}_{1/2,2} = \frac{1}{4}\ev{\left[\epsilon^{fg}\overline{\psi}_{L,f}(x_1)\,  W(x_1,x_2)\,\psi_{L,g}(x_2)\right]\, \left[\epsilon^{st}\overline{\psi}_{L,s}(x_3)\,  W(x_3,x_4)\,\psi_{L,t}(x_4)\right]^{\dagger}}
\end{eqnarray}
where $f,g,s,t$ are flavor indices. This is a topologically trivial configuration of charges, i.e. an $n = 0$ configuration, where the fields at $x_1, x_2$ induce an Abelian instanton and the fields at $x_3, x_4$ an Abelian anti-instanton around them. At the limit of infinite separation between these two clusters, the  overall correlator is a product of the two clusters, where each cluster is the 't~Hooft vertex corresponding to the instanton/anti-instanton. Following the same procedure as in the previous subsection
\begin{eqnarray}\label{eq:cdcor}
    C^{\rm CD}_{1/2,2} & = & \frac{1}{4}\Big\langle W(x_1, x_2)\, W(x_3, x_4)^* \, \epsilon^{fg}\epsilon^{st}\Big[\delta_{fs} \, \delta_{gt} \, G_{\eta\eta^\dagger}(x_1, x_3) \,  G_{\chi\chi^\dagger}(x_2,x_4) \nn\\
    &&~~~~~~~~~~~~~~~~~~~~~~~~~~~~~~~~~~~~~~~~~
    + \delta_{ft}\,  \delta_{gs} \, G_{\eta\chi^\dagger}(x_1, x_4) \,  G_{\chi\eta^\dagger}(x_2,x_3)  \Big] \Big\rangle \, ,
\end{eqnarray}
where $G(x,x')$ is the fermion propagator. In computing this propagator, we can drop all higher partial waves and keep only the s-wave. The reason behind this is that, in the infinite separation limit between the clusters, the effect we are isolating is an instanton--anti-instanton effect. We know that instanton zero modes only exist in the lowest partial wave, and so any higher partial waves contributing to \eqref{eq:cdcor} die-off in the limit of infinite cluster separation. For this reason, we take for $G(x,x')$ the s-wave expression \eqref{eq:grlowbet}.
By direct calculation
\begin{eqnarray}\label{eq:CDcorrres}
    C^{\rm CD}_{1/2,2} & = & \frac{1}{2} \frac{F^{\theta\varphi}_{1234}}{(2\pi)^2r_1r_2r_3r_4} \frac{1}{(\Delta t_{12} - i \Sigma r_{12})(\Delta t_{34} + i \Sigma r_{34})} \, \nn \\[5pt]
    && ~~ \times \pqty{1 -  \bqty{\frac{(\Delta t_{14} - i \Sigma r_{14})(\Delta t_{23} - i \Sigma r_{23})}{(\Delta t_{13} - i \Delta r_{13})(\Delta t_{24} + i \Delta r_{24})}}} \,,
\end{eqnarray}
where $F^{\theta\varphi}_{1234}$ is a tensor product over the $\Omega^{(3)}$ of the four fermions. \eqref{eq:CDcorrres} reproduces \cite{Polchinski:1984uw, Affleck:1993np}, which was calculated in the DBC model, and \cite{Craigie:1984pc}, which was calculated in the $n/2=0$ sector of the full EFT.
The sign differences from these references are alleviated by taking $r_{3,4} \rightarrow -r_{3,4}$ as was done in \cite{Polchinski:1984uw}.
Indeed, when $t_1,t_2\ll t_3,t_4$, the factor in the square brackets becomes $-1$, and cluster decomposition gives
\begin{eqnarray}
    \lim_{t_{1,2} \ll t_{3,4} } C^{\rm CD}_{1/2,2}  & = & \frac{1}{2\pi\,r_1r_2} \frac{F^{\theta\varphi}_{12}}{\Delta t_{12} - i\Sigma r_{12}} \frac{1}{2\pi r_3 r_4}\frac{F^{\theta\varphi\,*}_{34}}{\Delta t_{34} + i\Sigma r_{34}}=\,C^{flip}_{1/2,2}(x_1,x_2) \pqty{C^{flip}_{1/2,2}(x_3,x_4)}^* \, ,\nonumber\\
\end{eqnarray}
as shown in \cite{Rubakov:1982fp,Polchinski:1984uw,Affleck:1993np}. The interpretation is clear, the fermions at $x_{1,3}$ form a vortex whereas the anti-fermions at $x_{2,4}$ form an anti-vortex, leaving overall an overall winding number $n/2=0$. The factorization of the ``cluster decomposition'' correlator into a vortex and an anti-vortex closely resembles the analogous situation in $QED_2$, as was shown in \cite{Hortacsu:1979fg} following the cluster decomposition calculation of \cite{Rothe:1978hx} and the discovery of topological vortices in $QED_2$ in \cite{Nielsen:1976hs,Nielsen:1977qk}. 

\begin{figure}[ht]
    \begin{center}
    \includegraphics[width=0.48\linewidth]{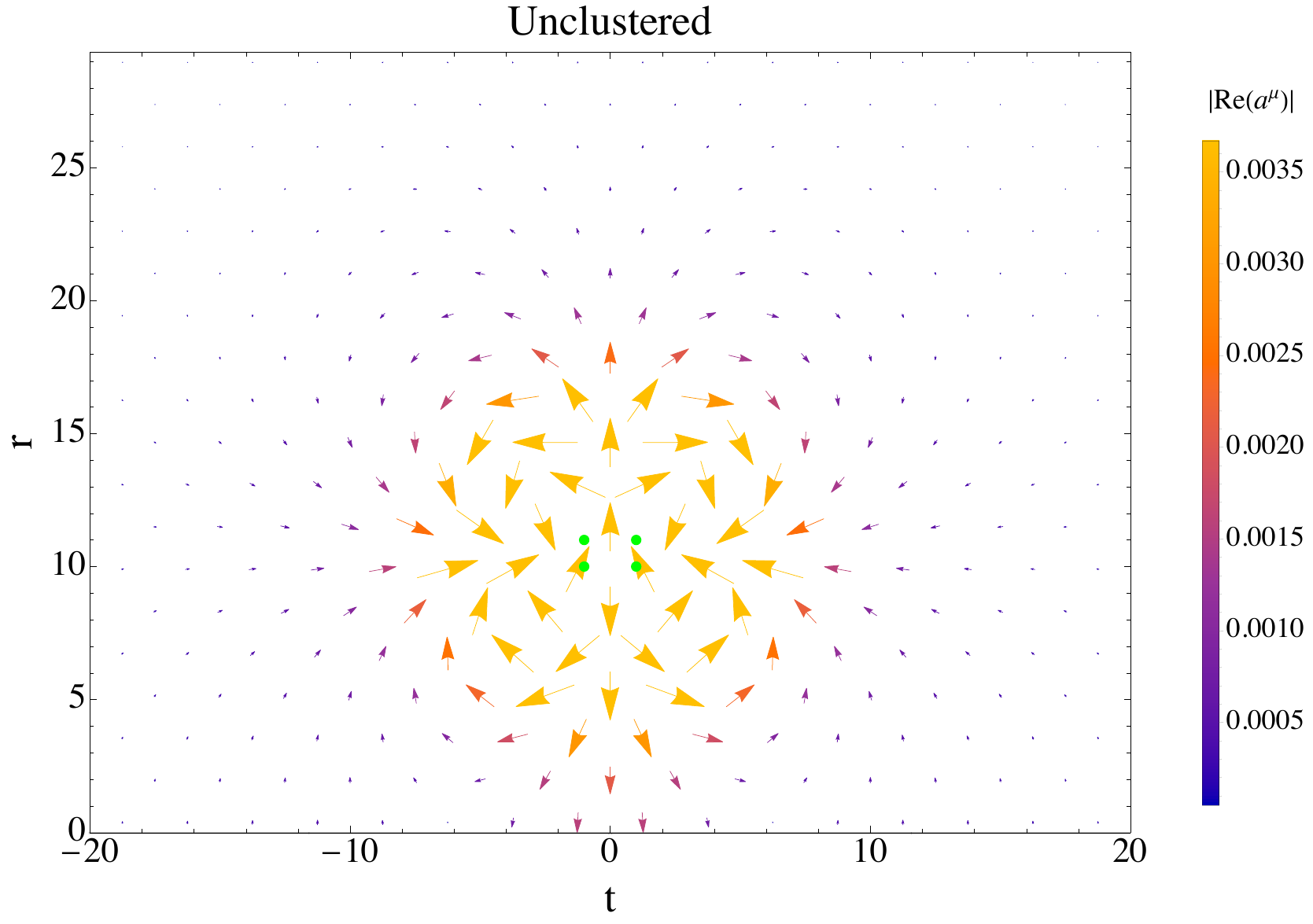} ~
    \includegraphics[width=0.48\linewidth]{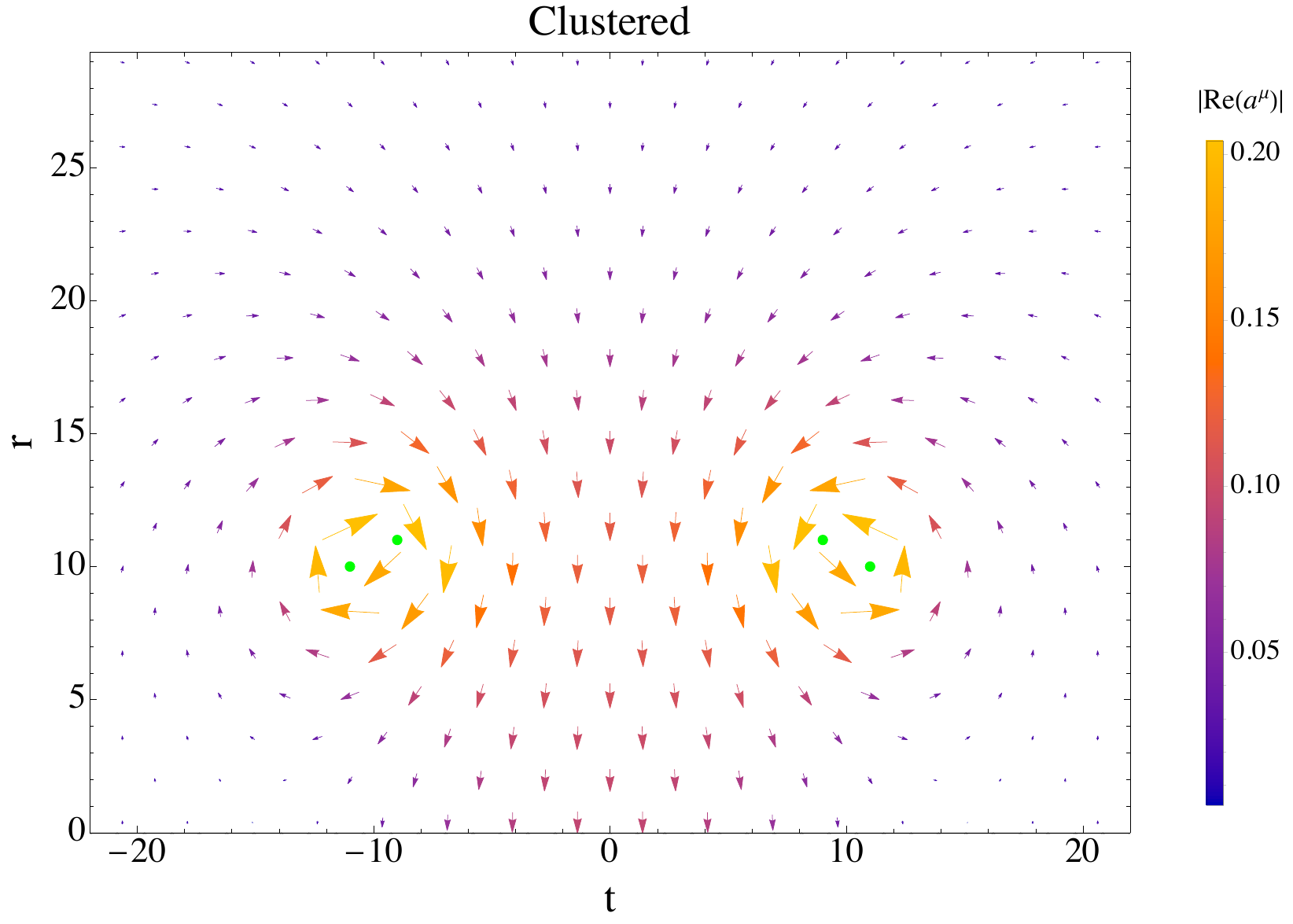}
    \caption{A vector plot of the leading order contribution to ${\rm Re}(a^{\alpha})$ for the cluster decomposition setup, the sources (fermion insertions) are denoted by green dots.
    \textbf{Left.} The sources are placed at $x_1 = (-1,10)\, ,~x_2 = (-1,11)\, ,~x_3 = (1,11)$ and $x_4 = (1,10)$ where it is manifest that this configuration has no winding around the fermion insertions, yeilding an $n/2=0$-vortex. 
    \textbf{Right.} The sources are placed at $x_1 = (-11,10)\, ,~x_2 = (9,11)\, ,~x_3 = (-9,11)$ and $x_4 = (11,10)$ where it is again manifest that this is, overall, an $n/2=0$ vortex. This corresponds to the cluster decomposition limit $t_{1,3} \ll t_{2,4}$ where we see the sources at $x_{1,3}$ form a $n/2=1/2$ vortex whereas the sources at $x_{2,4}$ form a $n/2=-1/2$ anti-vortex. Looking far from the sources we see that the overall configuration is an $n/2=0$ vortex.}\label{fig:cluster decomposition}
    \end{center}
\end{figure}

Fig.~\ref{fig:cluster decomposition} contains vector plots of the radial photon corresponding to both limits. The left panel shows the unclustered limit, where no winding is apparent. The right panel shows that in the cluster decomposition limit each pair of particles behaves as a vortex/anti-vortex.

\subsection{Monopole Catalysis}
%
We now revisit the Callan-Rubakov process \cite{Rubakov:1982fp, Callan:1982au, Callan:1982ah} 
\begin{eqnarray}
    u^1+u^2 + {\rm mon} \rightarrow e^\dagger+(d^3)^\dagger + {\rm mon} 
\end{eqnarray}
which we already discussed in the context of a topological configuration in Section~\ref{sec:intuition MC}. We emphasize that while the reasoning here is the same, the topological configuration is induced by the particle content of the Callan-Rubakov process.

\begin{figure}[ht]
    \begin{center}
    \includegraphics[width=0.85\linewidth]{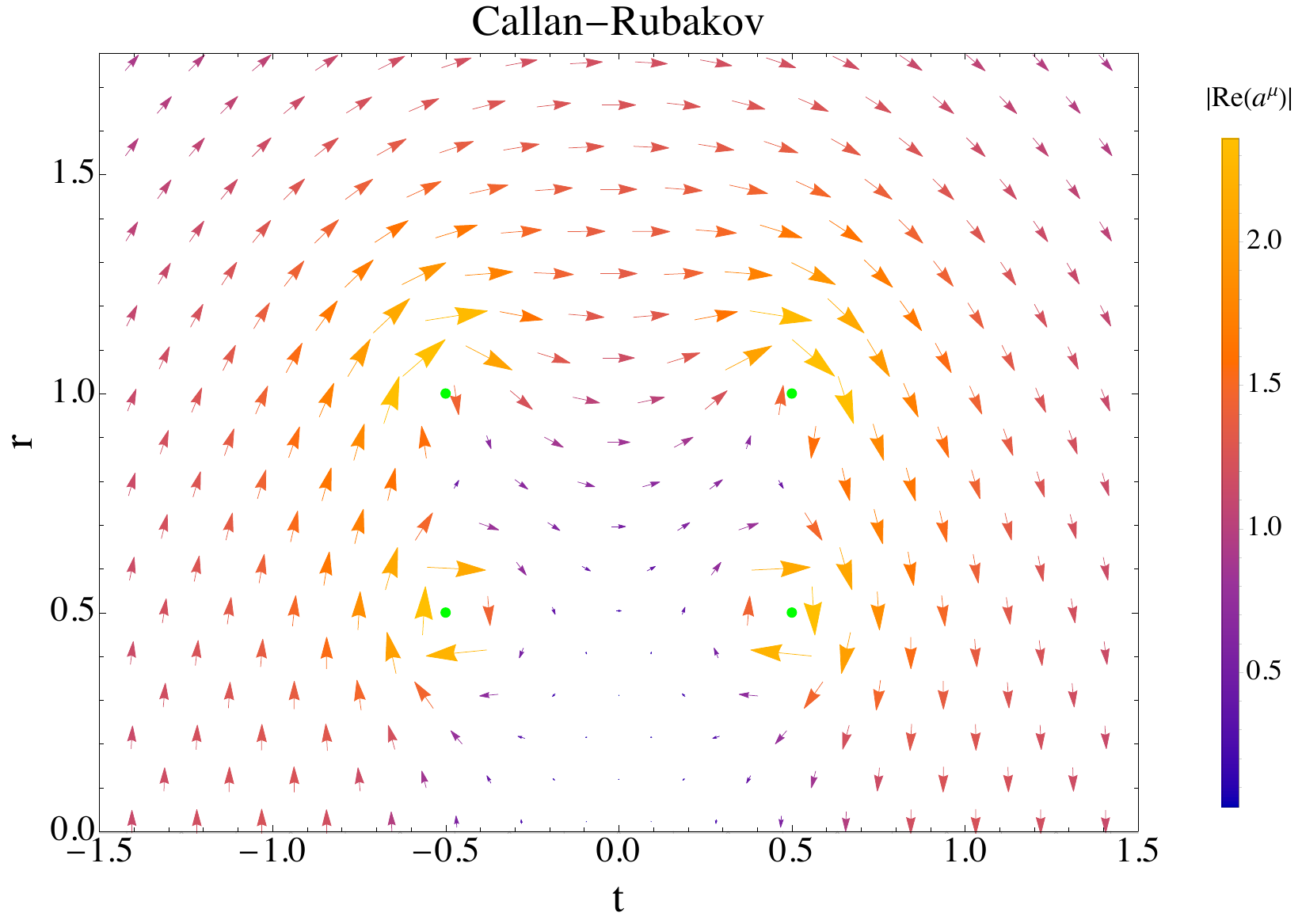} 
    \caption{A vector plot of ${\rm Re}(a^{\alpha})$ for a the Callan-Rubakov process. Sources are denoted by green dots, and their images which source the vortices by black dots. The $r<0$ region is shaded in gray.
    The sources are placed at $x_1 = (-1/2,1/2)\, ,~x_2 = (-1/2,1)\, ,~x_3 = (1/2,1/2)$ and $x_4 = (1/2,1)$ and the formation of a $n/2=1/2$ vortex is manifest. }\label{fig:CR}
    \end{center}
\end{figure}

The Callan-Rubakov process involves the reduction of an $SU(5)$ gauge theory with 4 fermion doublets in the representations of the $SU(2)$ subgroup where a 't~Hooft-Polyakov monopole ($q=1/2$ in our conventions) resides
\begin{eqnarray}\label{eq:MC emb}
    \colvec{-\overline{u}^2\\(u^1)^\dagger}~,~\colvec{\overline{u}^1\\(u^2)^\dagger}~,~\colvec{e\\-(\overline{d}_3)^\dagger}~,~\colvec{d^3\\(\overline{e})^\dagger}\, .
\end{eqnarray}
In the language of Section~\ref{sec:2D reduction} these are $N_f = 4$ different fermion flavors with $q=1/2$. The upper components are identified with $\psi_{i \, +}$ and the lower components with $\psi_{i \, -}$. In this language the monopole catalysis correlator is
\begin{eqnarray}
    C^{\rm CR}_{1/2,4} & = & \left\langle\,\tfrac{1}{4!}\,W(x_1,x_3)W(x_2,x_4)\,\,\epsilon_{4}\left(\overline{\psi}_{L1}(x_1),\overline{\psi}_{L2}(x_2),{\psi}_{L3}(x_3),{\psi}_{L4}(x_4)\right)\right\rangle\, ,
\end{eqnarray}
which gets its only contribution from the $n/2=1/2$ sector of the path integral. By direct calculation along the lines of the previous examples,
\begin{eqnarray}\label{eq:resCRfull}
    C^{\rm CR}_{1/2,4} & = & \frac{1}{(2\pi)^2}\,\frac{1}{\sqrt{(\Delta t_{13} - i \Sigma r_{13})(\Delta t_{14} - i\Sigma r_{14})(\Delta t_{23} - i \Sigma r_{23})(\Delta t_{24}- i\Sigma r_{24})}}
\end{eqnarray}
reproducing well-know results \cite{Polchinski:1984uw, Affleck:1993np, Craigie:1984pc}, and in the limit $r_1=r_2=r_3=r_4$, the original results of Rubakov \cite{Rubakov:1982fp} and Callan \cite{Callan:1982au, Callan:1982ah}. Note that the correlator in the 2D effective theory is completely flavor invariant. It is only the embedding of the fermions into doublets of the larger gauge group that creates a boundary condition which converts particular types of fermions into others, leading to monopole catalysis. In Fig.~\ref{fig:CR} we demonstrate the winding in the radial photon $a^\mu$ for the Callan-Rubakov process.

\subsection{General Lessons for Monopole Catalysis Correlators}\label{sec:gen}
Drawing on our full path integral and explicit calculations of the correlators in this section, we are now in the position to draw some general lessons about monopole catalysis.

First and foremost, we discover that \textit{every} charge conserving correlator with $2|qn|$ fermionic insertions per each flavor (for any half-integer $n/2$) leads to a 't~Hooft vertex with $2|qn|N_f$ legs. The instanton number for this correlator is $qn$. For charge conservation, half of the fermions in these correlators are $\psi$ and half are $\overline{\psi}$. All of these correlators are not only nonzero, but also saturate the unitarity bound lowest partial wave in the corresponding $|qn|N_f\rightarrow |qn|N_f$ process. For Callan-Rubakov we have $q=1/2,\,N_f=4,\,n/2=1/2$, so we have a 1/2-instanton process leading to $2\rightarrow 2$ scattering.

The second thing that we learn is that the $n$-vortex configuration generated in each monopole catalysis correlators is the combination of $2|qn|N_f$ ``fractional vortices'' centered around each one of the $2|qn|N_f$ insertion points in the correlator (see, e.g. Fig~\ref{fig:CR} for the Callan-Rubakov case). The fractional winding number of each one of these ``fractional vortices'' is $1/(4|q|N_f)$, which comes from the inverse Schwinger mass squared, and they always combine to an overall $n/2$-vortex.

Last, but not least, we learn that the solutions \eqref{eq:4Dzmnew} to the Dirac equation in the vortex background are not strictly left- or right-moving, as is the case in a monopole background without a vortex. As a matter of fact, the zero-modes are bound to the vortex through their explicit dependence on $E_r(t,r)$. Moreover, since the vortex is split into $2|qn|N_f$ fractional vortices, we infer that every fermion is evenly localized over all of these fractional vortices. It is easy to see that this had to be the case, otherwise flavor symmetry would have been broken. The picture that emerges is then of the fermions being \textit{entangled} by the vortex. It would be very interesting to consider the implications of this picture for the ``Monopole Unitarity Puzzle'' \cite{Callan:1983tm}. 

\section{Comparison to the ``Dyon Boundary Condition'' Model}\label{sec:DBC}
In this paper we showed that the correct 2D EFT for monopole catalysis is \textit{massless, axial} $QED_2$ in ${AdS}_2$ with the charge-depositing (chirality preserving)  boundary condition \eqref{eq:bcr0}, that descends directly from the 't-Hooft Polyakov UV completion \cite{Rubakov:1982fp,Callan:1982ah,Callan:1982au}. We have explicitly calculated the charge-conserving monopole-catalysis correlators of this theory and showed that they are consistent with the literature and saturate the lowest partial wave unitarity bound. The path that we chose in this paper was an explicit reduction of $QED_4$ in the background of a monopole to a solvable 2D theory. We had to invoke the UV physics only in two places: (a) by deriving the charge-depositing boundary condition \eqref{eq:bcr0}, that explicitly does not play a role in our correlators; and (b) by determining how the standard model fields embed into our 4D Dirac fermions. The advantage of this route is that it is a direct reduction of the 4D theory without any gaps.

In this section, we address a different route to represent the IR physics of monopole catalysis. This route, initiated by Polchinski in his seminal work \cite{Polchinski:1984uw}, aims to reproduce the physics of monopole catalysis in a 2D model without a dynamical photon, and hence no vortex configurations. The fermions of this $2D$ model are the lowest partial waves of the fermions of monopole-background $QED_4$. The role of the latter is then simulated by an interaction with a localized rotor degree of freedom at $r=0$, representative of the dyon collective coordinate of 't~Hooft Polyakov monopoles. Subsequent authors used techniques familiar from the study of the Kondo problem \cite{Kondo:1964nea, Affleck:1995ge, Maldacena:1995pqn} to show that the rotor degree of freedom could be integrated out, leaving behind a charge conserving, chirality violating boundary condition \cite{Affleck:1993np, Maldacena:1995pqn, Smith:2019jnh, Smith:2020nuf, Hamada:2022eiv,vanBeest:2023dbu,vanBeest:2023mbs}. Following these authors, we refer to this boundary condition as the DBC. Interestingly, the DBC cannot be formulated linearly on the $2D$ fermions. Instead, it is conventionally written in terms of $2D$ currents, or alternatively,  a linear relation on the bosons corresponding to the original fermions via $2D$ Abelian bosonization.  

The elements of the DBC model are $N_f$ flavors of massless 2D Dirac fermions $\chi^f$ with charge 1, defined on the half plane $(t,r)$ with $r\geq 0$. For clarity let us decompose these Dirac fermions into their left- and right-moving Weyl components, $\chi^f_L$ and $\chi^f_R$, respectively. The global symmetry of the model is then\footnote{The $U(1)$ here is the gauge symmetry, but since we omit the 2D photon, we will count it as part of the global symmetry.} $U(1)\times SU(N_f)_L\times SU(N_f)_R$, and we can group the fermions into representations of this symmetry as $\chi_L=(\chi^1_L,\ldots,\chi^{N_f}_L)^T~,~\chi_R=(\chi^1_R,\ldots,\chi^{N_f}_R)^T$, transforming as $\chi_L(1,\square,\mathds{1})$ and $\chi_R(-1,\mathds{1},\square)$, respectively.

Any possible boundary condition at $r=0$ can at most respect a diagonal $U(1)\times SU(N_f)_V$ subgroup of the bulk global symmetry \cite{Smith:2019jnh, Smith:2020nuf, vanBeest:2023dbu,vanBeest:2023mbs}. The DBC is the boundary condition respecting a particular $U(1)\times SU(N_f)_V$, under which $\chi_L$ transforms as $\chi_L(1,\square)$ while $\chi_R$ transforms as $\chi_R(-1,\square)$. This boundary condition is not linear in the fermions and is often formulated using \textit{Abelian bosonization}. The latter maps
\begin{eqnarray}\label{eq:fermbos}
    \chi^f_L=\sqrt{\frac{\mu}{2\pi}}\,e^{i\sqrt{\pi}\phi^f_L(t-ir)}~~~,~~~\chi^f_R=\sqrt{\frac{\mu}{2\pi}}\,e^{i\sqrt{\pi}\phi^f_R(t+ir)}
\end{eqnarray}
where $\phi^f_{L,R}=\int_r^{\infty}\,(\partial_t\pm i\partial_r)\,\phi^f(t,r)\,dr\,$, and the $\phi^f$ are massless $2D$ scalars. In \eqref{eq:fermbos}, $\mu$ is an arbitrary mass scale that drops out of all correlators. The DBC at $r=0$ is then
\begin{eqnarray}
    \phi^{f_1}_L(t)=\mathcal{R}^{f_1}_{~f_2}\phi^{f_2}_R(t)\,,
\end{eqnarray}
where $\mathcal{R}^{i}_{~j}=\delta^i_{~j}-\frac{2}{N_f}$.

Let us now focus on $N_f=4$ and illustrate how the DBC model reproduces the Callan-Rubakov monopole catalysis correlator. The latter is given by
\begin{eqnarray}\label{eq:toychi}
    C^{\rm CR,\,DBC}_{1/2,4}=\left\langle\,\frac{1}{4!}\,\epsilon_4\left(\chi^1_L(x_1),\chi^2_L(x_2),\chi^3_R(x_3),\chi^4_R(x_4)\right)\right\rangle\,,
\end{eqnarray}
which leads to the Callan-Rubakov process
\begin{eqnarray}
    \chi^1_L+\chi^2_L\rightarrow(\chi^3_R)^\dagger+(\chi^4_R)^\dagger\,,
\end{eqnarray}
with $(\chi^1_L,\chi^2_L,\chi^3_R,\chi^4_R)=(u_1,u_2,e,d_3)|_{s-wave}$.

To compute \eqref{eq:toychi}, it is useful to perform an $SU(N_f)$ transformation on the $\psi^f_{L,R}$ to a basis\footnote{No relation to the $\widetilde{\psi}$ that appeared earlier in this paper.} $\widetilde{\phi}^f_{L,R}$ that diagonalizes $\mathcal{R}_{ij}$ \cite{Maldacena:1995pqn}. This transformation is
\begin{eqnarray}
    \colvec{\phi^{1}_{L,R}\\\phi^{2}_{L,R}\\\phi^{3}_{L,R}\\\phi^{4}_{L,R}}=\colmatf{0 & 0 & -\frac{\sqrt{3}}{2} & \frac{1}{2} \\
    0 & \sqrt{\frac{2}{3}} & \frac{1}{2 \sqrt{3}} & \frac{1}{2} \\
    -\frac{1}{\sqrt{2}} & -\frac{1}{\sqrt{6}} & \frac{1}{2 \sqrt{3}} & \frac{1}{2} \\
    \frac{1}{\sqrt{2}} & -\frac{1}{\sqrt{6}} & \frac{1}{2 \sqrt{3}} & \frac{1}{2}}\colvec{\widetilde{\phi}^{1}_{L,R}\\\widetilde{\phi}^{2}_{L,R}\\\widetilde{\phi}^{3}_{L,R}\\\widetilde{\phi}^{4}_{L,R}}\,.
\end{eqnarray}
The boundary condition on the $\widetilde{\phi}$ at $r=0$ is then
\begin{eqnarray}\label{eq:tildebc}
    \widetilde{\phi}^{i}_L(t)&=&\widetilde{\phi}^{i}_R(t)~~,~~i\in\{1,2,3\}\nonumber\\[5pt]
    \widetilde{\phi}^{4}_L(t)&=&-\widetilde{\phi}^{4}_R(t)\,.
\end{eqnarray}
In terms of the $\widetilde{\phi}^f_{L,R}$, the fermions are expressed as
\begin{eqnarray}\label{eq:dbcferm}
    \chi^1_{L,R}&=&\sqrt{\frac{\mu}{2\pi}}\,\exp{i\sqrt{\pi}\,\left[-\frac{\sqrt{3}}{2}\widetilde{\phi}^3_{L,R}+\frac{1}{2}\widetilde{\phi}^4_{L,R}\right]}\nonumber\\[5pt]
    \chi^2_{L,R}&=&\sqrt{\frac{\mu}{2\pi}}\,\exp{i\sqrt{\pi}\,\left[\sqrt{\frac{2}{3}}\widetilde{\phi}^2_{L,R}+\frac{1}{2\sqrt{3}}\widetilde{\phi}^3_{L,R}+\frac{1}{2}\widetilde{\phi}^4_{L,R}\right]}\nonumber\\[5pt]
    \chi^3_{L,R}&=&\sqrt{\frac{\mu}{2\pi}}\,\exp{i\sqrt{\pi}\,\left[-\frac{1}{\sqrt{2}}\widetilde{\phi}^1_{L,R}-\frac{1}{\sqrt{6}}\widetilde{\phi}^2_{L,R}+\frac{1}{2\sqrt{3}}\widetilde{\phi}^3_{L,R}+\frac{1}{2}\widetilde{\phi}^4_{L,R}\right]}\nonumber\\[5pt]
    \chi^4_{L,R}&=&\sqrt{\frac{\mu}{2\pi}}\,\exp{i\sqrt{\pi}\,\left[\frac{1}{\sqrt{2}}\widetilde{\phi}^1_{L,R}-\frac{1}{\sqrt{6}}\widetilde{\phi}^2_{L,R}+\frac{1}{2\sqrt{3}}\widetilde{\phi}^3_{L,R}+\frac{1}{2}\widetilde{\phi}^4_{L,R}\right]}
\end{eqnarray}

The correlator \eqref{eq:toychi} is then given in the DBC theory by
\begin{eqnarray}\label{eq:DBCcor}
    C^{\rm CR,\,DBC}_{1/2,4}=\frac{1}{4!}\int\prod_{f=1}^4\,D\widetilde{\phi}^f\,\epsilon_4\left(\chi^1_L(x_1)\chi^2_L(x_2)\chi^3_R(x_3)\chi^4_R(x_4)\right)\,e^{-\int\,d^2x\,\frac{1}{2}\sum_f\,\partial_{\alpha}\widetilde{\phi}^f\partial^{\alpha}\widetilde{\phi}^f}\,,
\end{eqnarray}
where the fermionic insertions are interpreted according to \eqref{eq:dbcferm}, and the path integral is over $\widetilde{\phi}^f$ satisfying the boundary condition \eqref{eq:tildebc}. This path integral is Gaussian, and it is equal to\footnote{We checked that all of the different contractions in $\epsilon_{4}$ give the same result.} 
\begin{eqnarray}\label{eq:DBCcor2}
    C^{\rm CR,\,DBC}_{1/2,4}=\frac{\mu^2}{(2\pi^2)}\exp{\pi\int\,d^2x\,\left[\sum_{f=1}^3\rho^f(x)\mathcal{D}_N(x,x')\rho^f(x')+\rho^4(x)\mathcal{D}_D(x,x')\rho^4(x')\right]}\,,\nonumber\\
\end{eqnarray}
where $\mathcal{D}_{N}(x,x')$ ($\mathcal{D}_{D}(x,x')$) is the Neumann (Dirichlet) Green's function for a massless scalar in $2D$, given by
\begin{eqnarray}\label{eq:Ds}
    \mathcal{D}_{N}(x,x')&=&-\frac{1}{4\pi}\left[\log\left(\frac{1}{\mu^2(\Delta t^2+\Delta r^2)}\right)+\log\left(\frac{1}{\mu^2(\Delta t^2+\Sigma r^2)}\right)\right]\nonumber\\[5pt]
    \mathcal{D}_{D}(x,x')&=&-\frac{1}{4\pi}\left[\log\left(\frac{1}{\mu^2(\Delta t^2+\Delta r^2)}\right)-\log\left(\frac{1}{\mu^2(\Delta t^2+\Sigma r^2)}\right)\right]\,,
\end{eqnarray}
where $\mu$ is the same arbitrary mass scale from \eqref{eq:fermbos}. The sources $\rho$ are extracted from \eqref{eq:dbcferm} and are explicitly
\begin{eqnarray}\label{eq:rhos}
    \rho^1(x)&=&-\frac{1}{\sqrt{2}}R(x,x_3)+\frac{1}{\sqrt{2}}R(x,x_4)\nonumber\\[5pt]
    \rho^2(x)&=&\sqrt{\frac{2}{3}}L(x,x_2)-\frac{1}{\sqrt{6}}R(x,x_3)+\frac{1}{\sqrt{6}}R(x,x_4)\nonumber\\[5pt]
    \rho^3(x)&=&-\frac{\sqrt{3}}{2}L(x,x_1)+\frac{1}{2\sqrt{3}}L(x,x_2)+\frac{1}{2\sqrt{3}}R(x,x_3)+\frac{1}{2\sqrt{3}}R(x,x_4)\nonumber\\[5pt]
    \rho^4(x)&=&\frac{1}{2}L(x,x_1)+\frac{1}{2}L(x-x_2)+\frac{1}{2}R(x,x_3)+\frac{1}{2}R(x,x_4)\,,
\end{eqnarray}
where
\begin{eqnarray}\label{eq:LR}
    L(x,x')&\equiv&\delta^{(2)}(x-x')+\partial_t\delta(t-t')\Theta(r-r')\nonumber\\[5pt]
    R(x,x')&\equiv&\delta^{(2)}(x-x')-\partial_t\delta(t-t')\Theta(r-r')\,.
\end{eqnarray}
Putting \eqref{eq:Ds} and \eqref{eq:rhos} in the DBC path integral \eqref{eq:DBCcor2}, we get exactly
\begin{eqnarray}\label{eq:DBCrescmp}
    C^{\rm CR,\,DBC}_{1/2,4}=C^{\rm CR}_{1/2,4}|_{s-wave}\,,
\end{eqnarray}
where $C^{\rm CR}_{1/2,4}|_{s-wave}$ is the result from the full EFT, \eqref{eq:resCRfull}, stripped of its angular dependence. This had to be the case, as the DBC was tailored to conserve standard model charge. The latter requirement, together with angular momentum conservation (expressed covariantly using pairwise helicity \cite{Csaki:2020inw,Csaki:2020yei,Csaki:2022tvb,Mouland:2024zgk}), uniquely determine the Callan-Rubakov process \cite{ Csaki:2022qtz}. As stressed above, the true underlying physics for monopole catalysis is the Abelian instanton leading to \eqref{eq:resCRfull}, which can be simulated for certain processes in the DBC model.

\section{Conclusions}\label{sec:conc}
We showed that the Callan-Rubakov process is actually an Abelian instanton process in the low-energy, $U(1)$ effective gauge theory.  The instanton number is the product of two topological 1st Chern numbers, the magnetic charge and a half-integer vortex number  of the gauge field in the $(t,r\geq0)$ half-plane. As is standard in the literature, our derivation involved the truncation of all photon and fermion higher partial waves -- and as such it should only be fully trusted for minimal monopoles with $q=1/2$. Furthermore we have seen that every $2D$ vortex is decomposed into fractional vortices, one for each fermion in the scattering process. For the minimal instanton amplitude to be non-vanishing, there must be one member of each fermion species in either the initial or final state, in order to saturate the legs of the 't~Hooft vertex. The charge-depositing boundary condition merely serves to impose the correct global symmetry of the non-Abelian UV theory that we are matching to, but does not in fact lead to any charge deposition in monopole catalysis correlators. With this boundary condition we reproduce the behaviour expected from an instanton in a 't~Hooft-Polyakov monopole background. In fact, in the limit where the energies are much smaller than the VEV of the 't~Hooft-Polyakov scalar, and the correlator insertions are outside the core, our Abelian instantons can be trivially embedded into an $SU(2)$ theory with a 't~Hooft-Polyakov monopole. In this case the vortices are embedded in the unbroken $SU(2)$ generator outside the core.

We compared our results to the correlators of the DBC model, which simulate the instanton as the interaction with a rotor degree of freedom. The latter gives the correct monopole catalysis correlators by construction, as they are fixed by charge and angular momentum conservation. Nevertheless, it does not descend from the UV theory and does not identify Abelian instantons as the underlying mechanism for monopole catalysis.

Future applications of this work will include re-examining the ``Monopole Unitarity Puzzle'' discovered by Callan \cite{Callan:1983tm}, which has perplexed many physicists for over 40 years. In this process, a single massless positron in an s-wave falls in towards the monopole, but naively there does not seem to be a consistent asymptotic final state. 
A variety of solutions have been proposed \cite{Callan:1983tm,Polchinski:1984uw,Callan:1989em,Maldacena:1995pqn,Kitano:2021pwt,Brennan:2021ewu,Csaki:2022qtz,Hamada:2022eiv,vanBeest:2023dbu,Brennan:2023tae,Khoze:2023kiu,vanBeest:2023mbs,Khoze:2024hlb}, but the correct physics remains unclear.  One class of solutions argues that the UV physics imposes new, charge conserving dyon boundary conditions which lead to outgoing states in a topologically twisted sector \cite{vanBeest:2023dbu,vanBeest:2023mbs}. This approach, as far as we understand, does not explain the putative UV origins of the charge-conserving boundary condition. It is also not immediately clear how to relate the dyon boundary condition to the topological origins of monopole catalysis demonstrated in this work. 
Another proposal by some of the present authors \cite{Csaki:2022qtz} identified a unique entangled state of three quarks that preserved all quantum numbers. The criticism of this approach stems from the observation that the helicities of the quarks forbid them from ever reaching the monopole. However, we have seen that the Abelian instantons responsible for monopole catalysis are part of the Abelian EFT \textit{outside} the monopole core, and, furthermore, that the solutions \eqref{eq:4Dzm} of the Dirac equation in the \textit{vortex} background are not strictly left/right moving. It is a simple matter to check that the proposed one to three process
has a non-vanishing instanton number. Furthermore, the entanglement of the fermions by the fractional vortices, as discussed in Section~\ref{sec:gen}, is reminiscent of the entanglement seen in the amplitude/pairwise helicity picture of \cite{Csaki:2022qtz}.
What remains to be seen is whether this picture is consistent with the crossing relations of the theory. This is a non-trivial problem because the theory in the monopole background does not have crossing symmetry, so the crossing relations are not straightforward.

\section{Acknowledgements}
We thank Ofer Aharony, Hsin-Chia Cheng, Zohar Komargodski, Markus Luty, Riccardo Rattazzi, and David Tong for helpful discussions. OT would especially like to thank Jiji Fan, Katherine Fraser, Matthew Reece, and John Stout for many fruitful discussions regarding the QFT of monopole catalysis.
Part of the work was performed at Aspen Center for Physics, which is supported by NSF grant PHY-2210452.
CC is supported in part by the NSF grant PHY-2014071 and in part by the US-Israeli BSF grant 2016153. 
RO is supported by an ERC STG grant (“Light-Dark”, grant No. 101040019). 
The work of OT is supported in part by the NSF-BSF Physics grant No 2022713. 
JT was supported in part by the DOE under grant DE-SC-000999. 
SY is supported in part by the Israeli Science Foundation Excellence Center. 
This project has received funding from the European Research Council (ERC) under the European Union’s Horizon Europe research and innovation programme (grant agreement No. 101040019).  Views and opinions expressed are however those of the author(s) only and do not necessarily reflect those of the European Union. The European Union cannot be held responsible for them.

\appendix

\section{QFT in \texorpdfstring{${AdS}_2$}{}}\label{app:ads2}
The Euclidean ${AdS}_2$ metric is given by
\begin{eqnarray}
    g_{\alpha\beta} = \pqty{\frac{R^2}{r^2}, \frac{R^2}{r^2}} \, .
\end{eqnarray}
In this case the zweinbein is $e^{\alpha}_a = (r/R)\delta^{\alpha}_a$ and the spin connection is $\Omega_{t}^{01} = -\Omega_{t}^{10} = r^{-1}$. The scalar Laplacian is given by
\begin{eqnarray}
    \square \phi = \frac{r^2}{R^2} \partial_\alpha \partial^{\alpha} \phi \, ,
\end{eqnarray}
and the Dirac operator can be conveniently written as
\begin{eqnarray}
    \slashed{\nabla} \Psi =  \pqty{\frac{R}{r}}^{-1/2} \gamma^a \partial_a \pqty{\frac{R}{r}}^{1/2} \Psi \, ,
\end{eqnarray}
where $\gamma^a \partial_a$ is the flat space Dirac operator. Throughout we use the 2D chiral basis $\gamma_a = (\sigma_1, \sigma_2)$ giving $\gamma_5 = -\sigma_3$.

The massless scalar propagator in ${AdS}_2$, ${\cal D}(x,x')$, is the same as the flat space one in \eqref{eq:sch2G2}. The massless Dirac fermion propagators are related to this by ${\cal D}_{\Psi}(x,x') = i\slashed{\nabla} {\cal D}(x,x')$. Additionally, we provide the bulk-to-bulk propagator for a massive scalar in ${AdS}_2$, $\mathcal{P}(x,x';m)$, which is defined as 
\begin{eqnarray}\label{eq:GreenAds}
    \left(\square - m^2\right)\mathcal{P}(x,x';m)&=& \frac{1}{\sqrt{g_{\rm 2D}}}\delta^{(2)}(x-x') \, .
\end{eqnarray}
Explicitly, it is given by \cite{DHoker:1998ecp}
\begin{eqnarray}\label{eq:Dprop}
    \mathcal{P}(x,x'; m) & = & -\frac{2^{-\Delta}\Gamma(\Delta)\,\xi^{\Delta}}{\sqrt{4\pi} \Gamma\left(\Delta+\frac{1}{2}\right)}  {\,}_2 F_1\left(\frac{\Delta}{2}, \frac{\Delta+1}{2} ; \Delta+\frac{1}{2} ; \xi^2\right)\,,
\end{eqnarray}
where ${}_2F_1$ is the Gauss hypergeometric function, $\xi=\frac{2rr'}{r^2+r'^2+(t-t')^2}$ is the ${AdS}_2$ invariant distance, and $\Delta=\frac{1+\sqrt{1+4 m^2 R^2}}{2}$.

\section{Derivation of Charge-Depositing Boundary Condition from 't~Hooft-Polyakov UV Completion}\label{app:chvbc} 
%
\subsection{Infinite Mass Limit of 't~Hooft Polyakov Monopoles}

In this section we derive the charge-depositing boundary condition \eqref{eq:bcr0YL} from its 't~Hooft-Polyakov UV completion, along the lines of \cite{Besson:1980vu,Rubakov:1982fp, Callan:1982ah,Callan:1982au, Isler:1987xn}, and more concretely, \cite{Lam:1984sm}. Specifically, we consider the action for $SU(2)$ gauge theory with massless fermions in the background of a 't~Hooft-Polaykov monopole. We take our $SU(2)$ generators in the normalization
\begin{eqnarray}
\tau^1=\frac{1}{2}\colmatt{0&1\\1&0}~~
\tau^2=\frac{1}{2}\colmatt{0&-i\\i&0}~~
\tau^3=\frac{1}{2}\colmatt{1&0\\0&-1}\,,
\end{eqnarray}
so that ${\rm tr}(\tau^a\tau^b)=\frac{1}{2}\delta^{ab}$. We also define
\begin{eqnarray}
    \vec{\tau}&=&\left(\tau^1,\tau^2,\tau^3\right)\nonumber\\[5pt]
    \left(\tau_{r},\tau_{\theta},\tau_{\varphi}\right) &=& \left(\vec{\tau}\cdot\vu{r},\vec{\tau}\cdot\hat{\theta},\vec{\tau}\cdot\hat{\varphi}\right) \,.
\end{eqnarray}
We follow the careful analysis of \cite{Lam:1984sm}, which yields a gauge-invariant boundary condition for the fermions at $r=0$. 
To do this, we start with the action for $SU(2)$ gauge theory coupled to a triplet scalar,
\begin{eqnarray}\label{eq:GG}
    S_{SU(2)}&=&\int\,d^4x\,\left[-\frac{1}{2}{\rm tr}F_{\mu\nu}F^{\mu\nu}-{\rm tr}D_\mu\Phi D^\mu\Phi-V(\Phi)\right]\, ,
\end{eqnarray}
where
\begin{eqnarray}\label{eq:DPhi}
    D_\mu\Phi&=&\partial_\mu\Phi - ie[A_{\mu},\Phi]\nonumber\\[5pt]
    V(\Phi)&=&\frac{\lambda}{4}\left(2{\rm tr}\Phi^2-v^2\right)^2\,.
\end{eqnarray}
Note that here and only here, we work in Lorentzian signature $(-+++)$ rather than in Euclidean signature, for consistency with the literature. The conserved angular momentum operator of the theory is
\begin{eqnarray}
    \vec{J}=\vec{L}+\vec{S}+\vec{T}\,,
\end{eqnarray}
the sum of the orbital, spin, and isospin angular momentum generators. 
We now consider the most general rotationally invariant ($J^2=0$) ansatz \cite{Witten:1976ck,Lam:1984sm} for the gauge field and the scalar,
\begin{eqnarray}\label{eq:rotan}
    A_t&=&\frac{a_t(t,r)}{\sqrt{4\pi R^2}}\tau_r \nonumber\\[5pt]
    \vb{A}&=&\frac{\vu{r} \cross \vec{\tau}}{e r} \pqty{1 - K(t,r) \cos\phi(t,r)}+ \frac{\, \pqty{\tau_r\vu{r}-\vec{\tau}}}{e r} K(t,r)\sin\phi(t,r)+\tau_r \vu{r}\frac{a_r(t,r)}{\sqrt{4\pi R^2}}\nonumber\\[5pt]
    \Phi&=&\frac{ \tau_r}{e R} H(r)\, ,
\end{eqnarray}
where $H,K,\phi$ are dimensionless 2D scalars, $a_\alpha$ is a dimensionless 2D gauge field and $R$ is an arbitrary distance scale. 
Below $K$ and $H$ will be set to their classical solutions, making up the 't~Hooft-Polyakov monopole (with $g=1/2$). 
$a_\alpha=(a_t,a_r)$ is the massless photon outside of the monopole core, while $\phi$ is a degree of freedom  associated with the spacetime$\times\,SU(2)$ rotations broken by the monopole. 
Substituting this ansatz into \eqref{eq:GG} and integrating over $(\theta,\varphi)$, we get the 2D effective action
\begin{eqnarray}\label{eq:2da}
S_{\rm eff}=\int\,d^2x\,&&\left\{-\frac{1}{4}\,\frac{r^2}{R^2}\,f^{\alpha\beta}f_{\alpha\beta} + \frac{K^2}{R^2}\left(a_
\alpha+\frac{\sqrt{4\pi}R}{e}\partial_\alpha\phi\right)^2\right.\nonumber\\[5pt]
&&-\frac{4\pi}{e^2}(\partial_\alpha K)^2
-\frac{2\pi r^2}{R^2}(\partial_\alpha H)^2
-\frac{2\pi}{e^2 r^2}\left(1-K^2\right)^2\nonumber\\[5pt]
&&-\frac{4\pi}{R^2} K^2 H^2-\frac{r^2}{R^2} \frac{\pi\lambda}{R^2}(H^2-v^2R^2)^2\left.\right\}\,.
\end{eqnarray}
Here $d^2x=dtdr$ and the integration is over $r\geq0$. 
Note that \eqref{eq:rotan} is invariant with respect to $U(1)$ transformations accompanied with a shift in $\phi$. 
It is then tempting to treat $\phi$ as an unphysical gauge artifact. 
Nevertheless, we know that \textit{large gauge transformations}, i.e. $U(1)$ transformations that don't vanish at spatial infinity, are in fact physical. 
So the correct statement is that  $\gamma(t)=\lim_{r\rightarrow\infty}\phi(t,r)$ is, in fact, physical, and so we retain $\phi$ in our action. 
The function $\gamma$ is referred to in the literature as the dyon collective coordinate of the 't~Hooft-Polyakov monopole \cite{Mottola:1978pz, Adler:1979mg, Weinberg:1979zt}.

Since we are interested in QFT in a monopole background, we set $H,\, K$ to their time-independent classical 't~Hooft-Polyakov solutions $H(r),\, K(r)$, where
\begin{eqnarray}\label{eq:KHtP}
    K(0)&=&1~~,~~K(r)\sim e^{-Mr}\nonumber\\[5pt]
    H(0)&=&0~~,~~H(r)\sim vR(1-e^{-Mr})\,.
\end{eqnarray}
Here, $M$ is the mass of the monopole, which we take to be larger than any other mass/energy scale in our theory. Note that setting $K(r),\,H(r)$ ``higgses'' the $SU(2)$ symmetry down to the $U(1)$ gauge symmetry outside the monopole core, generated, in this gauge, by the $\tau_r$ generator.
We can now write down the 2D effective gauge action in the monopole background as
\begin{eqnarray}\label{eq:effa2d}
    S_{\rm eff}&=&S_{2D}+S_{boundary}\nonumber\\[5pt]
    S_{2D}&=&\int\,d^2x\,\left\{-\frac{1}{4}\,\frac{r^2}{R^2}\,f^{\alpha\beta}f_{\alpha\beta}\right\}\nonumber\\[5pt]
    S_{bag}&=&\int\,d^2x\,\left\{\frac{K(r)^2}{R^2}\left(a_\alpha+\frac{1}{e_{eff}}\partial_\alpha\phi\right)^2\right\}\,.
\end{eqnarray}
$e_{\rm eff}=\frac{e}{\sqrt{4\pi R^2}}$ is a mass-dimension-1 2D coupling constant. 
Here we used the labeling $S_{bag}$ to emphasize that this action is nonzero only for $r\lesssim 1/M$, and its only consequence is the imposition of a ``bag'' boundary condition on the fermions \cite{Besson:1980vu, Rubakov:1982fp, Callan:1982ah, Callan:1982au}, as we shall see below. We see that the photon is ``Higgsed'' \textit{only} at the boundary; this signals the possibility of charge exchange with the monopole, but only at rates suppressed by the large monopole mass. 
We alert the reader that this is \textit{not} the higgsing of $SU(2)$ to $U(1)$ outside the monopole core. We are now ready to add to our theory $N_f$ massless $LH$ Weyl fermions $\kappa_f$ in $SU(2)$ doublets, by adding to the $SU(2)$ action \eqref{eq:GG} the term
\begin{eqnarray}
    S \supset i\sum_f\,\overline{\kappa}_f\,\slashed{D}\,\kappa_f\,,
\end{eqnarray}
where
\begin{eqnarray}   \slashed{D}\,\kappa_f=\bar{\sigma}^\mu\left(\partial_\mu + i Q e A_\mu\right)\kappa_f\,.
\end{eqnarray}
Here $\bar{\sigma}^\mu=(1,-\vb*{\sigma})$ and $Q=1$. Note that the $\kappa_f$ are $SU(2)$ doublets, i.e.
\begin{eqnarray}   
    \kappa_f=\colvec{\chi_f\\\eta_f}\,,
\end{eqnarray}
where the $\chi_f$ and $\eta_f$ are LH Weyl fermions with charges $1$ and $-1$ under the diagonal $U(1)$ symmetry outside the monopole core. They are exactly the same fields which appear in the $4D$ Dirac fermion of Section~\ref{sec:ZM}, namely
\begin{eqnarray}   
    \psi_f=\colvec{\chi_f\\\eta^{\dagger}_f}\,,
\end{eqnarray}
where this time the $\chi_f$ and $\eta^\dagger_f$ denote the LH and RH Weyl components of the Dirac fermion $\psi_f$.
As we did in the gauge/scalar sector, we consider only the $j=0$ partial wave for the fermions. This is the famous Jackiw-Rebbi ansatz \cite{Jackiw:1975fn,Jackiw:1976xx}:
\begin{eqnarray}
    \kappa^{j=0}_f=\frac{1}{\sqrt{4\pi R^2}}\left(\frac{R}{r}\right)^{\frac{3}{2}}\,\left(\tau_L\,\chi^{j=0}_f(t,r)+\tau_R\,\eta^{j=0}_f(t,r)\right)\epsilon\,,
\end{eqnarray}
where $\tau_L=\frac{1}{2}I_2+\tau_r$ and $\tau_R=\frac{1}{2}I_2-\tau_r$ and $\epsilon=2i\tau_2$. We now gather the s-waves into a 2D Dirac fermion $\ell_f=(\chi^{j=0}_f,\eta^{j=0}_f)$, and define 2D gamma matrices $(\gamma^0,\gamma^1)=(-i\sigma_2,\sigma_1)$ and $\gamma^5=\sigma_3$.
The 2D effective action now becomes, in Euclidean signature,
\begin{eqnarray}\label{eq:effa2dflE}
    S_{2D}&=&\int\,d^2x\,\sqrt{g^{2D}}\left\{\frac{1}{4}\,f^{\alpha\beta}f_{\alpha\beta}-i\sum_f\,\overline{\ell}\slashed{D}\ell\right\}\nonumber\\[5pt]
    S_{bag}&=&\int\,d^2x\,\sqrt{g^{2D}}\left\{\frac{K(r)^2}{R^2}\left(a_\alpha+\frac{1}{e_{2D}}\nabla_\alpha\phi\right)^2+i\frac{K(r)}{R}\overline{\ell}e^{-i\gamma^5\phi}\ell\right\}\,,
\end{eqnarray}
where $K(r)\sim \Theta(M^{-1}-r)$, and $g^{2D}_{\alpha\beta}=\frac{\mathcal{R}^2}{r^2}{\rm diag}(1,1)$ is the metric of AdS$_2$. We eventually take $M\rightarrow\infty$ in all of our physical processes, as we are interested only in processes that don't decouple in the IR. We remind the reader that all indices in \eqref{eq:effa2dflE} are raised and lowered using the $AdS_{2}$ metric $g$, see Appendix~\ref{app:ads2}. We thus reach our final form of the 2D EFT for the lowest partial waves: is the \textit{massless axial Schwinger model in} ${\rm AdS}_2$. 
The necessity of the boundary action in \eqref{eq:effa2dflE} is apparent from the fact that left- and right-moving fermions have opposite charges under the 2D photon. 
Since left movers going into the boundary have to come back as right movers, this process involves depositing charge on the monopole, making it a dyon by exciting the dyon collective coordinate $\phi$ \cite{Callan:1982ah,Callan:1982au}. Clearly, such processes would be suppressed by the mass of the dyon---and practically irrelevant in the IR. 

\subsection{Derivation of Charge-Depositing ``Bag'' Boundary Condition}
We are now in position to derive the ``bag boundary condition imposed on the fermions at $r=0$ by $S_{bag}$, along the lines of \cite{Besson:1980vu, Rubakov:1982fp, Callan:1982ah, Callan:1982au}. As we shall see, this is a charge-depositing boundary condition. To do this, we solve the Dirac equation stemming from \eqref{eq:effa2dflE} for $0\leq r\leq M^{-1}$, taking $K(r)=1$ (the boundary condition is insensitive to the exact r-dependence of $K(r)$ in this region, and $K=0$ for $r>M^{-1}$). Since we are now considering the full 't~Hooft-Polyakov monopole with a finite core, the only boundary condition we are allowed to impose strictly at $r=0$ (not $r=M^{-1}$) is that of regularity, like at the center of the hydrogen-like atom. Solving the AdS$_2$ Dirac equation
\begin{eqnarray}\label{eq:AdsDiracbag}
    \left[\slashed{D}-\frac{1}{R}e^{-i\gamma^5\phi}\right]\,\ell=0 ~~~~~,~~~0\leq r\leq M^{-1}\,,
\end{eqnarray}
and keeping the solution regular at $r=0$, we get the modes
\begin{eqnarray}\label{eq:AdsDiracbagsol}
    \ell=\colvec{\chi\\\eta}=-e^{iEt}\,E\,\sqrt{r}\,\left\{J_{\frac{3}{2}}(Er)\,\colvec{e^{i\frac{\phi}{2}}\\-e^{-i\frac{\phi}{2}}}+iJ_{\frac{1}{2}}(Er)\,\colvec{e^{i\frac{\phi}{2}}\\e^{-i\frac{\phi}{2}}}\right\}\,,\nonumber\\
\end{eqnarray}
for $0\leq r\leq M^{-1}$. What we are interested in is $\ell(r=M^{-1})$ as $M\gg E$. In this limit, the $J_{1/2}$ dominates over the $J_{3/2}$, and we get
\begin{eqnarray}\label{eq:AdsDiracbagsolfull}
    \chi=e^{i\phi}\eta~~~,~~{\rm at}~r=M^{-1}\rightarrow0\,.
\end{eqnarray}
This is the charge-depositing ``bag'' boundary condition \eqref{eq:bcr0YL}.

\section{KK Decomposition}\label{app:partial wave decomp}
%
In this appendix we describe the 2D EFT reduction of the theory achieved by integrating out the $S^2$ submanifold of the 4D theory. We start by elaborating on the partial wave decomposition of the fermions and the photon, and subsequently laying out the full partition function containing all KK modes. Finally, we outline the implications of including higher partial waves in this work.

\subsection{Fermions}
The fermion KK decomposition is most naturally described in terms of Weyl fermions in an ${AdS}_2 \times S^2$ metric
\begin{eqnarray}
    g^{4D}_{\mu\nu} = {\rm diag}\pqty{\pqty{\frac{R}{r}}^2, \pqty{\frac{R}{r}}^2, R^2, R^2 \sin\theta}~~~,~~~ \mu,\nu \in \Bqty{t,r,\theta,\varphi} \, .
\end{eqnarray}
This corresponds to the Weyl rescaled $\psi\rightarrow (R/r)^{3/2} \psi$ left and right components of the 4D Dirac fermions in Section~\ref{sec:Abelian Instantons}. The 4D Weyl operator of a charged fermion in a magnetic monopole background \eqref{eq:dirac potential}, in ${AdS}_2 \times S^2$, is given by
\begin{eqnarray}\label{eq:LHGFrec}
    i \overline{\sigma}^\mu\pqty{\nabla_\alpha - i e Q A_{D \, \mu}} &=& \pqty{\frac{R}{r}}^{-1}\pqty{\pqty{\frac{R}{r}}^{-1/2} \pqty{- i \partial_t + \sigma_r \, \partial_r} \pqty{\frac{R}{r}}^{1/2} - \frac{\sigma_r (K + 1)}{r}} \, ,
\end{eqnarray}
where $K \equiv  \vb*{\sigma}\cdot\vb{L}+q \, \sigma_r$. Here $\sigma_r = \vb*{\sigma} \cdot \vu{r}$ and $\vb{L} \equiv \vb{r}\times\pqty{\vb{p}-e\vb{A}^{\rm mon}} 
 - q \vu{r}$ is the orbital angular momentum operator in the background of the Dirac monopole, or explicitly
\begin{eqnarray}\label{eq:L ang-mom}
    L_z &=& -i \partial_{\varphi}  - q\nonumber\\[5pt]
    L_x&=&\frac{1}{2}\left(L_++L_-\right)~~,~~L_y=\frac{1}{2i}\left(L_+-L_-\right)\, , \nn \\[5pt]
    L_\pm&=&e^{\pm i \varphi}\left[\pm \partial_\theta+i\cot\theta\partial_\theta-q\tan\left(\frac{\theta}{2}\right)\right]\,\nn, \\[5pt]
     L^2 & = & -\frac{1}{\sin^2 \theta} \bqty{ \sin\theta \partial_{\theta}\pqty{\sin\theta \partial_{\theta}} + \pqty{\partial_{\varphi} - i q (1-\cos\theta)}^2 } + q^2 \, ,
\end{eqnarray}
whereas the total angular momentum operator is defined as $\vb{J}\equiv\vb{L}+\frac{1}{2}\vb*{\sigma}$. Since $J^2$, $K$ and $J_z$ commute, they have a common eigenbasis. It is given by $\Omega^{(a)}_{qjm}\, ,~ a=1,2,3$, which are defined such that
\begin{eqnarray}\label{eq:Omega def}
    J^2\Omega^{(a)}_{qjm}&=&j(j+1)\,\Omega^{(a)}_{qjm}~~,~~J_z\Omega^{(a)}_{qjm}=m\,\Omega^{(a)}_{qjm} \, ,\nonumber\\[5pt]
    K\Omega^{(1)}_{qjm}&=&-(1-\tilde{\ell})\,\Omega^{(1)}_{qjm}~~,~~
    K\Omega^{(2)}_{qjm}=-(1+\tilde{\ell})\,\Omega^{(2)}_{qjm} ~~, ~~ j> j_{\rm min}\nn \\[5pt]
    &&K\Omega^{(3)}_{q j_{\rm min}m}=-\Omega^{(3)}_{q j_{\rm min}m} \, ,
\end{eqnarray}
where $\tilde{\ell}=\sqrt{(j+1/2)^2-q^2}$ and $ j_{\rm min}=|q|-1/2$. Furthermore, to diagonalize the kinetic term we perform a rotation of the higher partial waves to helicity $\sigma_r$ eigenstates
\begin{eqnarray}
   \Omega^{(\pm)}_{qjm} \equiv \frac{1}{\sqrt{2}} \pqty{ \Omega^{(1)}_{qjm} \mp \Omega^{(2)}_{qjm} } ~~, ~~
   \sigma_r \, \Omega^{(\pm)}_{qjm} = \pm \Omega^{(\pm)}_{qjm} ~~, ~~ \sigma_r \, \Omega^{(3)}_{qjm} = s_q \Omega^{(3)}_{qjm} \, .
\end{eqnarray}
where $s_q = {\rm sgn}(q)$. The explicit expressions for $\Omega^{(a)}_{qjm}$ are given in appendix~\ref{app:Omegas}. Note that they are not eigenfunctions of $L^2$.

The lowest partial wave $j=j_{\rm min}$ has a single Weyl spinor solution whereas the higher partial waves $j > j_{\rm min}$ have two Weyl spinor solutions. This determines the structure of the KK tower of the fermions, in particular each level $j > j_{\rm min}$ in the tower reduces to $2j+1$, Weyl fermions in the effective 2D theory, as opposed to the lowest partial wave, which reduces to $j_{\rm min}$ 2D Weyl fermions. The full partial wave expansion for $\psi_s$ a 4D Weyl fermion with charge $Q=s$ is
\begin{eqnarray}\label{eq:fermion PWD}
    \psi_s(x) = \sum_m P_q \chi_{m}(t,r) \frac{\Omega^{(3)}_{qjm}}{R} + \sum_{j > j_{\rm min}} \sum_m \bqty{\xi_{R \, s j m}(t,r) \frac{\Omega^{(+)}_{qjm}}{R} + \xi_{L \, sjm}(t,r) \frac{\Omega^{(-)}_{qjm}}{R} } \, ,
\end{eqnarray}
where $P_q =(1 - s_q \gamma_5)/2$.
Integrating over the angles we find the kinetic and mass terms of the ${AdS}_2$ Dirac fermions
\begin{eqnarray}
    S \supset \int d^2 x \sqrt{g_{\rm 2D}} \, \sum_m \, i P_q \overline{\chi}_{qm} \slashed{\nabla}_{\rm 2D} P_q \chi_{qm} + \sum_{j > j_{\rm min}} \sum_m i \overline{\xi}_{qjm} \pqty{\slashed{\nabla}_{\rm 2D} + \frac{\tilde{\ell}}{R}} \xi_{qjm} \, .
\end{eqnarray}

\subsubsection{Explicit Angular Eigenfunctions}\label{app:Omegas}
%
The eigenvalues of $L^2$ defined in \eqref{eq:L ang-mom} are monopole spherical harmonics 
\begin{eqnarray}
    Y_{qlm}(\theta,\varphi)=\sqrt{\frac{2l+1}{4\pi}}\mathcal{D}^{l*}_{m,-q}(\varphi,\theta,-\varphi)\,,
\end{eqnarray}
where the Wigner matrices ${\cal D}^l_{m,m'}$ are defined as \cite{Varshalovich:1988ifq}
\begin{eqnarray}
    \mathcal{D}^{l}_{m',m}(\upsilon,\sigma,\gamma)&=&e^{-i(m'\upsilon+m\gamma)}\,d^l_{m'm}(\sigma)\nonumber ~~~ , ~~~ d^l_{m'm}(\sigma) = \left\langle lm'|e^{-i\sigma J_y}|lm\right\rangle \, .
\end{eqnarray}
Using monopole spherical harmonics, one can form spinorial angular harmonics that are eigenstates of $J^2$ and $J_z$, as
\begin{equation}\label{eq: spinorial angular harmonics}
	\begin{aligned}
		\Phi^{(1)}_{q j m}(\theta, \varphi) = &  \pmqty{
			\sqrt{\frac{j+m}{2 j}} Y_{q, j-1 / 2, m-1 / 2}(\theta, \varphi) \\
			\sqrt{\frac{j-m}{2 j}} Y_{q, j-1 / 2, m+1 / 2}(\theta, \varphi)}\\
		\Phi_{q j m}^{(2)}(\theta, \varphi) = & \pmqty{
			-\sqrt{\frac{j-m+1}{2 j+2}} Y_{q, j+1 / 2, m-1 / 2}(\theta, \varphi) \\
			\sqrt{\frac{j+m+1}{2 j+2}} Y_{q, j+1 / 2, m+1 / 2}(\theta, \varphi)}\,.
	\end{aligned}
\end{equation}
The $Y_{q l m}$ are a complete orthonormal basis for wave functions of the spinless charge-monopole system \cite{Dray:1984gy}, and the coefficients in \eqref{eq: spinorial angular harmonics} are merely Clebsh-Gordan coefficients. Consequently, the set of all $\Phi^{(1)}_{q j m}, \Phi^{(2)}_{q j m}$ forms a complete orthonormal set \cite{Kazama:1976fm}.

The $\Phi^{(a)}_{q j m}$ are not yet eigenstates of $K$, but we can form linear combinations of them that are $K$ eigenstates,
\begin{equation}
	\begin{aligned}
			\Omega_{q j m}^{(1)} &=\frac{1}{2}\left(\sqrt{1+\frac{q}{j+1 / 2}}+\sqrt{1-\frac{q}{j+1 / 2}}\right) \Phi_{q j m}^{(1)} \\
			&-\frac{1}{2}\left(\sqrt{1+\frac{q}{j+1 / 2}}-\sqrt{1-\frac{q}{j+1 / 2}}\right) \Phi_{q j m}^{(2)},~~~~j>j_{\rm min} \\
			\Omega_{q j m}^{(2)} &=\frac{1}{2}\left(\sqrt{1+\frac{q}{j+1 / 2}}-\sqrt{1-\frac{q}{j+1 / 2}}\right) \Phi_{q j m}^{(1)} \\
			&+\frac{1}{2}\left(\sqrt{1+\frac{q}{j+1 / 2}}+\sqrt{1-\frac{q}{j+1 / 2}}\right) \Phi_{q j m}^{(2)},~~~~j>j_{\rm min} \\
			\Omega_{q, j_{\rm min}, m}^{(3)} & = \Phi^{(2)}_{q, j_{\rm min}, m}\,,
	\end{aligned}
\end{equation}
where $j_{\rm min}=\abs{q}-\frac{1}{2}$. Since $\Omega^{(1)}_{q j m}, \Omega^{(2)}_{q j m}, \Omega^{(3)}_{q j_{\rm min} m}$ are achieved via a unitary rotation from $\Phi^{(1)}_{q j m}, \Phi^{(2)}_{q j m}$, they remain a complete orthonormal set.

\subsection{Photon}\label{app:KKphot}
%
The partial wave expansion for a general vector field is given by
\begin{eqnarray}\label{eq:photon PWD}
    \pqty{4\pi R^2}^{-1/2} A^t(x) &=&  a^t(t,r)+ \sum_{j>0} \sum_m W^0_{jm}(t,r) Y_{jm}(\theta, \varphi) \, , \nn \\[5pt]
    \pqty{4\pi R^2}^{-1/2} \vb{A}(x) &=& a^r(t,r)\vu{r} + \sum_{j > 0} \sum_{m=-j}^j \sum_{l = j-1}^{j+1} W_{j l m}(t,r) \vb{Y}_{j l m}(\theta,\varphi) \, ,
\end{eqnarray}
where $\vb{Y}_{j l m}$ are vector spherical harmonics \cite{Arfken:2012:MMP:2331047, Hill:1954} and the factor $\sqrt{4\pi R^2}$ is put in by hand to ensure the 2D charge has the correct mass dimension. The vector spherical harmonics are defined as angular momentum eigenstates corresponding to the addition of the photon's spin and orbital angular momentum using Clebsch-Gordan coefficients
\begin{eqnarray}
    \vb{Y}_{j l m} = \sum_{m_l, m_h} \braket{j,m}{l,m_l;1,m_h} Y_{l m_l} \vu{e}_{m_h} ~~,~~ \vu{e}_{\pm 1} = \mp \frac{\vu{x} \pm i \vu{y}}{\sqrt{2}} ~~,~~ \vu{e}_0 = \vu{z} \, .
\end{eqnarray}
They are a complete orthonormal set, and obey the relation
\begin{eqnarray}
    \square \vb{Y}_{j l m} = - \frac{j(j+1)}{R^2} \vb{Y}_{j l m} = - \pqty{\mu^W_j}^2 \,\vb{Y}_{j l m}\, ,
\end{eqnarray}
where $\square \vb{A}$ is the full AdS$_2 \times S^2$ vector Laplacian. After the 2D reduction each $j>0,m$ in the tower contains two physical photon KK modes which we denote as $W^{(1,2) \, \alpha }_{jm}$.

\subsection{Kaluza-Klein Decomposition}\label{app:KK decomposition}

Plugging the partial wave decomposition for the fermions \eqref{eq:fermion PWD} and the photons \eqref{eq:photon PWD} into the action \eqref{eq: vanilla 4D action} and subsequently integrating over $S^2$ we find the full partition function in terms of the KK modes. This is given by
\begin{eqnarray}
    Z & = & \int {\cal D} \Phi \exp\Bqty{- (S + S_{\rm src})}\, , \nn \\[5pt]
    S & = & \int d^2 x \sqrt{g_{\rm 2D}} \, \Bqty{ 
    {\cal L}_a + {\cal L}_W + {\cal L}_\chi + {\cal L}_\xi + {\cal L}^{\rm int}_{a \chi} + {\cal L}^{\rm int}_{a \xi} + {\cal L}^{\rm int}_{W\chi} + {\cal L}^{\rm int}_{W\xi} } \, ,\\[5pt]
    {\cal D} \Phi & \equiv & {\cal D} a_\alpha \prod_{j m} {\cal D} W_{\alpha;jm} \prod_{m} {\cal D} \chi_m {\cal D} \overline{\chi}_m \prod_{sjm} {\cal D} \xi_{sjm} {\cal D} \overline{\xi}_{sjm}\, , \nn
\end{eqnarray}
where $g_{\rm 2D \, \alpha\beta} = (R/r)^2 \delta_{\alpha\beta}$ is the ${AdS}_2$ metric, ${\cal L}_X$ are the free Lagrangians of the KK modes, ${\cal L}^{\rm int}_{XY}$ are the interaction terms between $XY$ fields and $S_{\rm src}$ is the source term action. The Lagrangians are given by
\begin{eqnarray}
    {\cal L} & = & 
    \frac{1}{4} f_{\alpha\beta} f^{\alpha\beta} 
    + \sum_k \bqty{\frac{1}{4} F^{(k)\,\alpha\beta}_{jm} F^{(k)}_{jm\, \alpha\beta} + \frac{1}{2} \pqty{\mu_j^W}^2 W^{(k) \, \alpha}_{jm} W^{(k)}_{jm \, \alpha}} \nn \\[5pt]
    &&
    - i \sum_m \overline{\chi}_m \pqty{\slashed{\nabla} - i e_{\rm 2D}\gamma_5 \slashed{a}} \chi_m 
    - i \sum_{s\,j\,m} \overline{\xi}_{s j m} \pqty{\slashed{\nabla} + \mu^\xi_j - i s e_{\rm 2D}\slashed{a}} \xi_{sjm}
    \\[5pt]
    &&
    - e_{\rm 2D}\sum_{k \,j\, m_i} \, c_0(j, m_i) \overline{\chi}_{m_1} \gamma_5 \slashed{W}^{(k)}_{j m_2} \chi_{m_3} 
    - e_{\rm 2D}\sum_{s\, k\, j_i\, m_i}  s \, c(j_i,m_i)\,  \overline{\xi}_{s j_1 m_1} \slashed{W}^{(k)}_{j_2 m_2} \xi_{s j_3 m_3} \nn \\[5pt]
    {\cal L}_{\rm src} & = & J_{00}^\alpha a_\alpha + \sum_m \pqty{\overline{\eta}_{j_{\rm min} m} \chi_m + \overline{\chi}_m \eta_{j_{\rm min} m}}
    +\sum_{k jm} J^\alpha_{jm}  W^{(k)}_{jm \, \alpha} + \sum_{sjm} \pqty{\overline{\eta}_{sjm} \xi_{sjm} + \overline{\xi}_{sjm} \eta_{sjm} } \, ,\nn \\
\end{eqnarray}
where $f_{\alpha\beta} = \partial_\alpha a_{\beta} - \partial_\beta a_\alpha$, $F^{(k)}_{j m \, \alpha \beta} = \partial_\alpha W^{(k)}_{jm\, \beta}  - \partial_\beta W^{(k)}_{jm\, \alpha}$ and the sources $\eta_s, J^\alpha$ are decomposed to their respective partial wave expansions. The coefficients in the interaction terms involving $W^\alpha_{jm}$ are 3-j symbols
\begin{eqnarray}
    c(j_i, m_i) = \pmqty{j_1 & j_2 & j_3 \\ m_1 & m_2 & m_3} \, , ~~~~ c_0(j, m_i) = \pmqty{j_{\rm min} & j & j_{\rm min} \\ m_1 & m_2 & m_3}\, ,
\end{eqnarray}
which arise from the addition of angular momenta. Note that the 2D charge $e_{\rm 2D} = (4\pi R^2)^{-1/2} e$ has mass dimension $1$, reflecting the fact that it represents a charge in the 4D theory that is smeared over a 2-sphere of radius $R$.

Integrating out $W^{(k) \, \alpha}_{jm}$ yields high dimensional operators the lowest of which are $4$-fermi interactions. For the minimal monopole $c_0(j,m_i) = 0$, and thus no $\chi$ quartic interactions are generated, whereas for non-minimal monopoles such terms are generated. As discussed in Section~\ref{sec:setup and trunc} we remind the reader that our results can be fully trusted only for the minimal monopole since such quartics can spoil the topology. Neglecting the multi-fermion interactions generated by integrating out the photon KK modes we are left with
\begin{eqnarray}
    Z  & \simeq & \int {\cal D} a^\alpha \, \exp\pqty{\int d^2 x \sqrt{g_{\rm 2D}} \Bqty{- \frac{1}{4} f^{\alpha \beta} f_{\alpha\beta} + J_{00}^\alpha a_\alpha}} \, Z_\chi[a^\alpha, \eta_{j_{\rm min} m}, \overline{\eta}_{j_{\rm min} m}] \, Z_\xi[a^\alpha, \eta_{sjm}, \overline{\eta}_{sjm}] \, , \nn \\[5pt]
    Z_\chi & = & \int \prod_m {\cal D} \chi_m {\cal D} \overline{\chi}_m \, \exp\pqty{\int d^2 x \sqrt{g_{\rm 2D}} \, \sum_m \pqty{i  \overline{\chi}_m \,  \slashed{D}_5 \,  \chi_m  + \overline{\eta}_{j_{\rm min} m} \chi_m + \overline{\chi}_m \eta_{j_{\rm min} m } } } \, ,\nn \\[5pt]
    Z_\xi & = & \int \prod_{sjm}{\cal D} \xi_{sjm} {\cal D} \overline{\xi}_{sjm} \, \exp\pqty{\int d^2 x \sqrt{g_{\rm 2D}} \, \sum_{sjm} \pqty{i  \overline{\xi}_{sjm} \pqty{ \slashed{D}[s] + \mu^\xi_j} \xi_{sjm}  + \overline{\eta}_{sjm} \xi_{sjm} + \overline{\xi}_{sjm} \eta_{sjm} } } \, , \nn \\
\end{eqnarray}
where
\begin{eqnarray}
    \slashed{D}_5 = \slashed{\nabla} - i e_{\rm 2D}\gamma_5 \slashed{a}  \, , ~~~~~~~~ \slashed{D}[s] = \slashed{\nabla} - i s e_{\rm 2D}\slashed{a} \, ,
\end{eqnarray}
and $Z_\chi$ and $Z_\xi$ are the fermionic partition functions in the presence of a 2D gauge field. Since the $\xi_{sjm}$ spectrum does not contain zero modes, it does not contribute to the topological effects responsible for monopole catalysis. Integrating the $\xi_{sjm}$ modes leads to a standard Euler-Heisenberg expansion for massive fermions, generating perturbative $O(e^2)$ corrections to the processes we consider in this work. In contrast, the spectrum of $\chi_m$ contains zero modes which are integral to the non-perturbative physics of Abelian instantons.

All told, after neglecting the high partial waves we are left with the partition function of the \emph{Massless Axial Schwinger Model in} AdS$_2$.

\section{Fermion Determinants with Zeta Function Regularization}~\label{app:fermdet}
Here we provide the detailed computation of the non-zero mode fermion determinant ${\rm det}^{\prime}(i \slashed{D})$ for the axial Schwinger model in ${AdS}_2$. The anomaly of the axial Schwinger model is equal to that of the vector one. To see this, note that the Schwinger mass is generated by a self-energy-type fermion loop proportional to the square of the fermion charge. For this reason, switching to the axial model by changing the sign of the charge of the right-moving fermion does not alter the Schwinger mass or the anomaly. We thus only have to recap the standard computation of the Schwinger mass in the vector model, which we do using zeta-function techniques following \cite{Hortacsu:1979fg}, as well as the lecture notes \cite{Wipf:2015ln}. By definition we have
\begin{eqnarray}
    {\rm det}' (i\slashed{D}) = \prod_{\lambda_i \neq 0} \lambda_i\,.
\end{eqnarray}
Here the $\lambda_i$ are the nonzero eigenvalues of ${\slashed{D}}$, 
\begin{eqnarray}\label{eq:Dhatapp}
    i{\slashed{D}} u_i = \lambda_i u_i\,.
\end{eqnarray}
These eigenfunctions are orthonormal with respect to the inner product
\begin{eqnarray}
    \left\langle u_i | u_j\right\rangle&=&\delta_{ij} 
    ~~~ , ~~~ \left\langle a | b \right\rangle \equiv \int\,d^2x\, \sqrt{g_{\rm 2D}} \, \bar{a}_i(x)b_j(x)\,.
\end{eqnarray}
One may check that $\{i {\slashed{D}}, \gamma_5\} = 0$ meaning for every $\lambda_i >0$ we also have the eigenstate $\gamma_5 u_i$ with the eigenvalue $- \lambda_i$. For this reason, the determinant is given by
\begin{eqnarray}
    \log{\rm det}' (i{\slashed{D}}) =\log\left( \prod_{\lambda_i > 0} \lambda^2_i\right)=\frac{1}{2}\log{\rm det}' (-{\slashed{D}}^2)\,.
\end{eqnarray}
To calculate the determinant, we define an interpolating Dirac operators
\begin{eqnarray}\label{eq:Dkappa}  \slashed{D}_\kappa=e^{\gamma_5\sigma\kappa}\slashed{\nabla}e^{\gamma_5\sigma\kappa}\,,
\end{eqnarray}
where $a^\alpha \equiv e^{-1}\pqty{\epsilon^{\alpha\beta} \nabla_{\beta} \sigma + \nabla^{\alpha} \upsilon}$, so that $\slashed{D}_0=\slashed{\nabla}$ and $\slashed{D}_{1}=\slashed{D}$ up to an unphysical gauge rotation in $\upsilon$. We would like to evaluate the derivative $\frac{1}{2}\frac{d}{d\kappa}\log{\rm det}'\left(-{\slashed{D}}^2_\kappa\right)$ and consequently integrate it over $\kappa$ from $0$ to $1$. To this end, we use the zeta-function regularized version of this determinant (see, e.g. \cite{Vassilevich:2003xt}), namely
\begin{eqnarray}
    \frac{1}{2}\frac{d}{d\kappa}\log{\rm det}' \left(-{\slashed{D}}^2_\kappa\right)=-\frac{1}{2}\lim_{s\rightarrow0}\frac{d}{d\kappa}\frac{d}{ds}\zeta_{{\slashed{D}}^2_\kappa}(s)\,,
\end{eqnarray}
where the zeta function is defined as 
\begin{eqnarray}
\zeta_{{\slashed{D}}^2_\kappa}(s)\equiv\sum_{\lambda_i>0} \pqty{\lambda^2_{i,\kappa}}^{-s} \,.
\end{eqnarray}
Note that we put the $\kappa$ label on the eigenvalues since they correspond to ${\slashed{D}_\kappa}$. Acting with $\frac{d}{d\kappa}$ on the zeta function, we have
\begin{eqnarray}\label{eq:a9}
    \frac{d}{d\kappa}\zeta_{{\slashed{D}}^2_\kappa}(s)=
    -2s\sum_{\lambda_i>0} \lambda^{-(2s+1)}_{i,\kappa}\, \frac{d}{d\kappa}\lambda_{i,\kappa}=-2s\sum_{\lambda_i>0}\lambda^{-(2s+1)}_{i,\kappa} \,\left\langle u^{\kappa}_i|\pqty{\frac{d}{d\kappa} \,  i{\slashed{D}}_\kappa} |u^{\kappa}_i\right\rangle\,.
\end{eqnarray}
The last equality stems from the famous Feynman-Hellman formula. From the definition
\eqref{eq:Dkappa} we have
\begin{eqnarray}
    \frac{d}{d\kappa}\, i{\slashed{D}}_\kappa&=&\sigma\left\{\gamma_5,i\slashed{D}_{\kappa}\right\}\,,
\end{eqnarray}
so that
\begin{eqnarray}
    \frac{d}{d\kappa}\zeta_{{\slashed{D}}^2_\kappa}(s)=-4\sigma s\sum_{\lambda_i>0}\pqty{\lambda^2_{i,\kappa}}^{-s}\,\left\langle u^{\kappa}_i|\gamma_5|u^{\kappa}_i\right\rangle\,.
\end{eqnarray}
The zeta function is now evaluated in a standard way, by presenting it as the Mellin transform of the heat kernel,
\begin{eqnarray}\label{eq:hkp}
    \frac{d}{d\kappa}\zeta_{{\slashed{D}}^2_\kappa}(s)&=&-\frac{4\sigma s}{\Gamma(s)}\int_0^\infty\,dt\,t^{s-1}\,\sum_{\lambda_i>0}e^{t\lambda^2_i}\,\left\langle u^{\kappa}_i|\gamma_5|u^{\kappa}_i\right\rangle\nonumber\\[5pt]
    &=&-\frac{4 \sigma s}{\Gamma(s)}\int_0^\infty\,dt\,t^{s-1}\,{\rm tr}^\prime\,\left[e^{t{\slashed{D}}^2}\gamma_5\right]\,.
\end{eqnarray}
Here the trace on the first term on the right is taken over all eigenstates and spinor indices. Differentiating with respect to $s$, and taking the $s\rightarrow 0$ limit, we have
\begin{eqnarray}\label{eq:hk}
    \frac{1}{2}\frac{d}{d\kappa}\log{\rm det}' (-{\slashed{D}}^2_\kappa)&=&-\frac{1}{2}\lim_{s\rightarrow0}\frac{d}{d\kappa}\frac{d}{ds}\zeta_{{\slashed{D}}^2_\kappa}(s)=-\frac{1}{2}\lim_{s\rightarrow 0}\frac{d}{ds}\left\{-\frac{4 \sigma s}{\Gamma(s)}\int_0^\infty\,dt\,t^{s-1}\,{\rm tr}^\prime\,\left[e^{t{\slashed{D}}^2_{\kappa}}\gamma_5\right]\right\}\nonumber\\[5pt]
    &=&\lim_{s\rightarrow 0}\frac{d}{ds}\left\{\frac{2\sigma s}{\Gamma(s)}\int_0^\infty\,dt\,t^{s-1}\,{\rm tr}\,\left[e^{t{\slashed{D}}^2_{\kappa}}\gamma_5\right]\right\}-2\sigma\sum_{i}\left\langle u^{\kappa \, (0)}_i|\gamma_5|u^{\kappa \, (0)}_i\right\rangle\, , \nonumber \\
\end{eqnarray}
where $u^{\kappa \, (0)}_i$ are the zero-modes of the operator $i\slashed{D}_\kappa$. A standard textbook evaluation of the first term via the heat-kernel expansion gives
\begin{eqnarray}\label{eq:hk2}
   \lim_{s\rightarrow 0}\frac{d}{ds}\Bqty{\frac{2\sigma s}{\Gamma(s)}\int_0^\infty\,dt\,t^{s-1}\,{\rm tr}\,\left[e^{t{\slashed{D}}^2_{\kappa}}\gamma_5\right]} & = & - \frac{\sigma \kappa e_{\rm 2D}}{4\pi}\int d^2x \, \sqrt{g_{\rm 2D}} \,\epsilon_{\alpha\beta}f^{\alpha\beta} \nn \\[5pt]
   && ~~~~~~~~~ \times \bqty{\kappa-{\rm independent~terms}}\,.
\end{eqnarray}

The zero mode term on the right-hand side of \eqref{eq:hk} can be rearranged into a more familiar form by converting the $u^{\kappa \, (0)}_i$ to the known non-orthonormal zero modes 
\begin{eqnarray}
\chi^{\kappa \, (0)}_i = e^{\kappa \gamma_5 \sigma} v^{(0)}_i = \sum_l B^\kappa_{il} \, u^{\kappa \, (0)}_{l} 
\end{eqnarray}
where $v^{(0)}_i$ are $\kappa$ independent spinors and $B_{il}$ relates these spinors to the orthonormal basis of zero-modes. These definitions implies that
\begin{eqnarray}
    \frac{d}{d\kappa} \chi^{\kappa \, (0)}_l = \sigma\gamma_5\chi^{\kappa \, (0)}_l ~~~~ , ~~~~ \braket{\chi^{\kappa \, (0)}_l}{\chi^{\kappa \, (0)}_m} = \pqty{B^{\kappa \, \dagger} B^\kappa}_{lm}\, .
\end{eqnarray}
For this reason, we present the last term of \eqref{eq:hk} as
\begin{eqnarray}
    2\sigma\sum_{i}\left\langle u^{\kappa \, (0)}_i|\gamma_5|u^{\kappa \, (0)}_i\right\rangle &=& \sum_{l m} \pqty{B^{\kappa \, \dagger} B^\kappa}^{-1}_{lm} \bra{\chi_l^{\kappa (0)}} 2 \sigma \gamma_5 \ket{\chi_m^{\kappa (0)}}
    \nn \\[5pt]
    &=&\frac{d}{d\kappa} \tr \log (B^{\kappa\,\dagger} B^\kappa)  = \frac{d}{d\kappa}\log{\rm det}\,(B^{\kappa\,\dagger} B^\kappa)\,,
\end{eqnarray}
where $u^{\kappa \, (0)}_i=B^{\kappa}_{il}\chi^{\kappa \, (0)}_l$\,. Putting everything together and integrating over $\kappa$ from $0$ to $1$, we have
\begin{eqnarray}
    \log\pqty{\frac{{\rm det}' (i{\slashed{D}})}{{\rm det}' (i{\slashed{\nabla}})}}
    & = & -\frac{e_{\rm 2D}}{4\pi} \int d^2x \,\sigma \, \epsilon_{\alpha\beta}f^{\alpha\beta}-\log{\rm det}\,(B^\dagger B)\,.\nonumber\\
\end{eqnarray}
All-in-all, we get
\begin{eqnarray}\label{eq:hk3}
    \frac{{\rm det}' ({i \slashed{D}})}{{\rm det}' (i{\slashed{\nabla}})}&=&\frac{1}{{\det}(B^\dagger B)}e^{ - \frac{e_{\rm 2D}}{4\pi} \int d^2x \, \sqrt{g_{\rm 2D}}\,\sigma \, \epsilon_{\alpha\beta}f^{\alpha\beta}} = \frac{1}{{\det}(B^\dagger B)}e^{\frac{1}{2\pi}\int d^2x \,\sqrt{g_{\rm 2D}}\,\sigma\square\sigma}\nonumber\\[5pt]
    &=&\frac{1}{{\det}(B^\dagger B)} e^{-\int d^2x \,\sqrt{g_{\rm 2D}}\,\frac{e^2_{\rm 2D}}{2\pi}(\epsilon_{\alpha\gamma}a^\gamma)\pqty{g^{\alpha\beta}_{\rm 2D} -  \frac{\nabla^{\alpha}\nabla^{\beta}}{\square}} (\epsilon_{\gamma\delta}a^\sigma)}\,.
\end{eqnarray}
Note that the factor $1/\det(B^\dagger B)$ cancels out with the Jacobian of the non-orthonormal zero modes in the 't~Hooft vertex.

\bibliographystyle{JHEP}
\bibliography{Draft.bib}{}

\end{document}